\documentclass[11pt,Chicago]{uuthesis2e}
\usepackage {amsthm}

\fourlevels

\usepackage {amsmath}
\usepackage {amssymb}
\usepackage {bm}
\usepackage{natbib}
\usepackage{cases}
\usepackage{subcaption}
\usepackage{caption}
\usepackage {graphicx}
\usepackage{pdfpages}   %

\usepackage{url}
\usepackage{tabularx}
\usepackage{multirow}
\newcolumntype{L}{>{\raggedright\arraybackslash}X}

\newcounter{reprintpage}
\makeatletter
\newcommand{\includepaper}[1]{%
  \setcounter{reprintpage}{0}%
  \includepdf[pages=-, scale=0.795, offset=-6pt 0pt,
              pagecommand={\thispagestyle{headings}%
                           \stepcounter{reprintpage}%
                           \uuthesis@reprintlabel}]{#1}%
}
\makeatother
\usepackage{rotating}
\usepackage{titletoc}

\usepackage {uuthesis-2016-h}  %

\usepackage {mythesis}

\usepackage {uuthesis-chapterbib}

\makeatletter
\def \thebibliography #1%
{%
    \section{References}%
    \par \removelastskip
    \singlespace
    \par \removelastskip
    \vskip 7.5pt
    \list{[\arabic{enumi}]}%
    {%
        \settowidth \labelwidth {[#1]}%
        \leftmargin = \z@
        \itemsep = 3.5pt
        \parsep = 5pt
        \usecounter{enumi}%
    }%
    \def \newblock {\hskip .11em plus .33em minus -.07em}%
    \sloppy
    \clubpenalty = 4000
    \widowpenalty = 4000
    \sfcode`\. = 1000
    \relax
}
\makeatother

\makeatletter
\let\uuthesis@orig@float\@float
\def\@float#1{\@ifnextchar[{\uuthesis@float@p{#1}}{\uuthesis@orig@float{#1}[p]}}
\def\uuthesis@float@p#1[#2]{\uuthesis@orig@float{#1}[p]}
\makeatother

\newlength{\portraitwidth}
\makeatletter
\def\@CHAPTERdottedtocline#1#2#3#4#5{\pagebreak[3]%
  \ifnum #1>\c@tocdepth\relax \else
  \vskip .75em plus 1pt
  {%
    \leftskip #2\relax
    \rightskip \@tocrmarg plus 2em
    \parfillskip -\@tocrmarg
    \parindent #2\relax\@afterindenttrue
   \interlinepenalty\@M
   \leavevmode
   \@tempdima #3\relax \advance\leftskip \@tempdima \hbox{}\hskip -\leftskip
   \normalsize\bfseries
    #4\nobreak\normalsize\rmfamily
    \leaders\hbox{$\m@th \mkern \@dotsep mu.\mkern \@dotsep
       mu$}\hfill\nobreak \hbox to\@pnumwidth{\hfil\rmfamily #5}\par}\fi}
\renewcommand{\@pnumwidth}{2.4em}
\makeatother

\makeatletter
\newsavebox{\uuthesis@tabbox}
\newsavebox{\uuthesis@heldtab}
\newif\ifuuthesis@heldtab
\def\uuthesis@tabletype{table}
\let\uuthesis@orig@tabular\tabular
\let\uuthesis@orig@endtabular\endtabular
\def\tabular{\begin{lrbox}{\uuthesis@tabbox}\uuthesis@orig@tabular}
\def\endtabular{\uuthesis@orig@endtabular\end{lrbox}%
  \@ifundefined{@captype}{\uuthesis@puttab\uuthesis@tabbox}{%
    \ifx\@captype\uuthesis@tabletype
      \uuthesis@flushtab
      \global\setbox\uuthesis@heldtab\box\uuthesis@tabbox
      \global\uuthesis@heldtabtrue
    \else
      \uuthesis@puttab\uuthesis@tabbox
    \fi}}
\def\uuthesis@puttab#1{%
  \ifdim\wd#1>\linewidth
    \resizebox{\linewidth}{!}{\usebox{#1}}%
  \else
    \usebox{#1}%
  \fi}
\def\uuthesis@flushtab{%
  \ifuuthesis@heldtab
    \global\uuthesis@heldtabfalse
    \par\vskip\baselineskip
    \uuthesis@puttab\uuthesis@heldtab
    \par
  \fi}
\let\uuthesis@orig@endtable\endtable
\def\endtable{\uuthesis@flushtab\uuthesis@orig@endtable}
\@ifundefined{endsidewaystable}{}{%
  \let\uuthesis@orig@endsidewaystable\endsidewaystable
  \def\endsidewaystable{\uuthesis@flushtab\uuthesis@orig@endsidewaystable}}
\let\uuthesis@orig@tabularx\tabularx
\def\tabularx{%
  \let\tabular\uuthesis@orig@tabular
  \let\endtabular\uuthesis@orig@endtabular
  \uuthesis@orig@tabularx}
\makeatother

\makeatletter
\def\uuthesis@reprintlabel{%
  \edef\uuthesis@tmp{reprint:\thechapter:\thereprintpage}%
  \expandafter\label\expandafter{\uuthesis@tmp}}
\newcommand{\lotline}[3]{\l@table{\numberline{#1}#2}{#3}}
\newcommand{\reprinttable}[4]{\lotline{#1}{#4}{\pageref{reprint:#2:#3}}}
\def\uuthesis@pagefield#1#2#3\@nil{#2}
\newcommand{\uuthesis@pagenum}[1]{%
  \expandafter\ifx\csname r@#1\endcsname\relax ??\else
    \expandafter\expandafter\expandafter\uuthesis@pagefield\csname r@#1\endcsname\@nil
  \fi}
\newcommand{\reprintsection}[4]{%
  \addtocontents{toc}{%
    \protect\contentsline{section}{\protect\numberline{#1}#4}%
      {\uuthesis@pagenum{reprint:#2:#3}}{}}}
\newcommand{\reprintcredit}[2]{%
  \addtocontents{toc}{\protect \ifuuthesis@needtocspace
    \vspace{\protect \uuthesis@chaptersectionspace}
    \protect \fi \protect \global \protect \uuthesis@needtocspacefalse}%
  {\centering\normalsize\normalfont\doublespace
   Reprinted with permission from #1\par
   An ADA Accessible version of this chapter can be found at this link:
   \url{#2}\par}}
\def\listoftables{%
  \newpage
  \thispagestyle{empty}%
  \addcontentsline{toc}{chapter}{LIST OF TABLES}%
  \mainheading{LIST OF TABLES}\vskip -5.8pt
  
\reprinttable{2.1}{2}{7}{Fits from the hypothetical scenario where cloud
  areas are measured within the $100\times 100$ km GOES subdomains and fits
  are obtained over a subjectively defined linear region (Fig.~7).}
\reprinttable{2.2}{2}{8}{Estimated values of $\alpha$ (denoted as
  $\hat{\alpha}$) to cloud areas measured within the full domain and
  subdomains over the region where $n_\mathrm{truncated}/n_\mathrm{total}
  < 0.5$ as a function of the choice of including or excluding truncated
  clouds in the fit.}
\reprinttable{2.3}{2}{12}{Kolmogorov--Smirnov $p$ values, as in Fig.~3, for
  more combinations of bin location, sample size and $\alpha$.}
\reprinttable{2.4}{2}{14}{Rate of reliable estimates of the power law
  exponent $\hat{\alpha}$ for different linear-regression-based estimation
  methods (``estimators'') for 200 samples.}
\reprinttable{2.5}{2}{14}{Continuation of Table D1 for minimum bin counts of
  30 and 50.}

\reprinttable{3.1}{3}{13}{Comparison of the individual fractal dimension
  $D_\mathrm{i}$ with the three methods of calculating the ensemble fractal
  dimension $D_\mathrm{e}$: the box dimension, correlation dimension, and the
  product $D_\mathrm{e} = \beta D_\mathrm{i}$ (Eq.~7).}
\reprinttable{3.2}{3}{16}{Comparison of calculated values of $\beta$ for two
  methods of treating cloud holes as a function of reflectance threshold
  $R$.}
\reprinttable{3.3}{3}{17}{As in Table 1, but for a wider range of
  reflectance thresholds $R$.}

\lotline{4.1}{Various theories of atmospheric turbulence and corresponding
  parameter values for Eqn.~\ref{eq:2D structure function}.}
  {\pageref{tab:turbulence theories}}

\lotline{5.1}{Cloud geometric exponents as a function of reflectance
  threshold $R$ for STEAM, the giga-LES hydrodynamic model SAM, and MODIS
  observations.}{\pageref{tab:cloud geometry}}

\lotline{A.1}{Inputs to the STEAM algorithm.}{\pageref{tab:steam inputs}}
\lotline{A.2}{Compensation factors for the resolution dependence of mean
  absolute turbulon amplitudes, where $\Delta x$ is the final output grid
  resolution in the horizontal direction.}
  {\pageref{tab:interpolation compensation}}
\lotline{A.3}{Per-regrid retention factors in the anisotropic and isotropic
  regimes, measured under the cell-consistent interpolation convention.}
  {\pageref{tab:hop retention}}
\lotline{A.4}{Contents of a STEAM output file.}{\pageref{tab:output file}}
\lotline{A.5}{STEAM run sets of the comparison analysis.}
  {\pageref{tab:steam runs}}
}
\makeatother
\makeatletter
\@addtoreset{paragraph}{subsubsection}
\makeatother

\pdfcatalog{/Lang (en-US) /ViewerPreferences << /DisplayDocTitle true >>}
\input{glyphtounicode}
\pdfglyphtounicode{Digamma}{03DD}  %
\makeatletter
\newskip\uuthesis@titlegap
\def\titlepage{%
  \thispagestyle{empty}%
  \null\vfill
  \noindent\hspace{1em}%
  \begin{minipage}[c]{\minilength}%
    \centering
    \baselineskip=21pt
    {\HFmainhead\bfseries
     \begin{minipage}[c]{4.7in}\centering\@title\end{minipage}\par}%
    \vskip\uuthesis@titlegap
    by\par
    \@author\par
    \vskip\uuthesis@titlegap
    {\uuthesis@smallsinglespace
     A \@thesistype~submitted to the faculty of\\
     The University of Utah\\
     in partial fulfillment of the requirements for the degree of\par}%
    \vskip\uuthesis@titlegap
    \@degree\par
    \vskip\uuthesis@titlegap
    \@department\par
    The University of Utah\par
    \@submitdate\par
  \end{minipage}%
  \vfill
  \newpage}
\makeatother

\usepackage[bottom]{footmisc}

\makeatletter
\def\uuthesis@displayskips{%
  \abovedisplayskip 12pt\relax
  \belowdisplayskip 12pt\relax
  \abovedisplayshortskip 12pt\relax
  \belowdisplayshortskip 12pt\relax}
\g@addto@macro\normalsize{\uuthesis@displayskips}
\uuthesis@displayskips
\makeatother

\newlength{\textparindent}
\makeatletter
\long\def\@makefntext#1{\parindent\textparindent\noindent
  \hskip\textparindent\hbox{$^{\@thefnmark}$}#1}
\def\footnoterule{\hrule width 1.8in\kern-0.4\p@}
\makeatother

\sbox0{\ttfamily x}%
\DeclareFontShape{OT1}{cmtt}{m}{n}{<-9.5> fixed * [9.6] cmtt9 <9.5-11.5> cmtt10 <11.5-> cmtt12}{}

\makeatletter
\def\HFsectionHT{21pt}
\def\HFsubsectionHT{21pt}
\def\HFsubsubsectionHT{21pt}
\def\HFsubsectionSKIP{13.5pt}
\def\HFsubsubsectionSKIP{13.5pt}
\def\uuthesis@nohang#1{\noindent #1}
\def\section{\@Ustartsection{section}{3}{\z@}{15.5pt}{1sp}%
  {\centering\HFsection\bfseries\nohyphenation\let\@hangfrom\uuthesis@nohang}}
\def\subsection{\@Ustartsection{subsection}{4}{\z@}{\HFsubsectionSKIP}{1sp}%
  {\centering\HFsubsection\bfseries\nohyphenation\let\@hangfrom\uuthesis@nohang}}
\def\subsubsection{\@Ustartsection{subsubsection}{5}{\z@}{\HFsubsubsectionSKIP}{1sp}%
  {\HFsubsubsection\bfseries\nohyphenation\hsize=\mainheadingwidth}}
\def\HFchapterSKIP{0.16pt}                 %
\newskip\uuthesis@labeltitlefix            %
\newskip\uuthesis@consecutiveskip          %
\newif\ifuuthesis@afterchapter
\def\@Ustartsection#1#2#3#4#5#6{%
   \@testsizetrue\ifnum\c@secnumdepth=#2\relax \@testsizefalse \fi
   \if@noskipsec \leavevmode \fi
   \par\@tempskipa #4\relax
   \@afterindenttrue
   \ifdim \@tempskipa <\z@ \@tempskipa -\@tempskipa \@afterindentfalse\fi
   \ifuuthesis@afterchapter
     \global\uuthesis@afterchapterfalse
   \else
     \if@nobreak \everypar{}%
       \addvspace{\uuthesis@consecutiveskip}%
     \else
       \addpenalty{\@secpenalty}\addvspace{\@tempskipa}%
     \fi
   \fi
   \@ifstar
     {\@Ussect{#3}{#4}{#5}{#6}}{\@dblarg{\@Usect{#1}{#2}{#3}{#4}{#5}{#6}}}}
\def\uuthesis@chaptertitle#1{%
  \begin{center}%
    \vskip\uuthesis@labeltitlefix
    \parbox{\mainheadingwidth}{%
      \begin{center}%
        \HFchapter\bfseries
        \uppercase{{\nohyphenation #1}}%
      \end{center}}%
  \end{center}}
\def\@chapter[#1]#2{%
    \cleardoublepage
    \thispagestyle{empty}%
    \global \@topnum \z@
    \if@oneappendix
        \uuthesis@noise{\@chapapp:}%
        \refstepcounter{chapter}%
        \addcontentsline{toc}{chapter}{\thesisTOC{\@chapapp: #1}\protect \global \protect \uuthesis@needtocspacetrue}%
        \mainheading{\@chapapp}%
    \else
        \refstepcounter{chapter}%
        \edef \@tmp{\thechapter}%
        \ifx \@tmp \@optionA
          \addtocontents{toc}{\leavevmode \\[-1.2ex] \kern -1.5em\protect \contentsline{}{\textbf{APPENDICES}}{}{}}%
        \fi
        \ifx \@tmp \@optionONE
          \addtocontents{toc}{\leavevmode \\[-1.2ex] \kern -1.5em\protect \contentsline{}{\textbf{CHAPTERS}}{}{}}%
        \fi
        \addcontentsline{toc}{chapter}{\protect\numberline{\thechapter.}\thesisTOC{#1}\protect \global \protect \uuthesis@needtocspacetrue}%
        \uuthesis@noise{\@chapapp\space\thechapter.}%
        \mainheading{\@chapapp\space\thechapter}%
    \fi
    \uuthesis@chaptertitle{\MakeUppercase #2}%
    \@restorepar \@endpefalse
    \vskip \HFchapterSKIP
    \nobreak
    \@afterheading
    \global\@nobreakfalse
    \global\uuthesis@afterchaptertrue
    \everypar{\global\uuthesis@afterchapterfalse\clubpenalty\@clubpenalty\everypar{}}%
}
\def\prefacesection#1{\newpage
  \thispagestyle{empty}%
  \normalspace
  \addcontentsline{toc}{chapter}{\uppercase{#1}}%
  \mainheading{#1}\vskip\HFchapterSKIP}
\def\tableofcontents{%
    \newpage
    \thispagestyle{empty}%
    \mainheading{CONTENTS}%
    \par \removelastskip
    \singlespace
    \par \removelastskip
    \vskip -8.1pt
    \@starttoc{toc}%
    \uuthesis@noise{Contents.}}
\makeatother

\author                 {Thomas D. DeWitt}
\title                  {Weather and Climate Without Fluid Mechanics}
\thesistype             {dissertation}

\degree                 {Doctor of Philosophy}

\approvaldepartment     {Atmospheric Sciences}
\department             {Department of Atmospheric Sciences}
\graduatedean           {Darryl P. Butt}   
\departmentchair        {Anna Gannet Hallar}   

\committeechair         {Timothy J. Garrett}
\firstreader            {Steven K. Krueger}
\secondreader           {Court Strong}
\thirdreader            {Yi Zhang}
\fourthreader           {Graham Feingold}
\chairtitle             {Professor}

\submitdate             {December 2026}  
\copyrightyear          {2026}

\chairdateapproved      {}
\firstdateapproved      {}
\seconddateapproved     {}
\thirddateapproved      {}
\fourthdateapproved     {}

\begin{document}

\frontmatterformat
\titlepage
\copyrightpage
\dissertationapproval
\setcounter {page}     {2}             %
\preface    {abstract} {Abstract}
\tableofcontents
\setcounter{tocdepth}{3}   %
\listoftables    %

\maintext       %

\pagestyle{headings}

\chapter{Introduction} \label{sec:intro}

Few would argue that a fluid would be best modeled via the interactions of each individual molecule. What would be the point, given that the laws of fluid mechanics may be used? Fluid mechanics allows one to ignore almost all of the information that would be necessary to fully describe the state of a system---the positions and momenta of each individual particle---while, somehow, preserving the ``important'' information about bulk fluid properties such as pressure. Even with this loss of information, we retain the ability to predict future bulk properties of the system with a high degree of accuracy.

This situation, where the complete state of the system contains dramatically more information than we care about for any practical application, is broadly similar to our position when attempting to make weather or climate forecasts. The climate at a particular location in 2100 may be specified using just a few numbers for each variable: the mean, standard deviation, and perhaps a few higher order moments. But at present our only means of forecasting these statistics is to run a climate model over a 30-year interval, producing millions of simulated timesteps whose only purpose is to be averaged. This is analogous to running a molecular dynamics simulation and calculating each individual particle's momentum, only to average the momenta to obtain the bulk fluid pressure.

We should therefore ask whether any higher-level laws exist for the climate system that preserve the climatologically relevant information without requiring the specific sequence of meteorological states to be modeled. Such a relationship would be an example of ``emergence,'' analogous to how fluid mechanics emerges out of a sufficient number of interacting particles. Emergent relationships are found throughout science, typically consisting of a more complete, detailed, or small-scale ``micro'' description and a simplified, coarse-grained, large scale ``macro'' or ``emergent'' description. In some cases, a theoretical relationship between the small-scale description and the emergent description is known, such as the fact that pressure is proportional to the average kinetic energy of the molecules in an ideal gas. In other cases, an emergent description may be well understood and offer predictive power, even without any known coarse-graining relationship. 
Even the Navier-Stokes equations were known roughly two centuries before any explicit coarse-graining relationship was known that links particles to fluids. %

The simplification conferred by an emergent theory is its most important property. Broadly speaking, a core objective of science is to make the universe interpretable to human and human-like minds. Given that such minds have a bounded computational capability, any interpretable theory inevitably represents a lossy simplification of the universe. The question, then, is how can we best simplify a given system such that the resulting description is as simple as possible while also explaining a wide range of observations with as little error as possible?

The simplifications made by an emergent theory are not restricted to approximations in the mathematical laws. Often, the most important simplifications lie within the very concepts we use to describe the state of the system, which is called the \emph{ontology}. The ontology of a theory is just as important as the laws but can often be taken for granted. For fluids, the ontology is a set of continually varying fields representing temperature, pressure, velocity, and so on. There is no mathematical law that forces us to choose the specific concept of ``fluids''; rather, the laws assume their existence and properties, such as differentiability, and then specify their temporal evolution.

Both the ontology and the laws that make up an emergent theory can have a very different feel than their microscopic counterparts. Fields and particles are very different as purely mathematical objects: fields fill space with a continuous set of values, while a set of particles is a discrete collection of identical objects, each with only a two numbers (position and momentum) to characterize its state. The laws are correspondingly different, specifying either continuous, differentiable temporal evolution or a discrete set of elastic collisions. And furthermore, the Navier-Stokes are not a unique set of coarse-grained laws. If the system is in thermal equilibrium, thermodynamic fluctuation theory provides a more useful description of the likelihood of minute temperature fluctuations, this time using an entirely different mathematical framework. Instead of a time-dependent and deterministic theory, thermodynamic fluctuation theory describes a statistical ensemble over all time using only probabilities. Given the wide range of possible theories, we should not necessarily expect an emergent theory for weather and climate to have the same properties as fluids. Emergent laws describing Earth's albedo as a function of climate state, for example, may not even contain time as a variable.

One might wonder whether emergent laws are needed given the success of numerical weather prediction and general circulation models (GCMs), which numerically integrate the equations of fluid motion. With these models being ubiquitous in the atmospheric sciences, it is perhaps easily forgotten that it is not possible to use the equations of fluid mechanics in isolation, even if we only cared to simulate the wind field. For a hydrodynamic simulation to be numerically stable, it must have a small enough grid size that nonlinear interactions are negligible at the grid scale. In the atmosphere, this scale is typically millimetric---implying that today's most advanced kilometer-scale models resolve less than half of the full range of scales over which nonlinear dynamics occur in the atmosphere, which is between $\sim 10^{7}$\,m and $\sim 10^{-3}$\,m.

To overcome this limitation, all atmospheric models include an additional set of physics, one that models large numbers of unresolvable small-scale interactions using only large-scale properties. This is called a ``parameterization''. Even a next-generation GCM with grid resolution of 3\,km horizontally and 0.2\,km vertically \citep{shukla2009} must parameterize $\sim 10^{18}$ millimeter-scale turbulent circulations within each individual grid box. The parameterization must predict the overall effect of these small-scale circulations on the grid scale flow using only grid scale quantities. In general, GCMs also must parameterize numerous other small-scale interactions, ranging from cloud droplet interactions to radiative transfer.

As a representation of the overall effect of some large number of small-scale interactions, parameterizations are in fact a set of emergent laws. But if the details of why or how the parameterization works are viewed as instrumental, tuned simply to ensure the large-scale resolved features match observations, there is no guarantee that the resulting emergent laws will serve to make the atmosphere more interpretable to the human mind. Although interpretability and predictive power of a theory have often aligned historically, there is no guarantee that a theory that makes better predictions will necessarily be more interpretable \citep{krakauer2023}.
Modern machine learning-based approaches have already begun to decouple predictive power from human interpretability, a trend that will likely continue as the information processing capabilities of neural networks outpace those of the human mind. %

Of course, prediction is a worthy goal in itself, and a society may reasonably choose a less interpretable but more predictive method for day-to-day weather forecasts. But we should be careful to not wholly define science as the practice of developing methods to enable prediction. Equally important is the goal of understanding and interpretability for human minds. Emergent laws for weather and climate will be increasingly crucial for interpretability even if they are not competitive on predictive power.

\section{Multiplicative cascades}

Though it is still nascent, perhaps the most well developed emergent theory of atmospheric dynamics grew out of classical turbulence theory in the 1980s. The goal was to adapt and generalize the concept of a turbulent cascade to account for atmospheric motion. A classical turbulent cascade, as first articulated by \citet{richardson1926}, envisions a turbulent flow as a superposition of a large number of turbulent circulations of varying sizes, shapes, and strengths. The circulations are imagined to be continually breaking up into smaller and smaller circulations or ``eddies'', a process called a ``cascade''. By appealing to conservation laws, the strength and number of circulations may be constrained as a function of their size. In the simplest case, the relevant conserved quantity is the kinetic energy dissipation rate, which leads to the famous $-5/3$ spectrum of kinetic energy proposed originally by \citet{kolmogorov1941}. In this early theory, the power spectrum was assumed to be isotropic, i.e. to follow the $-5/3$ law in all directions, and the distribution of dissipation rates was assumed to be uniform, which is classically termed ``homogeneous'' turbulence.\footnote{Later generalizations allowed for thin-tailed, e.g. Gaussian, distributions of dissipation rates, but these still produce nonintermittent and homogeneous flow. ``Homogeneous'' is used here in its classical sense of nonintermittent flow, as opposed to the more precise definition of statistical translational invariance.}

Although atmospheric turbulence is still often assumed to be homogeneous and isotropic at the small scale, the observations we present here suggest that neither assumption is accurate. As an alternative to Kolmogorov's theory, in a series of papers through the 1980s and 1990s Shaun Lovejoy and Daniel Schertzer proposed and empirically validated a mathematical model that drops both assumptions. Anisotropy, rather than isotropy, can be understood using a mathematical framework called ``Generalized Scale Invariance,'' (GSI) which redefines the notion of distance that is felt by atmospheric motions. Effectively, local energy dissipation rates can be thought of as stretching the space in which atmospheric dynamics operate, analogous to how mass stretches space-time in General Relativity. Intermittency, where the distribution of field values are heavy-tailed rather than Gaussian, replaces homogeneity through an interpretation of Richardson's cascade using a mathematical construction called a ``multiplicative cascade'' and an associated mathematical formalism termed ``Universal Multifractals'' (UM).

Taken as a whole, the GSI and UM framework is a set of emergent laws for atmospheric statistics. The framework does not explicitly reference the underlying laws of fluid mechanics but nonetheless constrains the statistics and even offers some predictive power over Earth's atmosphere, much as the laws of fluid mechanics predict Earth's weather without reference to particles. Many of the predictions of GSI and UM have been observed in the real atmosphere. Anisotropy following GSI predictions has been found for aerosol concentrations \citep{lilley2004}, the wind field \citep{pinel2012}, and cloud aspect ratios \citep[][their results are interpreted in Appendix \ref{sec:guillame}]{guillaume2018}. Multiplicative intermittency following UM predictions has been found for the wind field \citep{Hovde2011}, GOES imagery \citep{lovejoy1990}, and rain rates \citep{schertzer1987}.\footnote{For a more complete review, see \citet{lovejoy2023}.} Here, in Chapters \ref{sec:metrics} and \ref{sec:sondes}, we come to a similar conclusion: the predictions of multiplicative and anisotropic cascades often serve as a close fit to many observables.

Although the mathematics appear to reproduce realistic statistics, the frameworks remain difficult to interpret physically. One issue is that anisotropic universal multifractals lack an explicit ontology. To illustrate the problem, we should first overview the basic mathematical motivation behind the GSI and UM frameworks at a high level and defer specifics to Chapters \ref{sec:sondes} and \ref{sec:steam}.

\begin{figure}[htbp]
\centering
\includegraphics[width=\textwidth]{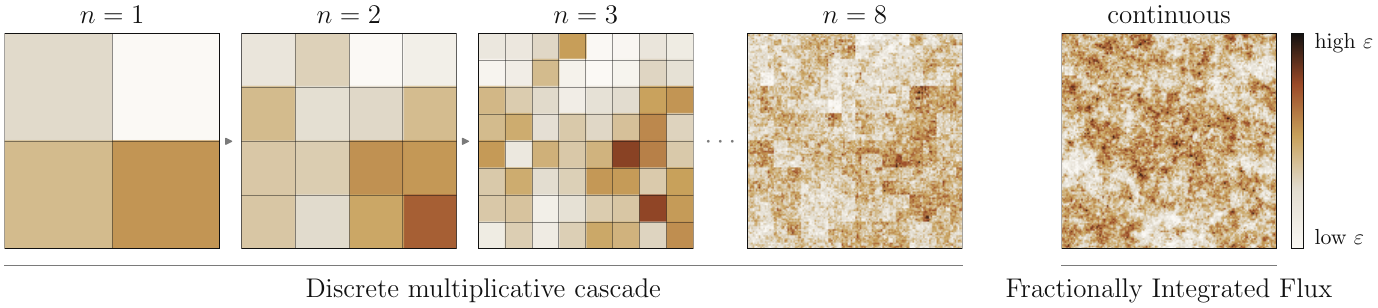}
\caption[Discrete multiplicative cascade and its continuous counterpart]{A
multiplicative cascade in two dimensions. The first four panels show the
conserved flux $\varepsilon$ after $n = 1$, 2, 3, and 8 steps of the
recursive construction described in the text. At each step, every block is
divided into four quadrants, and each quadrant is multiplied by an
independent random factor $\mu_i$ drawn from a lognormal distribution with
unit mean.
By $n = 8$ individual quadrants are too fine to see, but the rectilinear
patchwork they leave behind remains visible, motivating \citet{schertzer1987} to introduce the Fractionally Integrated Flux construction
shown in the rightmost panel. }
\label{fig:cascade}
\end{figure}

The simplest starting point is a one-dimensional discrete multiplicative cascade, representing some scalar field such as the kinetic energy dissipation rate $\varepsilon$. The cascade is constructed as follows. First, start with a uniform field with all values equal to 1 and spanning from 0 to $L$. Divide up the field into two equal segments of size $L/2$, and then draw two independent random factors $\mu_i$ from an appropriate distribution of choice, for example a lognormal with unit mean and standard deviation. Multiply each segment by the value drawn for $\mu_i$. Then, repeat the process recursively by dividing each new segment in half again, such that each new segment has length $L/4$, draw new random values $\mu_i$ from the same distribution for each of the four segments, multiply their values and repeat. At each step $n$, the next field values are $\varepsilon_{n+1} = \varepsilon_n\mu_i$ at each point. The process is continued over some wide range of scales that represent the inertial range of scales in turbulence.
This process may be easily extended to multiple dimensions as shown in Fig. \ref{fig:cascade}. For two dimensions, simply divide a square field into four equal-sized quadrants and multiply each quadrant by a random factor $\mu_i$ and then proceed recursively for each quadrant.

It is easy to see the analogy between a multiplicative cascade and Richardson's conceptual cascade. The square multiplicative perturbations represent the effect of an eddy at a particular scale, and the cumulative field is a superposition of many such ``eddies'' of varying size and strength. However, resulting fields are highly artificial in appearance due to the ``sharpness'' of the eddy shape. This sharpness appears because, at any given level during the cascade, field values are uniform within a quadrant but abruptly jump at the quadrant boundaries. To eliminate these discontinuities, \citet{schertzer1987} proposed the Fractionally Integrated Flux model, which replaces the discrete cascade construction by a fractional integral. Although the results are more visually realistic (Fig. \ref{fig:cascade}, rightmost panel) and the statistics match observations, the construction method is even more divorced from Richardson's intuitive idea of a superposition of eddies.

A key property of multiplicative cascades is that the field values are \emph{multiplied} by the random factors. This multiplicative aspect is how the resulting fields become heavy-tailed and intermittent, matching observations. But consider the ontological implications of multiplication. In the simplest example, we can consider the starting field as representing the mean state of the system, and a given realization of the multiplicative cascade as representing a particular state of the field at one instant in time. This is because the multiplicative factors all have mean value 1, so the multiplicative perturbations do not change the ensemble mean over a sufficient number of individual realizations. 

If the mean state is to retain the same dimensions as a given realization, then the multiplicative perturbations must be dimensionless. The ``eddies'' of a multiplicative cascade cannot have units of velocity, contradicting the dimensional arguments used by Richardson.
We must instead interpret an eddy as an \emph{operation} done \emph{to} the field, not a constituent \emph{of} the field. A multiplicative cascade does not supply any constituent element at all, leaving the question of what might cause the multiplicative perturbations to occur unanswered.
Physical interpretability becomes worse with the Fractionally Integrated Flux model, because the cascade is replaced by an integration of fractional order, a mathematical operation that is well known to lack any clear physical interpretation even when the resulting statistics match observations \citep{podlubny1999}.

Anisotropy is also quite challenging. In GSI, anisotropy is implemented by redefining the notion of distance such that a multiplicative cascade operates in a non-Euclidian space. In effect, deep convection warps space-time in a mathematically similar way to how a black hole warps space-time. Despite the obvious disanalogy, the predictions of GSI do, in fact, appear supported by observations of the real atmosphere, as we show in Chapter \ref{sec:sondes}. Although non-Euclidian metrics are mathematically elegant, they are highly counter-intuitive, and this seems to have dissuaded theory-building on top of GSI despite its empirical success: Chapter \ref{sec:sondes} was the first empirical study of the predictions of GSI by someone other than the originators of the theory and their collaborators, even though the theory was first proposed 41 years ago.

Anisotropic multiplicative cascades are simultaneously the most predictive set of laws that currently exist for atmospheric science, while also lacking any plausible physical interpretation as an emergent theory. One could call this a pedagogical problem, as any framework that offers no physical picture is difficult to learn. But such a theory is also difficult to build upon because it is difficult to develop intuition for. A theory that lacks an intuitive foundation not only makes learning the theory difficult, but also limits its theoretical development and, consequently, its predictive power.

The path forward is neither to ignore anisotropic and multiplicative cascades nor to insist on keeping their mathematical formulation in its current form. We might plausibly build new, emergent theories for atmospheric dynamics that have a clear and physically interpretable ontology, but whose construction is motivated or, ideally, provably equivalent to UM and GSI constructions in at least some idealized cases. In this way, decades of mathematical and empirical work can be built upon rather than discarded, and progress might be made toward making emergent laws more practically useful in the atmospheric sciences. The goal of this Dissertation is to show that such a theory is plausible and to take the first steps toward building one.

\section{Arc of the dissertation}

This dissertation contains published and unpublished work that builds on a foundation laid by the M.S. thesis \citep{dewitt2023}. The work splits into two general strands. The first argues that scale invariance exists as a broad emergent regularity across multiple aspects of the atmospheric system, and that such a regularity may serve as a foundation for an emergent theory of atmospheric dynamics. This strand began in the thesis \citet{dewitt2023} and publication \citet{dewitt2024}, which found that cloud size distributions are scale invariant independently of climate state. Here, Chapter \ref{sec:finite-domains}, published in Atmospheric Chemistry and Physics \citep{dewitt2024b} and reproduced here, builds on this cloud size scaling work by showing that, even when the underlying cloud size distribution is scale invariant, a finite measurement domain could introduce a spurious scale break in the observed distribution. Although the mechanism is straightforward, it is typically not corrected and may explain past inconsistencies in reported size distributions. 

Chapter \ref{sec:metrics}, published in Atmospheric Chemistry and Physics \citep{dewitt2026} and reproduced here, moves beyond size distributions to fractal metrics which quantify cloud shape in addition to cloud size. Measurement biases of a more subtle nature, this time relating to the treatment of cloud holes, appear again to obfuscate observations of scale invariance even when the underlying clouds themselves respect scaling symmetries. We argue that the proposed fractal metrics are superior to more common methods used for comparing meteorological regimes and evaluate simulations, as they are based on fundamental scaling symmetries rather than subjectively defined and discrete categories that are more popular.

Chapter \ref{sec:sondes} moves beyond clouds to examine scale invariance in radiosonde and dropsonde observations of the wind field \citep{dewitt2025preprint}. We compare observed wind spectra to a range of distinct theories, including the dominant ``mesoscale transition'' paradigm, where distinct small-, medium-, and large-scale regimes are expected to break scale invariance in the atmospheric kinetic energy spectrum. Using a compilation of three stringently quality-controlled dropsonde and radiosonde datasets, we show that observations are in fact scale invariant over much of the observed range, and further that observed exponents are incompatible with this dominant paradigm. Calculated exponents appear consistent with the anisotropic theory of turbulence put forward by Lovejoy and Schertzer \citep{lovejoy1985,schertzer1985}. We emphasize that prior analyses that considered purely horizontal kinetic energy spectra cannot easily distinguish the two paradigms due to subtle measurement issues. When evidence derived from both horizontally separated measurements and vertically separated measurements, the ``transition'' paradigm becomes difficult to square with observations.

In the past, scale invariance has largely been a purely descriptive property of Earth's atmosphere and can therefore be seen as an observational curiosity rather than an important physical constraint on the dynamics. Multiplicative cascades already show that scale invariance is not limited to description but can be used as a principle around which explicit atmospheric simulations can be made, as shown in Fig. \ref{fig:cascade}. However, such simulations remain simplistic and are typically univariate, translationally invariant, and do not have explicit boundary conditions. In Chapter \ref{sec:steam} we propose a model that addresses these limitations and allows for simulations of a tropical convective atmosphere. The central idea is to turn Richardson's concept of an ``eddy'' into a formal constituent element of the model, which we term a ``turbulon.'' Unlike an eddy, a turbulon has an exact mathematical form and contains components that represent perturbations in all fields making up a turbulent flow. The resulting ``Superposition of Turbulons and Eddies Atmospheric Model'' (STEAM) produces three-dimensional fields of cloud condensate, temperature, water vapor, and pressure which broadly reproduce statistics obtained from state-of-the-art hydrodynamic LES models, and at a vastly reduced computational expense. 

What is not shown here is a suite of computational and visual tools developed alongside the publications below. The Python package \verb|objscale| \citep{dewitt2026objscale} implements methods developed in Chapters \ref{sec:finite-domains} and \ref{sec:metrics} for the computation of scale invariant statistics of cloud objects in two-dimensional images. A second Python package \verb|scaleinvariance| \citep{dewitt2026scaleinvariance} is the first general-purpose toolkit for simulation and analysis of geophysical multifractal fields, containing GPU- and CPU- optimized algorithms for simulation\footnote{An interactive visualization of the simulation capabilities of \texttt{scaleinvariance} is available at \texttt{thomasddewitt.com/thought-cloud/multifractal-explorer}.} and analysis of multifractal fields. The cloud visualization toolkit \verb|cloudyview| \citep{dewitt2026cloudyview} allows for real-time cloud field visualization in a browser,\footnote{The \texttt{cloudyview} Soar visualization engine is available at \texttt{thomasddewitt.com/soar}.} enabling one to fly through STEAM- or LES- simulated three-dimensional cloud fields as if in a video game. These packages and web-based visualizations were instrumental for developing the intuition behind the formal methods described here, and could serve as invaluable pedagogical tools to others as well.

\setlength{\bibhang}{0in}
\bibliographystyle{ametsocV6}
\bibliography{sources}

\chapter{Finite domains cause bias in measured and modeled distributions of cloud sizes} \label{sec:finite-domains}

\reprintcredit{DeWitt, T.~D. and Garrett, T.~J.: Finite domains cause bias
  in measured and modeled distributions of cloud sizes, \emph{Atmospheric
  Chemistry and Physics}, 24, 8457--8472, 2024,
  \url{https://doi.org/10.5194/acp-24-8457-2024}.
  \copyright~Author(s) 2024. Distributed under the Creative Commons
  Attribution 4.0 License.}{https://doi.org/10.5194/acp-24-8457-2024}

\reprintsection{2.1}{2}{1}{Introduction}
\reprintsection{2.2}{2}{3}{Fitting power law distributions to empirical data}
\reprintsection{2.3}{2}{5}{How a finite domain changes measured size distributions}
\reprintsection{2.4}{2}{9}{Conclusions}
\reprintsection{2.5}{2}{11}{Appendix A: Statistical variability in histogram bin counts}
\reprintsection{2.6}{2}{12}{Appendix B: Correction algorithms for domain truncation effects}
\reprintsection{2.7}{2}{13}{Appendix C: Validation of exponential distributions of percolation clusters}
\reprintsection{2.8}{2}{14}{Appendix D: Tables of linear regression failure rates}
\reprintsection{2.9}{2}{15}{References}

\includepaper{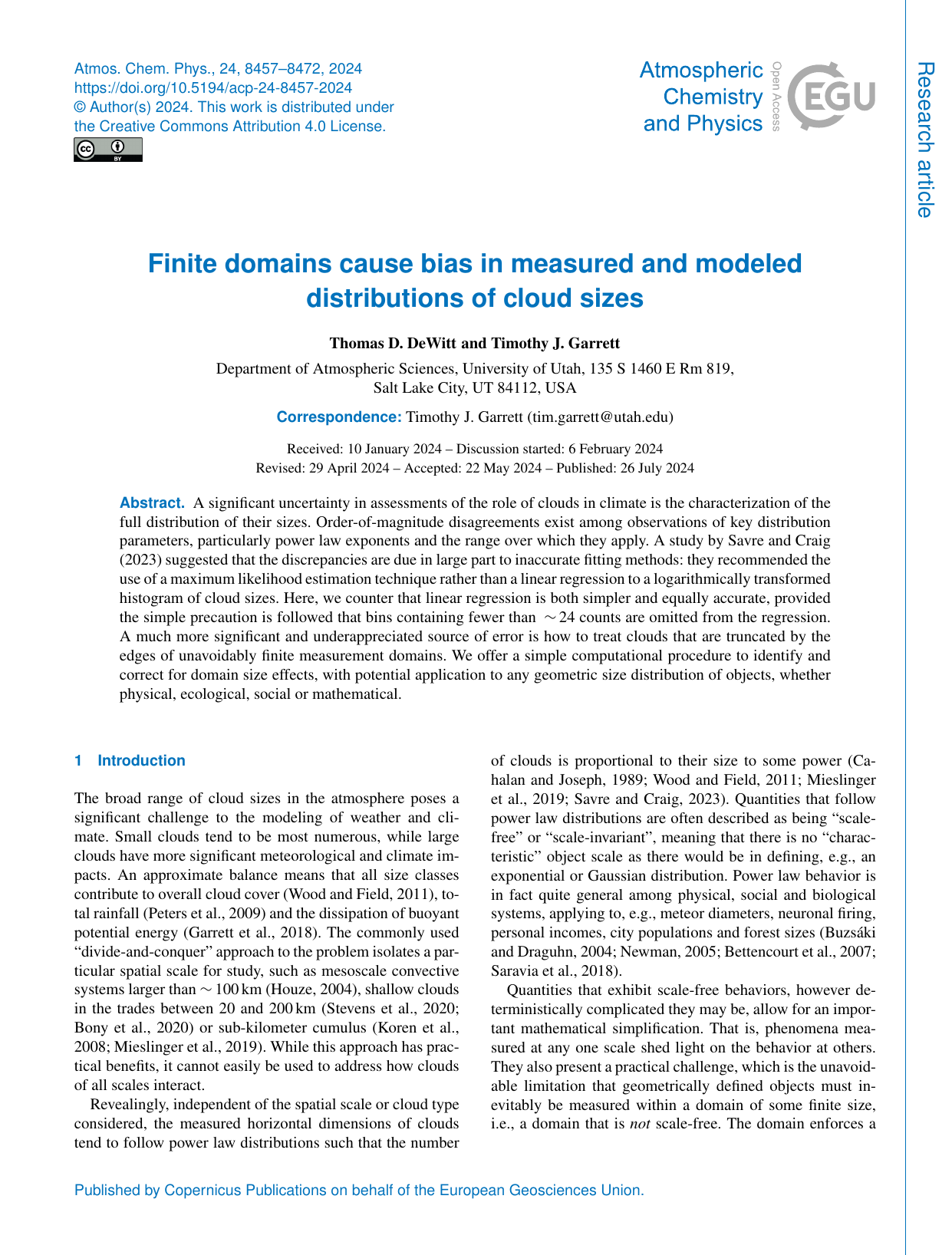}

\chapter{Toward less subjective metrics for quantifying the shape and organization of clouds} \label{sec:metrics}

\reprintcredit{DeWitt, T.~D., Garrett, T.~J., and Rees, K.~N.: Toward less
  subjective metrics for quantifying the shape and organization of clouds,
  \emph{Atmospheric Chemistry and Physics}, 26, 6951--6971, 2026,
  \url{https://doi.org/10.5194/acp-26-6951-2026}.
  \copyright~Author(s) 2026. Distributed under the Creative Commons
  Attribution 4.0 License.}{https://doi.org/10.5194/acp-26-6951-2026}

\reprintsection{3.1}{3}{1}{Introduction}
\reprintsection{3.2}{3}{2}{Datasets}
\reprintsection{3.3}{3}{3}{Determination of the individual fractal dimension $D_\mathrm{i}$}
\reprintsection{3.4}{3}{7}{The ensemble fractal dimension}
\reprintsection{3.5}{3}{12}{Conclusions}
\reprintsection{3.6}{3}{12}{Appendix A: The relationship between the individual and ensemble fractal dimension}
\reprintsection{3.7}{3}{15}{Appendix B: Comparison of the perimeter distribution exponent with and without cloud holes}
\reprintsection{3.8}{3}{17}{Appendix C: Parameters for a wider range of reflectance thresholds}
\reprintsection{3.9}{3}{19}{References}

\includepaper{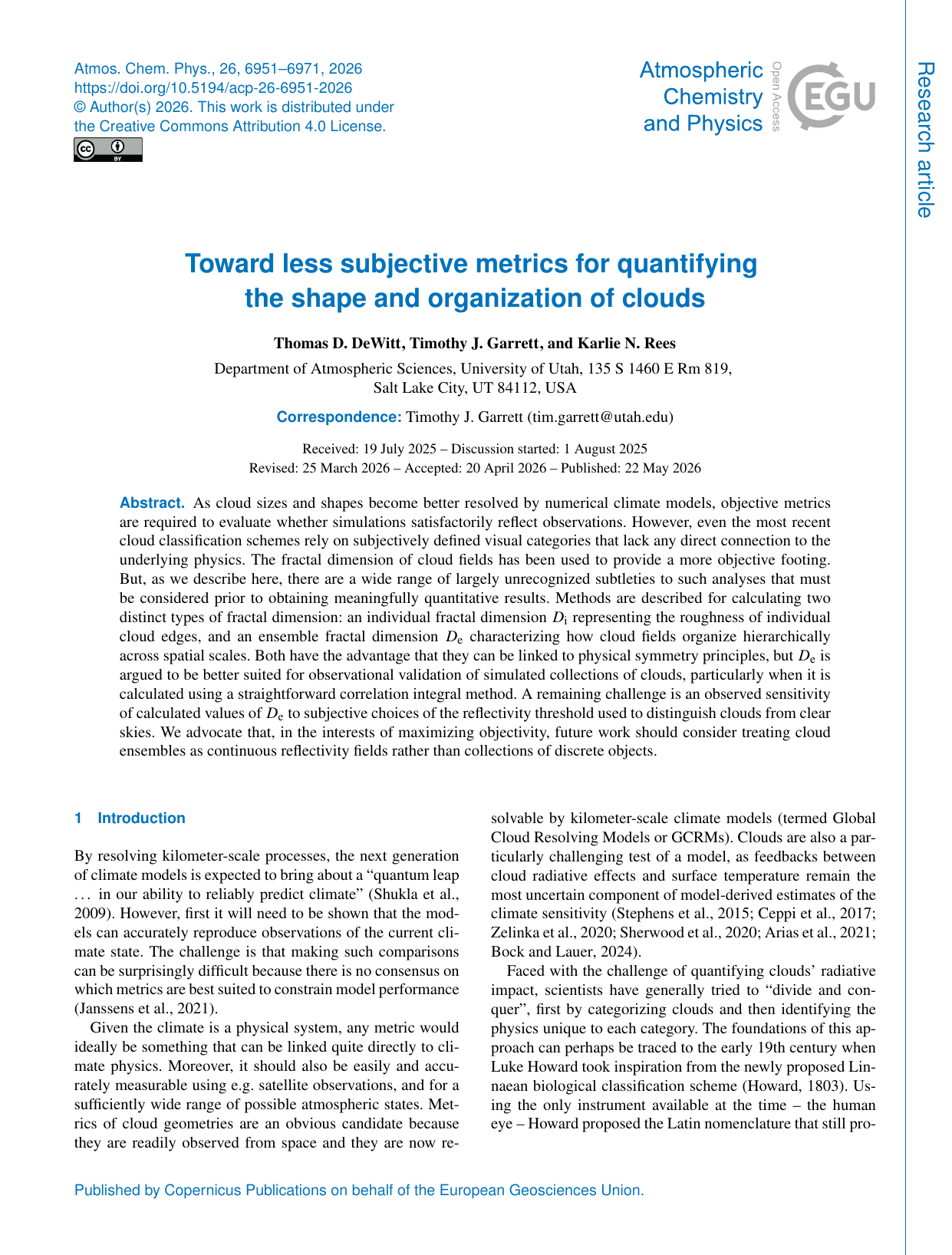}

\renewcommand*{\mainheadingwidth}{5in}
\chapter[Global sonde datasets do not support a mesoscale transition in the turbulent energy cascade]{Global sonde datasets do not\\ support a mesoscale transition\\ in the turbulent energy cascade} \label{sec:sondes}
\renewcommand*{\mainheadingwidth}{4.5in}
\section{Introduction}
\label{sec:introduction}

The dynamics of Earth's atmosphere are commonly characterized as being governed  by three-dimensional Kolmogorov turbulence at the smallest dynamical scales, gravity waves at the mesoscale, and quasi-geostrophic turbulence at the largest scales \citep{charney1971,gage1986,lindborg1999}. Combined, this ``transition'' paradigm partitions these regimes  according to their respective wavenumber spectra for kinetic energy, where an isotropic -5/3 exponent at small scales (e.g. $E\left(k\right) \propto k^{-5/3}$; \citet{kolmogorov1941})  gives way to a two-dimensional -3 exponent at large scales (e.g. $E\left(k\right) \propto k^{-3}$; \citet{charney1971}).
\citet{schertzer1985} proposed a very different paradigm that  represents the full range of atmospheric scales, from the  millimeter to the planetary, as being governed by a single theory of anisotropic turbulence. Here, motions are separable according to the direction of the flow, given that the gravitational force acts in only one  direction, rather than the scale of the flow. The spectral exponents  were  theoretically predicted to be $-5/3$ in the horizontal and $-11/5$ in vertical.

The challenge observationally has been that aircraft measurements are most easily performed along isobars where the exponents may differ from their isoheight counterparts. When calculated along isobars, both paradigms predict a transition in the spectral exponent from $-5/3$ to some larger value at scales of hundreds of kilometers. For quasi-geostrophic turbulence, the large-scale isobaric exponent is -3 \citep{charney1971}, while for Lovejoy-Schertzer turbulence, it is -2.4 \citep{lovejoy2009}. 

The question of which paradigm is best supported by observed isobaric spectra has been the subject of considerable debate \citep{lovejoy2009,lindborg2010,frehlich2010b,schertzer2012}. Although there is a consensus that some type of transition is present \citep{nastrom1983,nastrom1984,nastrom1985,gao1998,cho2001,nielsen1967,julian1970,boer1983}, %
a quantitative analysis that includes a statistical  fit to observations is rarely performed, and so it is not possible to conclusively discriminate which theory more closely reflects the true nature of turbulence. Beyond  this isobaric debate, the Lovejoy-Schertzer paradigm has been almost entirely overlooked. Setting aside studies by the authors who originated the idea \citep{lovejoy2007,pinel2012}, the key prediction of the directional dependence of the exponent has not  been tested --  despite it having been  proposed 40 years ago.

Fundamentally, the dichotomy is between two very different conceptual understandings of how air moves in the atmosphere. The commonly assumed transition paradigm argues  that the dynamics are governed by physics that depends on spatial scale, whereas the Lovejoy-Schertzer paradigm proposes a single dynamical mechanism that governs atmospheric dynamics regardless of spatial scale.  If the atmosphere can be shown to obey Lovejoy-Schertzer scaling, the tantalizing possibility is offered that observations at one scale and a simple  scaling transformation may be all that is necessary to model atmospheric motions at all scales. Determining which paradigm is most closely reflected by observations is the goal of the study presented here.

Because the details are not widely known, in Section \ref{sec:theory} we provide an overview of the theory behind Lovejoy-Schertzer turbulence as well as the various scale-dependent alternatives.
Using multiple dropsonde and radiosonde datasets described in Section \ref{sec:methods}, we test the theoretical predictions made by both paradigms in Section \ref{sec:results}. We consider a wide range of scales ranging from  $200\,\mathrm{m}$ to $20,000\,\mathrm{km}$ with emphasis placed on careful examination of spectra calculated along both the horizontal and vertical directions, the directional dependence being the distinctive prediction of Lovejoy-Schertzer turbulence. 

\section[Theories of isotropic and anisotropic atmospheric turbulence]{Theories of isotropic and\\* anisotropic atmospheric turbulence} \label{sec:theory}

In general, turbulence laws serve to  constrain how kinetic energy is distributed across spatial scales. Although these laws are most commonly represented through the power spectrum $E(k)$ of wind velocities as a function of the wavenumber $k$, there are other methods for performing a scale decomposition. Here, we consider the ``structure function'' as it is more robust to irregularly spaced sonde data \citep{lovejoy2007}. The structure function may be thought of as a real space version of the wavenumber spectrum where $k$ is replaced by a separation vector $\mathbf{\Delta r}$ of variable length and direction. Turbulence laws often hold for individual components of the wind vector $\mathbf{v}$, but for simplicity here we only consider the squared magnitude of the vector differences
\begin{align}
    \Delta v^2 \equiv \left(\mathbf{v}(\mathbf{r})-\mathbf{v}(\mathbf{r}+\mathbf{\Delta r})\right)^2. \label{eq:general turbulence law}
\end{align}

Eqn. \ref{eq:general turbulence law} is very general and it points to two basic questions.\footnote{A third basic question is how do the statistics of $\Delta v^2$ depend on the spatial location $\mathbf{r}$? The turbulence theories  considered in Table \ref{tab:turbulence theories} assume translational invariance, i.e. that any statistics do not depend on location. Translational invariance is unlikely in the atmosphere due to, for example, the altitude and latitude dependence of large scale dynamic features such as the jet stream. Here, we only consider velocity increments averaged over many spatial locations.}
\begin{table}
    \centering
    \renewcommand{\arraystretch}{1.2} %
    \caption{Various theories of atmospheric turbulence and corresponding parameter values for Eqn. \ref{eq:2D structure function}. Here, $N$ is the Brunt-V\"ais\"al\"a frequency, $\chi$ enstrophy flux, $\phi$ buoyancy variance flux, and $\varepsilon$ kinetic energy flux. ``Undefined'' means that the parameter has no meaning within the context of the theory, while ``not specified'' indicates the parameter has some value but it is not specified by the theory.
    Note that Lindborg's theory \citep{lindborg2006} suggests a transition between $H_v=1$ to $H_v=1/3$ at the Ozmidov length scale of order $3\,\mathrm{m}$. Given that our measurements are at larger scales, we only consider the value $H_v=1$.}\label{tab:turbulence theories}
    \begin{tabularx}{\textwidth}{l L L L}
        \hline
        \textbf{Type} & \textbf{Case} & \textbf{Exponents} & \textbf{Constants} \\
        \hline
        \multirow{4}{*}{Isotropic} & 3D \citep{kolmogorov1941} & $H_h = H_v = 1/3$ & $\varphi_h = \varphi_v=\varepsilon^{2/3}$ \\
                                   & \citet{bolgiano1959,obukhov1959} & $H_h = H_v = 3/5$ & $\varphi_h = \varphi_v=\phi^{2/5}$ \\
                                   & 2D \citep{kraichnan1967} & $H_h = 1$, $H_v$ undefined & $\varphi_h=\chi^{2/3}$, $ \varphi_v$ undefined \\
                                   & Gravity waves \citep{vanzandt1982} & $H_v=1$, $H_h$ not specified & $\varphi_v=N^2$, $\varphi_h$ not specified \\
        \hline
        \multirow{4}{*}{Anisotropic}
                                    & Quasi-Geostrophic \citep{charney1971}& $H_h = H_v=1$ & $\varphi_h =\chi_h^{2/3}, \varphi_v=\chi_v^{2/3}$, $\chi_v\ll \chi_h$ \\
                                    & \citet{schertzer1985} & $H_h=1/3, H_v=3/5$ & $\varphi_h=\varepsilon^{2/3}, \varphi_v=\phi^{2/5}$ \\
                                    & \citet{lindborg2006} & $H_h=1/3, H_v=1$ & $\varphi_h=\varepsilon^{2/3}, \varphi_v=N^2$ \\
                                    & Anisotropic 3D & $H_h = H_v=1/3$ & $\varphi_h =\varepsilon_h^{2/3}, \varphi_v=\varepsilon_v^{2/3}$, $\varepsilon_h\ne \varepsilon_v$ \\
        \hline
    \end{tabularx}
\end{table}
First, how do the statistics of $\Delta v^2$ depend on the length of the separation vector $\mathbf{\Delta r}$? Second, how do the statistics of $\Delta v^2$ depend on the direction of $\mathbf{\Delta r}$?
The answer to the second question of separation \emph{direction} is the main subject of this paper. Many of the foundational theories, such as those proposed by Richardson, Kolmogorov, and Obukhov, assume that turbulence statistics are isotropic, or that there is no directional dependence for the statistics of the flow. Before addressing the alternative, anisotropic turbulence, we must first consider how the statistics vary as a function of separation \emph{distance}.

Common to nearly all turbulence laws is the property of ``scale invariance", which requires that  fluctuations, when averaged over many potential realizations of the flow, follow a power-law function with respect to separation distance $\Delta r \equiv |\mathbf{\Delta r}|$ such that
\begin{align}
    \langle\Delta v^2\rangle = \varphi \Delta r^{2H} \label{eq:general structure function}
\end{align}
where $H$ is a constant termed the Hurst exponent and $\varphi$ is some dimensionally relevant quantity that is conserved throughout the turbulent ``cascade'' from one scale to the next.
Note that the kinetic energy spectrum in wavenumber space is obtained from  Eqn. \ref{eq:general structure function} via a Fourier transform to convert the real-space $\Delta x$ into wavenumber $k$ \citep{lovejoy2013}. The kinetic energy spectrum then becomes $E(k)\propto k^{ -(2H+1)}$. 

The next step is to identify the physical quantity $\varphi$ that is conserved during the turbulent cascade. This is the primary aspect by which various theories of turbulence are distinguished. We consider four theories. The first and most widely known was proposed by \citet{kolmogorov1941} where the relevant cascade quantity is the kinetic energy dissipation rate $\varepsilon$, with units of energy per mass per time ($\mathrm{m^2/s^3}$). 
This theory is also known as ``three-dimensional'' turbulence because, in its most basic form, it assumes isotropy, or that the statistics of the flow are identical in all three directions. 

Three-dimensional turbulence has limited relevance for atmospheric motion given that it neglects buoyancy forces. The second theory we consider, proposed independently by \citet{bolgiano1959} and \citet{obukhov1959}, addresses this concern by supposing the conserved cascading quantity relates to buoyancy forces rather than kinetic energy. For a Boussinesq flow, it can be shown that the conserved cascade quantity becomes $\phi=\partial f^2/\partial t$ where $f$ is thermal buoyancy \citep{lovejoy2013}. The quantity $\phi$ is the ``buoyancy variance flux'' with dimensions of acceleration squared per time ($\mathrm{m^2/s^5}$).

Bolgiano and Obukhov's theory was not widely adopted, and it was eventually replaced by various approaches founded on the basic assumption that any feedback between the flow and the stratification is negligible. In these theories,  stratification can influence the flow but the flow cannot modify the stratification--an idea encapsulated by the term ``background'' stratification. To this end, the third theory we consider here was proposed by \citet{charney1971} who adapted the theory of ``two-dimensional'' turbulence \citep{kraichnan1967} to the atmosphere. Traditional two-dimensional turbulence holds that the conserved cascade quantity is ``enstrophy flux'' $\chi$ with units vorticity squared per time ($\mathrm{s^{-3}}$). However, this theory was originally applied only to flows restricted to a plane such as a soap film. Charney modified the theory to account for some limited vertical flow, although the dominant set of dynamics remain a horizontal two-dimensional enstrophy cascade.

The original justification for neglecting feedbacks between the flow and stratification came from gravity wave theory, and so it is not surprising that gravity waves were later adopted directly to explain observed kinetic energy spectra, this time along the vertical direction \citep{vanzandt1982,dewan1997,lindborg2006}. In this fourth theory, the wave frequency, termed the Brunt-V\"ais\"al\"a frequency $N$ with units $\mathrm{s}^{-1}$, takes the place of the conserved cascade quantity. Although $N$ is not a typical cascade quantity, it nonetheless serves as the dimensionally relevant parameter for arguments based on dimensional analysis.

\begin{figure}
    \includegraphics{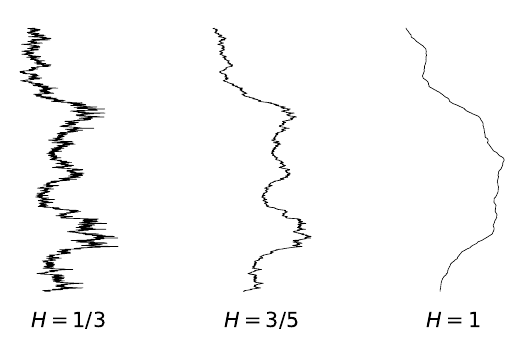}
    \centering \caption{Simulations of a synthetic stochastic process with varying $H$ \citep{lovejoy2010FIF}, representing example wind profiles that might be observed in hypothetical atmospheres where the theories represented by Eqns. \ref{eq:kolmogorov law}-\ref{eq:GW spectrum} apply. The profiles are generated with the same random seed but have varying degrees of ``smoothness'' as specified by the value of $H$.} \label{fig:example multifractals}
\end{figure}

For all the above theories, specification of the relevant cascade quantity is sufficient to determine the value of the Hurst exponent $H$. The value of $H$ is also the main testable prediction that can be used to discriminate between the various theories. Simply by using dimensional analysis,  the kinetic energy fluctuation becomes  a function of the relevant conserved cascade quantity $\varphi$ and the separation distance $r$. In this case, dimensional consistency requires, respectively,
\begin{align}
    &\langle\Delta v^2\rangle = \varepsilon^{2/3} \Delta r^{2H}, \qquad &H=1/3 \qquad &\text{(Kolmogorov spectrum)} \label{eq:kolmogorov law}\\ 
    &\langle\Delta v^2\rangle = \phi^{2/5} \Delta r^{2H}, \qquad &H=3/5 \qquad &\text{(Bolgiano-Obukhov spectrum)} \label{eq:Bolgiano-Obukhov law} \\
    &\langle\Delta v^2\rangle = \chi^{2/3} \Delta r^{2H}, \qquad &H=1 \qquad &\text{(Kraichnan spectrum)}  \label{eq:kraichnan law}\\
    &\langle\Delta v^2\rangle = N^{2} \Delta r^{2H}, \qquad &H=1 \qquad &\text{(Gravity wave spectrum)}  \label{eq:GW spectrum}
\end{align}
Visually, the value of $H$ imparts a unique character to any given transect or profile of $v$. A larger value of $H$ implies a smoother profile, as illustrated in Fig. \ref{fig:example multifractals}.

With respect to the directional dependence of the statistics defined by Eqn. \ref{eq:general structure function}, the simplest case is isotropy, where the statistics are identical for all directions of $\mathbf{\Delta r}$. However, it might instead be expected that the statistics of $\Delta v^2$ vary with direction in the atmosphere given that gravitational stratification has a strong directional dependence. In this case, it is conceivable that multiple theories could hold, even for the same spatial scale. 

One such proposal, first made by \citet{schertzer1985}, considers that gravity, and therefore buoyancy, operate only in the vertical  direction, and so the Bolgiano-Obukhov law (Eqn. \ref{eq:Bolgiano-Obukhov law}) is likely to hold only in the vertical direction. Horizontally, the flow is not bound to a quasi-two dimensional layer as the Kraichnan law requires, so the horizontal statistics are likely to follow the Kolmogorov law (Eqn. \ref{eq:kolmogorov law}):
\begin{align}
    \begin{cases}
    \langle \Delta v^2(\Delta x)\rangle\equiv \langle \Delta v(|\mathbf{\Delta x}|)^2\rangle = \varphi_h \Delta x^{2H_h}, 
    \qquad &\varphi_h  =\varepsilon^{2/3},\qquad H_h=1/3,\\ 
    \langle \Delta v^2(\Delta z)\rangle\equiv \langle \Delta v(|\mathbf{\Delta z}|)^2\rangle = \varphi_v \Delta z^{2H_v}, 
    \qquad &\varphi_v  =\phi^{2/5},\qquad H_v=3/5,
    \end{cases}\label{eq:Lovejoy-Schertzer turbulence, horiz/vert}
\end{align}
where $\mathbf{\Delta x}$ represents a purely horizontal separation vector and $\mathbf{\Delta z}$ represents a purely vertical one. There is no distinction made between the two horizontal directions in this theory, and so  $\mathbf{\Delta x}$ points in any horizontal direction. Together, we term Eqn. \ref{eq:Lovejoy-Schertzer turbulence, horiz/vert} the ``Lovejoy-Schertzer'' theory of turbulence. The two steps of dropping the isotropy assumption and proposing that two different laws hold simultaneously represent a significant shift from how turbulence is normally conceptualized. 

\begin{figure}
    \includegraphics[trim=0 0 22.5bp 0]{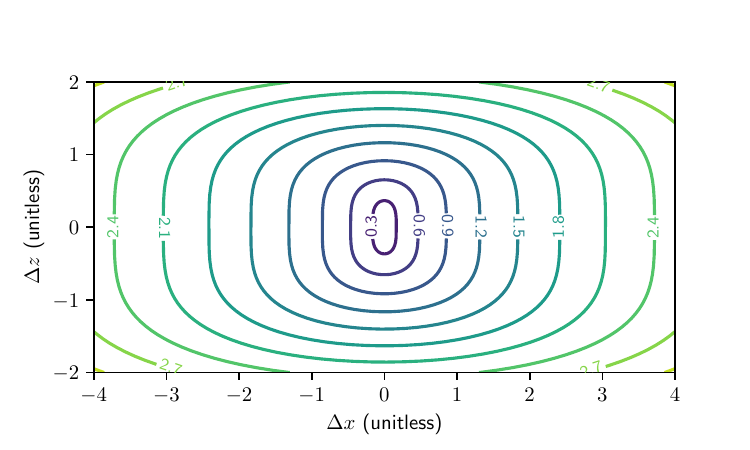}
    \centering \caption{Plot of the Lovejoy-Schertzer turbulent structure function (Eqn. \ref{eq:2D structure function}), representing the average sizes, shapes, and strengths of turbulent circulations. The function is shown in nondimensional form, i.e. $\phi = \varepsilon=1$, and using the theoretical values $H_h=1/3$ and $H_v=3/5$. Note that the empirical structure functions plotted in Section \ref{sec:results} consider absolute values for $\Delta x$ and $\Delta z$, which correspond to the first quadrant of this plot.} \label{fig:example 2D structure function}
\end{figure}

It is worth considering the nonobvious implications of this conceptual shift in more detail. One way to view a turbulent velocity field is as a superposition of geometrically simple circulations of varying sizes and wind speeds. The turbulence laws Eqns. \ref{eq:kolmogorov law}-\ref{eq:kraichnan law} may then be interpreted as a relationship between the cross-sectional length and the characteristic velocity of each of these simplified circulations. This picture works because the differences $\Delta v$ used in the structure function effectively isolate the kinetic energy perturbations that are due to circulations of a particular size $\Delta r$.
Thus, isolines of constant $\Delta v^2$ in $\Delta x, \Delta z$ space can be interpreted as describing the shapes and strengths of the simplified circulations from which the overall flow is ``built''. For the special case that turbulence is isotropic, the shapes of the circulations are spherical --  any direction has identical statistics as any other direction. If the turbulence is anisotropic, then the circulations are no longer spherical, as is illustrated in Fig. \ref{fig:example 2D structure function}.

In the Lovejoy-Schertzer theory of turbulence, the isolines are obtained by setting $\langle \Delta v^2(\Delta x)\rangle =\langle \Delta v^2(\Delta z)\rangle $ from Eqn. \ref{eq:Lovejoy-Schertzer turbulence, horiz/vert}. Solving for the aspect ratios of the circulations, we obtain
\begin{align}
    \frac{\Delta x}{\Delta z} = l_s^{-4/5}\Delta z^{4/5};\qquad l_s=\frac{\varepsilon^{5/4}}{\phi^{3/4}}. \label{eq:spheroscale}
\end{align}
Eqn. \ref{eq:spheroscale} is the first nontrivial implication of the Lovejoy-Schertzer theory. It implies that the mean aspect ratio of circulations systematically changes with circulation size (Fig. \ref{fig:example 2D structure function}). Indeed, this scaling of aspect ratio with size aligns with our intuitive notion that large-scale atmospheric circulations such as the Hadley cell are highly elongated in the horizontal direction, whereas small-scale circulations such as small convective elements are more spherical or even elongated vertically. 
Because the aspect ratios systematically change with scale, there is a unique scale for which the aspect ratio is equal to unity and circulations are nearly spherical, termed the ``spheroscale'' $l_s$ \citep{lovejoy1985,schertzer1985b}.

Another way to examine the shape of the turbulent circulations is to consider the full two-dimensional structure function $\langle \Delta v(\Delta x, \Delta z)^2\rangle$. The simple requirement that $\langle \Delta v(\Delta x, \Delta z)^2\rangle$ reduces to Eqns. \ref{eq:Lovejoy-Schertzer turbulence, horiz/vert} when $\Delta x = 0$ or $\Delta z=0$ is not enough to specify a unique function. If we further require that isotropic turbulence is recovered when $H_h=H_v$ and $\varphi_v=\varphi_h$, then a unique function is specified, as shown  in Fig. \ref{fig:example 2D structure function}:
\begin{align}
    \langle \Delta v(\Delta x, \Delta z)^2\rangle  = \left(\varphi_h^{1/H_h} \Delta x^2  +\varphi_v^{1/H_h} \Delta z^{2H_v/H_h}\right)^{H_h}. \label{eq:2D structure function}
\end{align}

We now fit the two-dimensional structure function given by Eqn. \ref{eq:2D structure function} to observed wind statistics, allowing for both exponents $H_v$ and $H_h$, as well as the coefficients $\varphi_v$ and $\varphi_h$, to be determined empirically. With these four free parameters, Eqn. \ref{eq:2D structure function} becomes very general. It includes as special cases each of the theories for turbulence mentioned so far. An additional case worth mentioning is that the exponents are equal but the constants $\varphi_h$ and $\varphi_v$  have different values. In this case, circulations have nonunitary aspect ratios but the aspect ratio does not change with circulation size. This type of anisotropy has been termed ``trivial anisotropy'' by \citet{lovejoy2013}. Studies investigating turbulent anisotropy based on the anisotropy stress tensor limit themselves to trivial anisotropy if they assume the cascade remains controlled by kinetic energy with $H_h=H_v$, as is sometimes done \citep{tennekes1972}. Table \ref{tab:turbulence theories} summarizes how Eqn. \ref{eq:2D structure function} relates to various turbulent theories.

\section{Methods} \label{sec:methods}

We calculate structure functions from three datasets of wind velocity. The first contains dropsonde measurements from the ACTIVATE field campaign \citep{vomel2023}, which took place over the North Atlantic Ocean between 2020 and 2022. Drops occurred during 169 flights spread over a variety of meteorological conditions and seasons.
The second dropsonde dataset considered here was obtained from NOAA hurricane reconnaissance flights that took place between 1996 and 2012, mainly over the Gulf of Mexico and the Atlantic Ocean.
Profiles were not considered if the data quality was marked as degraded or if the profile did not span the entire layer considered. We analyze a total of 683 ACTIVATE and 2325 hurricane profiles. Both dropsonde datasets contain vertical wind profiles $w(z)$ derived from the measured fall speed, but the uncertainty in the measurements is of order $\sim 1\,\mathrm{m/s}$ \citep{wang2015,vomel2023} -- too large for the purposes of calculating a structure function. For this reason, we only consider structure functions calculated using horizontal vector differences $\Delta v \equiv |\mathbf{v_h}(\mathbf{r})-\mathbf{v_h}(\mathbf{r}+\mathbf{\Delta r})|$ where $\mathbf{v_h}$ is the horizontal component of the full wind vector $\mathbf{v}$.

The third dataset we consider is the horizontal wind data from the Integrated Global Radiosonde Archive (IGRA) \citep{durre2006,durre2018}, a composite created from global  balloon-borne soundings obtained from thousands of stations and spanning many decades. 
A limitation of the dataset is that measurement methods and techniques vary temporally and spatially. For example, many modern sensors use GPS to measure geopotential height, but height inferred from pressure is also used, particularly for older data. 
Accordingly, we only consider measurements from the period between the years 2010 and 2025 when GPS data can be assumed to be in sufficiently widespread use. Measurement uncertainties for the two models of radiosonde widely used during this period, the Vaisala RS41 \citep{vaisala_rs41_docs,dirksen2014}
and the Graw DFM-17, \citep{graw_dfm17_docs, dirksen2014} 
are of order $10\,\mathrm{m}$ (geopotential height), $1\,\mathrm{hPa}$ (pressure), and $0.1\,\mathrm{m/s}$ (horizontal wind). 
The variety of measurement techniques and sensors used to compose the IGRA dataset introduces a source of uncertainty that is difficult to quantify, given that the techniques are not reported in the dataset. As an example, in Appendix \ref{sec:individual stations} we show a near perfect split in calculated Hurst exponents based on sounding nationality, a result likely caused by processing techniques being standard within a country but differing between countries. Any uncertainties we report below should therefore be interpreted with significant caution. 

\begin{figure}
    \includegraphics[width=.5\linewidth]{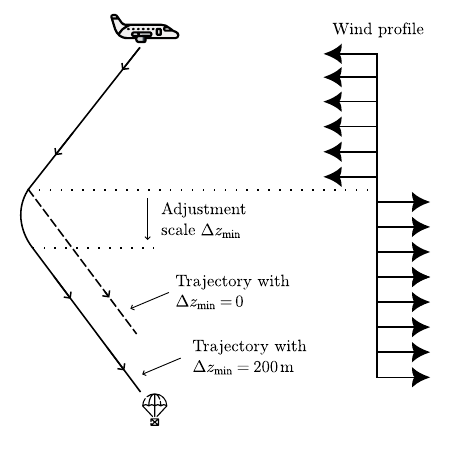}
    \centering \caption{Illustration of how dropsonde measurements of horizontal wind fluctuations $\Delta v(\Delta z)$ are effectively smoothed by sonde inertia. Wind fluctuations that occur over a smaller spatial scale than the sonde's adjustment scale $\Delta z_\text{min}$ cannot be reliably measured.} \label{fig:sonde inertia sketch}
\end{figure}

Shear measurements calculated from radiosonde and dropsonde horizontal wind profiles also have uncertainties related to sonde inertia (Fig. \ref{fig:sonde inertia sketch}).
If a sonde passes between layers with different mean wind speeds, the sonde does not immediately adjust to the new wind speed due to its inertia. Since horizontal wind velocity is approximated by the velocity of the sonde itself, shear cannot be measured over any distance smaller than the distance over which the sonde adjusts to a different horizontal wind. To account for this adjustment scale, and to remove additional spurious high-frequency wind variability due to e.g. oscillations caused by the payload swinging underneath the balloon, a smoothing is commonly applied to the wind profiles during data processing. The smoothing applied to ACTIVATE and hurricane sondes had a timescale of $5\,\mathrm{s}$ \citep{vomel2023} %
and $10\,\mathrm{s}$, \citep{durre2006}   %
respectively, while a typical radiosonde smoothing timescale is $40\,\mathrm{s}$ \citep{dirksen2014}.
Therefore, assuming typical descent/ascent rates of $20\,\mathrm{m/s}$ (dropsondes; \citet{vomel2023, wang2015}) and $5\,\mathrm{m/s}$ (radiosondes; \citet{dirksen2014}), the spatial scales for sonde adjustment  are of order $200\,\mathrm{m}$ in both cases. We therefore limit our analysis of vertical wind fluctuations to scales larger than $\Delta z_\text{min} = 200\,\mathrm{m}$.

To obtain $H$ in Eqns. \ref{eq:general structure function} and \ref{eq:2D structure function}, only second order structure functions are calculated here, although structure functions of other orders have also been considered by other studies. For comparison with prior results, we reproduce in the Supplement our analysis for first- and third-order structure functions $\langle \Delta v \rangle$ and $\langle \Delta v^3 \rangle$, respectively. All structure functions are calculated from all possible point pairs within a given profile, time period, or region, and fitted values are obtained using a least-squares regression. Uncertainties are reported as 95\% confidence intervals for the least-squares fit and do not include systematic bias originating from processing methodologies such as dataset smoothing.

For the IGRA data set, the structure functions are calculated for both vertical and horizontal separation directions. Purely vertical structure functions are calculated from individual sondes, each released for the 00z and 12z launch times between 2010 and 2025. Structure functions calculated along the horizontal direction are obtained using observations from different devices, first by identifying nearly simultaneous sonde launches. Then, each point observation is paired with each other point observation to obtain a list of observation pairs with varying horizontal and vertical separations. To be included in the analysis, observation pairs are required to have taken place within 2 hours of each other. For one-dimensional structure functions calculated along the horizontal direction,  observations are also required to have vertical separations no larger than $50\,\mathrm{m}$.
Due to the volume of data, it is not possible to consider all observation pairs from all possible sounding launches, and so  only soundings launched at 00z or 12z from every tenth day between 2010 and 2025 are considered for any structure function calculated along the horizontal direction.  

\subsection{Effect of noninstantaneous sonde measurements}

Here, we ignore temporal variability in the statistics for turbulent wind fluctuations in order to isolate the spatial statistics as given by Eqn. \ref{eq:Lovejoy-Schertzer turbulence, horiz/vert}. However, the sondes do not in fact measure wind profiles instantaneously. Dropsondes obtain profiles over approximately one hour, while radiosondes require two to three hours.

To estimate whether time differences between observation pair measurements impact our analyses, consider that, for isotropic turbulence, a turbulent circulation of size $l$ has a lifetime of order $\tau \sim l/\Delta v$ where $\Delta v$ is the wind speed associated with the circulation. For the spatial statistics of the circulation to be accurately sampled, the sonde must cross the circulation within a time interval that is shorter than the circulation lifetime. For sonde velocity $V$, this requires $l/V < \tau$ or equivalently $V>\Delta v$, implying that the sonde can only measure turbulent velocity fluctuations of a magnitude smaller than the sonde velocity. Given that  dropsondes with vertical velocities between 15 and 20 $\mathrm{m/s}$ satisfy this condition for most of the measured range of  scales, and that their measurements also support anisotropy, this isotropic criterion is less relevant.

For an anisotropic circulation, the horizontal and vertical circulation sizes are not identical, and the circulation's lifetime must be estimated using its horizontal length given that we are considering horizontal velocity components. In this case, accurate measurements of the vertical profile require $V>v l_z/l_x$ where $l_z$ is the vertical circulation length or the distance the sonde must pass through to sample a single circulation. From Eqn. \ref{eq:spheroscale}, large eddies are most anisotropic with aspect ratios $l_z/l_x\ll 1$. The largest measured velocity difference among spatially separated measurements in the dataset is  $\sim 20\,\mathrm{m/s}$, associated with eddies that are the most stratified with the smallest values of $l_z/l_x$ according to Eqn. \ref{eq:spheroscale}. These velocity differences are nonetheless only approximately four times faster than the slowest sonde velocity of $5\,\mathrm{m/s}$ (for the radiosondes). The sonde measurements may therefore be assumed effectively instantaneous.

We also include in our analysis horizontally separated measurement pairs that are separated in time by up to two hours. 
A typical circulation of horizontal size $l_x$ has a typical lifetime of $\tau = l_x/\Delta v$ where $\Delta v$ is the wind speed associated with the circulation. The minimum distance in our horizontal measurement pairs is set to $2\times 10^5\,\mathrm{m}$, so that the minimum circulation lifetime is of order $\tau = 2\times 10^5\,\mathrm{m} /10\,\mathrm{m/s} = 2\times 10^4\,\mathrm{s}$, or about 5.5 hours, implying that observation pairs separated by a maximum of two hours may be considered effectively instantaneous. 
As a final check that our thresholds are sufficient, we also computed a one-dimensional horizontal structure function for observation pairs that had vertical separations less than $5\,\mathrm{m}$ and were taken within $5\,\mathrm{min}$ of each other (see Supplement S4). Results for both sets of thresholds ($5\,\mathrm{m}$, $5\,\mathrm{min}$) and ($50\,\mathrm{m}$, $120\,\mathrm{min}$) indicated Hurst exponents that were nearly identical.

\section{Results} \label{sec:results}

\begin{figure}
    \includegraphics{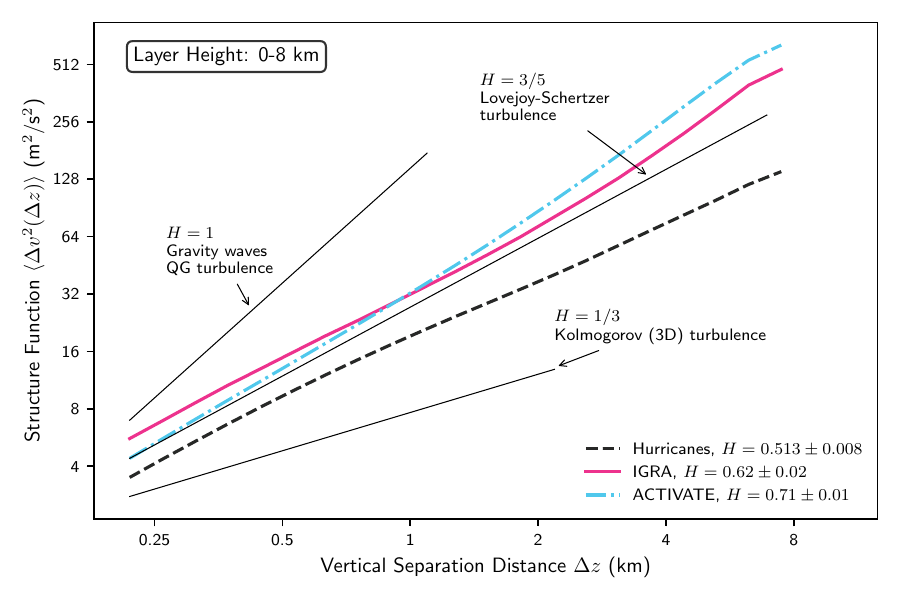}
    \centering \caption{Vertical structure functions for the IGRA radiosonde dataset (pink solid), the NOAA hurricane dropsonde dataset (black dashed) and the ACTIVATE dropsonde dataset (blue dot-dashed). Structure functions and Hurst exponents (Eqn. \ref{eq:general structure function}) are calculated for the lowest 8km of the troposphere, which is the layer measured by all three datasets. }\label{fig:8km vertical spectrum}
\end{figure}

First, we evaluate the structure function along the vertical direction (Eqn. \ref{eq:Lovejoy-Schertzer turbulence, horiz/vert})  for the lowest 8\,km of the atmosphere, which is the layer measured by both dropsonde datasets and IGRA radiosondes. As shown in Fig. \ref{fig:8km vertical spectrum}, the structure functions closely follow a power-law relationship over the full range of observed scales $0.2\,\mathrm{km} \le \Delta z \le 8\,\mathrm{km}$. Calculated values of $H_v$ range from $0.513 \pm 0.008$ for the hurricane dataset to $0.71 \pm 0.01$ for ACTIVATE. 

\begin{figure}
    \includegraphics{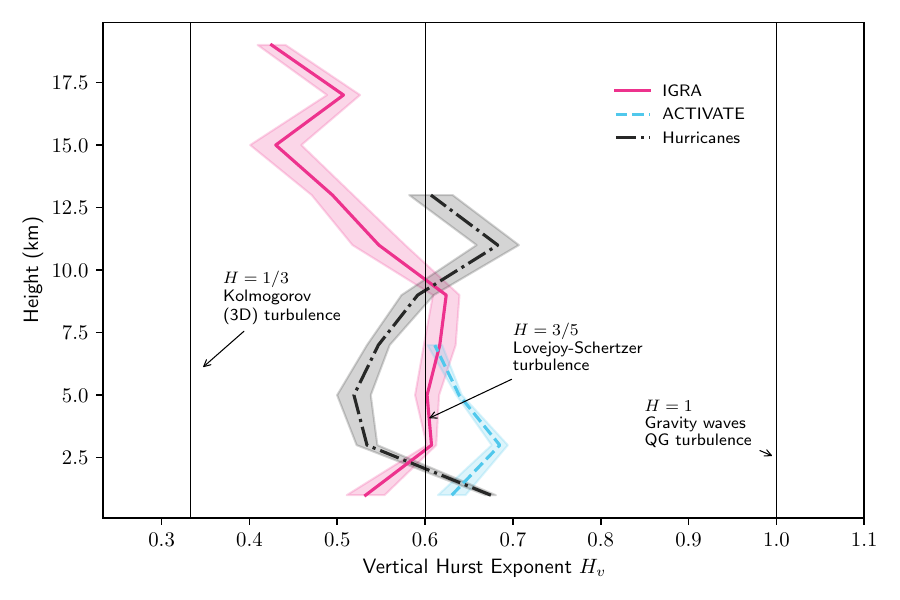}
    \centering \caption{Hurst exponents and 95\% confidence (shaded) calculated for structure functions as shown in Fig. \ref{fig:8km vertical spectrum} but evaluated within  stacked layers $2\,\mathrm{km}$ thick. }\label{fig:exponents with height}
\end{figure}

To investigate the dependence of the vertical Hurst exponent with height, wind measurements are divided into layers of thickness $2\,\mathrm{km}$. A Hurst exponent $H_v$ is calculated for the vertical structure function for $\Delta z$ ranging between $0.2\,\mathrm{km}$ and $2\,\mathrm{km}$ for each layer. Fig. \ref{fig:exponents with height} highlights how the calculated Hurst exponents lie close to the Lovejoy-Schertzer predicted value at each level within the troposphere. The structure functions for individual layers that are used to construct Fig. \ref{fig:exponents with height} are provided in the Supplement.  Above an altitude of roughly 18 to $20\,\mathrm{km}$, $H_v$ decreases to a value lying between approximately 0.4 and 0.5, which does not precisely agree with  any of the theories for turbulence described in Table \ref{tab:turbulence theories}.  As shown in the Supplement, tropospheric Hurst exponents for first- and third-order structure functions ($\langle \Delta v\rangle$ and $\langle \Delta v^3 \rangle$, respectively) are also consistent with the Lovejoy-Schertzer prediction.

\begin{figure}
    \includegraphics{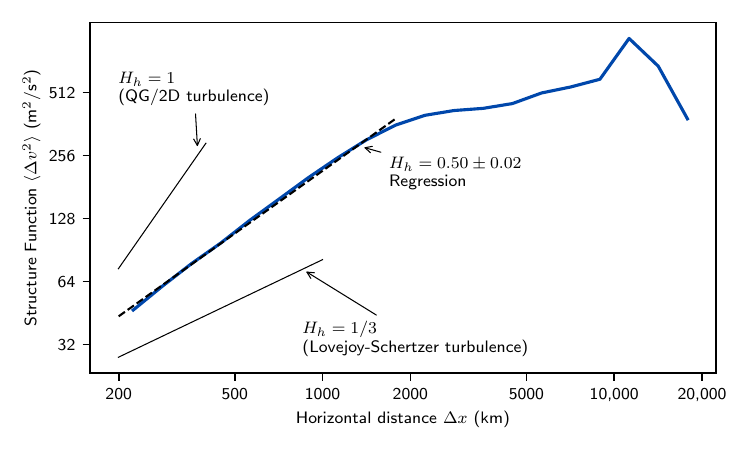}
    \centering \caption{Horizontal structure function $\langle \Delta v (\Delta x)^2\rangle$ calculated from IGRA radiosonde data and the associated theoretical power-law relationships for different turbulence theories shown in  Table \ref{tab:turbulence theories}. }\label{fig:horizontal structure function}
\end{figure}

Horizontal structure functions $\langle \Delta v (\Delta x)^2\rangle$ calculated using IGRA data are shown in Fig. \ref{fig:horizontal structure function}. There data show clear power-law behavior for separations smaller than about $1800\,\mathrm{km}$, while for larger scales the Hurst exponent approaches 0 between approximately $3000\,\mathrm{km}$ and the planetary half-circumference of $20000\,\mathrm{km}$. Regressions to the steepest portion of the slope between $200\,\mathrm{km}$ and $1800\,\mathrm{km}$ indicate a value for $H_h$ of $0.50 \pm 0.02$. 

\subsection{Two-dimensional structure functions}

That the values of the vertical and horizontal Hurst exponents are different supports the view that atmospheric turbulence is nontrivially anisotropic, or that the aspect ratios of atmospheric circulations may systematically change with scale as implied by Eqn. \ref{eq:spheroscale}. To address this possibility, we now examine in detail the full two-dimensional structure functions represented by Eqn. \ref{eq:2D structure function}. 
Note that Eqn. \ref{eq:2D structure function} cannot be easily logarithmically transformed, so what follows is limited to analyses calculated in linear space.

\begin{figure}
    \includegraphics{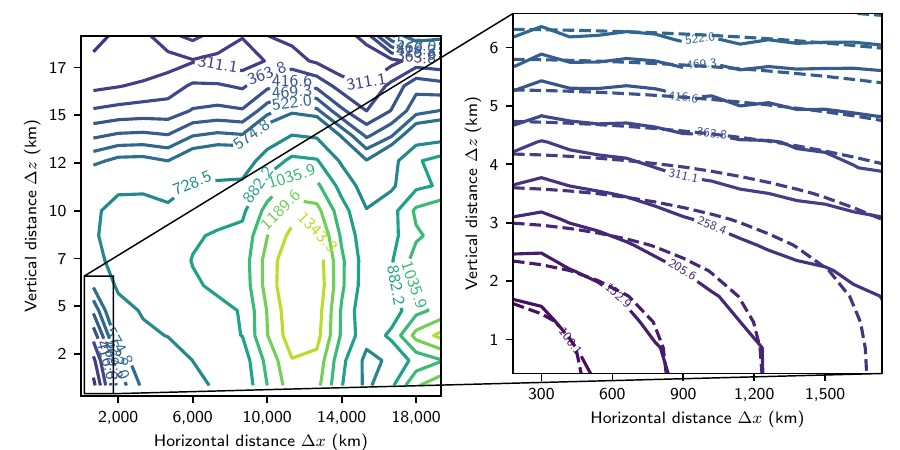}
    \centering \caption{Empirical two-dimensional horizontal/vertical structure function calculated from IGRA radiosonde data (solid; labels in $\mathrm{m^2 s^{-2}}$). For the inset, a fit (dashed) was performed using Eqn. \ref{eq:2D structure function}, with best-fit parameters described in text.}\label{fig:2Dvertical-horizontal structure function}
\end{figure}

Fig. \ref{fig:2Dvertical-horizontal structure function}
 shows the empirical two-dimensional structure function $\langle \Delta v(\Delta x, \Delta z)^2\rangle $ plotted for two ranges of scale.
As was also seen in Fig. \ref{fig:horizontal structure function},  the largest scales up to $20000\,\mathrm{km}$ horizontally and $20\,\mathrm{km}$ vertically show no clear power-law dependence of $\Delta v^2$ on $\Delta x$ or $\Delta z$, indicative of a regime that is not dominated by any turbulence theory listed in Table \ref{tab:turbulence theories}. By contrast, for horizontal separations between $200\,\mathrm{km}$ and $1800\,\mathrm{km}$, and for vertical separations smaller than $7\,\mathrm{km}$, the empirical structure function is well approximated by a least-squares fit to Eqn. \ref{eq:2D structure function}, with empirical values $H_h = 0.37 \pm 0.01$, $H_v = 0.63 \pm 0.01$, $\varphi_h=0.006 \pm 0.002 \,\mathrm{m^{2-2H_h} s^{-2}}$, and $\varphi_v=0.009 \pm 0.002\,\mathrm{m^{2-2H_v} s^{-2}}$. 

From Eqn. \ref{eq:spheroscale} and the theoretical relations $\varphi_h = \varepsilon^{2/3}$ and $\varphi_v = \phi^{2/5}$ (Table \ref{tab:turbulence theories}), empirical values for $\varphi_h$ and $\varphi_v$ imply a spheroscale of order $1\mathrm{m}$. Values calculated for $H_h$ and $H_v$ are nearly consistent whether they are calculated from the two-dimensional structure functions shown in Fig. \ref{fig:2Dvertical-horizontal structure function} or from the one-dimensional structure functions shown in Figs. \ref{fig:8km vertical spectrum}-\ref{fig:horizontal structure function}.

\begin{figure}
    \includegraphics{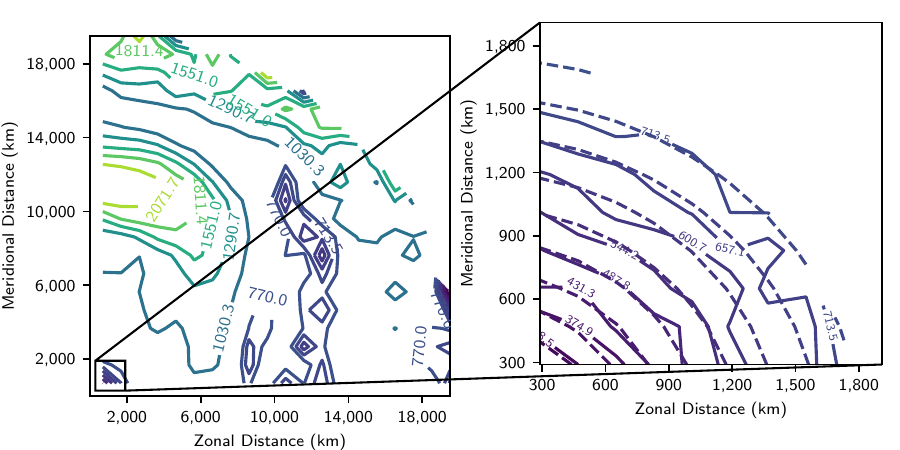}
    \centering \caption{As in Fig. \ref{fig:2Dvertical-horizontal structure function}, but for isoheight statistics calculated as a function of meridional and zonal direction. }\label{fig:2D horizontal structure function}
\end{figure}

Although we are unaware of any theory that predicts different Hurst exponents for the two horizontal directions, for completeness an isoheight two-dimensional structure function for the zonal ($x$) and meridional ($y$) directions is calculated from the sounding observations as shown in Fig. \ref{fig:2D horizontal structure function}. 
The structure function displays a maximum in $\Delta v^2$ near $\Delta x\sim 0$ and $\Delta y \sim 10,000\,\mathrm{km}$, which might be speculated to correspond to the jet stream. Otherwise, there is a ``flattening'' with $H_h\to 0$ at horizontal separation scales larger than $\sim 1800\,\mathrm{km}$ as in Fig. \ref{fig:horizontal structure function}. 

For separations smaller than $1800\,\mathrm{km}$, the empirical structure function is fit to a functional form that is analogous to Eqn. \ref{eq:2D structure function} but that applies in the $x$ and $y$ two horizontal directions:
\begin{align}
    \langle\Delta v(\Delta x, \Delta y)^2\rangle = \left(\varphi_x^{1/H_x} \Delta x^2  +\varphi_y^{1/H_x} \Delta y^{2H_y/H_x}\right)^{H_x}. \label{eq:2D h-h structure function}
\end{align}
Values for the least-squares fit  are $H_x = 0.35 \pm 0.05$, $H_y = 0.33 \pm 0.03$, $\varphi_x=0.03 \pm 0.04\,\mathrm{m^{2-2H_x} s^{-2}}$, and $\varphi_y=0.05 \pm 0.04\,\mathrm{m^{2-2H_y} s^{-2}}$. These values are consistent with the turbulence being either horizontally isotropic or ``trivially anisotropic'', with $H_x\simeq H_y$ but $\varphi_x\ne \varphi_y$, as noted also by \citet{lovejoy2011} based on an examination of reanalysis datasets. 
The mean aspect ratio of the circulations is poorly constrained when fitting all four parameters simultaneously, but constraining the exponents to the theoretical value $H_x=H_y=1/3$ yields $\varphi_x/\varphi_y = 0.90\pm 0.03$.

\section{Discussion} \label{sec:discussion}

For structure functions calculated along the vertical direction, the prediction of the  prevailing ``transition'' paradigm is that $H_v=1$ for the largest vertical scales, whether the model is a quasi-two-dimensional enstrophy cascade \citep{charney1971} or gravity waves \citep{dewan1997}. At smaller scales there is a transition to Kolmogorov turbulence where $H_v=1/3$. Neither value of $H_v$ is supported by Figs. \ref{fig:8km vertical spectrum} and \ref{fig:exponents with height}, at least for vertical separations down to $200\,\mathrm{m}$.

It is especially notable that the Kolmogorov value of $H_v=1/3$ is not supported even within the boundary layer below $2\,\mathrm{km}$ in altitude (Fig. \ref{fig:exponents with height}). This challenges the prevailing viewpoint that Kolmogorov turbulence, in either its isotropic or anisotropic forms (Table \ref{tab:turbulence theories}), characterizes  kilometer-scale boundary layer turbulence. If Kolmogorov turbulence does apply to the boundary layer,  Fig. \ref{fig:exponents with height} suggests that it can only exist for vertical separation scales smaller than $200\,\mathrm{m}$ that are not resolved here.

For large-scale structure functions calculated along the horizontal direction (Fig. \ref{fig:horizontal structure function}), the calculated value of $H_h=0.50\pm 0.02$ is a little higher than the $H_h=1/3$ value predicted by Lovejoy-Schertzer turbulence. Nonetheless, it lies closer to 1/3 than the value of $H_h=1$ expected for a two-dimensional turbulent enstrophy cascade as suggested by \citet{charney1971} and \citet{nastrom1983}, and therefore these results also appear to invalidate the transition paradigm. 

Overall, Hurst exponents for both the horizontal and vertical separation directions appear more strongly supportive of the Lovejoy-Schertzer paradigm of stratified turbulence (Eqn. \ref{eq:Lovejoy-Schertzer turbulence, horiz/vert}) than the transition paradigm. The two dimensional structure function $\Delta v (\Delta x, \Delta z)^2$ (Fig. \ref{fig:2Dvertical-horizontal structure function})  provides what is perhaps the most compelling support as it is not restricted to the orthogonal horizontal and vertical directions and therefore was calculated using many more observation pairs. A fit using Eqn. \ref{eq:2D structure function} approximately reproduces the empirical isolines of constant $\Delta v$, and the best-fit exponent values $H_h=0.37 \pm 0.01$, $H_v=0.63 \pm 0.01$ are again
close to the theoretical values predicted by the Lovejoy-Schertzer theory of turbulence of $H_h=1/3$ and $H_v=3/5$. 

\subsection[The effect of vertical smoothing on the Hurst exponents]{The effect of vertical smoothing\\* on the Hurst exponents} \label{sec:vertical exponent smoothing}

The largest discrepancy between the  empirically derived exponents obtained here and those predicted by the Lovejoy-Schertzer theory of turbulence is the value $H_h=0.50\pm 0.02$ obtained from the one-dimensional horizontal structure function in Fig. \ref{fig:horizontal structure function}. In fact, it does not clearly match any of the turbulence theories shown in Table \ref{tab:turbulence theories}. 

The discrepancy is reminiscent of the isobaric spectrum controversy discussed in the introduction -- even if $H = 1/3$ along isoheights it is possible that $H>1/3$ along isobars.
\citet{lovejoy2009}  argued that the basic reason that isobaric and isoheight structure functions differ is that isobars gently slope over large horizontal distances. As an example consider a sloping trajectory where $\Delta z = c \Delta x$ and $c\ll 1$. In this case, from Eqn. \ref{eq:2D structure function}, the observed structure function would follow
\begin{align}
    \langle\Delta v(\Delta x)^2\rangle = \left(\varphi_h^{1/H_h} \Delta x^2  +\varphi_v^{1/H_h} (c \Delta x)^{2H_v/H_h}\right)^{H_h}.\label{eq:sloping 2D structure function}
\end{align}
Given that $H_v/H_h>1$ (Fig. \ref{fig:2Dvertical-horizontal structure function}), at small scales the $\Delta x^{2}$ term dominates so that the observed structure function scales as $\langle\Delta v(\Delta x)^2\rangle \sim \Delta x^{2H_h}$, while at larger scales the $\Delta x^{2H_v/H_h}$ term dominates so that $\langle\Delta v(\Delta x)^2\rangle \sim \Delta x^{2H_v}$.
The implication is that a sloping trajectory will transition from $H\approx 1/3$ at smaller scales to $H\approx 3/5$ at some scale depending on the value of $c$. Such a transition is consistent with the well-known spectrum observed by \citet{nastrom1983} that helped motivate acceptance of the transition paradigm. 

Crucially, the implication is that any observed large-scale isobaric Hurst exponent cannot be simply assumed to be equal to the large-scale isoheight Hurst exponent, even though isobars are nearly flat. The more general lesson is that $H_h>1/3$ can be observed in the large-scale horizontal structure function if the observations depart even slightly from lying on a perfect isoheight surface.

Although our isoheight measurements are not isobaric, there is reason to believe that they also do not represent perfectly horizontal separations, mainly because radiosonde measurements are smoothed over a characteristic vertical distance of order $\sim 200\,\mathrm{m}$.\footnote{A second reason measurements may not represent perfect isoheights is measurement error in the vertical location, which would also tend to ``smooth'' the mean statistics in a similar manner. But given that measurement uncertainty in the GPS radiosondes is of order $\sim 10\,\mathrm{m}$, we assume vertical smoothing due to sonde inertia is more important.} Smoothing implies that a wind measurement at a given height is a function of the wind at nearby heights, plausibly causing vertically-separated fluctuations to influence any calculated horizontally-separated fluctuations. As a note, such smoothing cannot necessarily be removed by simply reprocessing the data. All sondes have a finite timescale of adjustment to the local wind in the presence of vertical wind shear. Such a timescale introduces an effective smoothing regardless of how the data are processed. 

To investigate the effect of vertical smoothing on horizontal statistics, we performed an experiment in a numerical hydrodynamic simulation using the System for Atmospheric Modeling (SAM) \citep{khairoutdinov2003}. The simulation is of a tropical atmosphere and uses a numerical grid with $100\,\mathrm{m}$ spacing along all three directions in the layer where we performed our analysis, which was between $2$ and $10\,\mathrm{km}$. The domain size is $204\,\mathrm{km}\times 204\,\mathrm{km}$. Details of the simulation are provided in \citet{dazlichGigaLarge2013}.

\begin{figure}
    \includegraphics{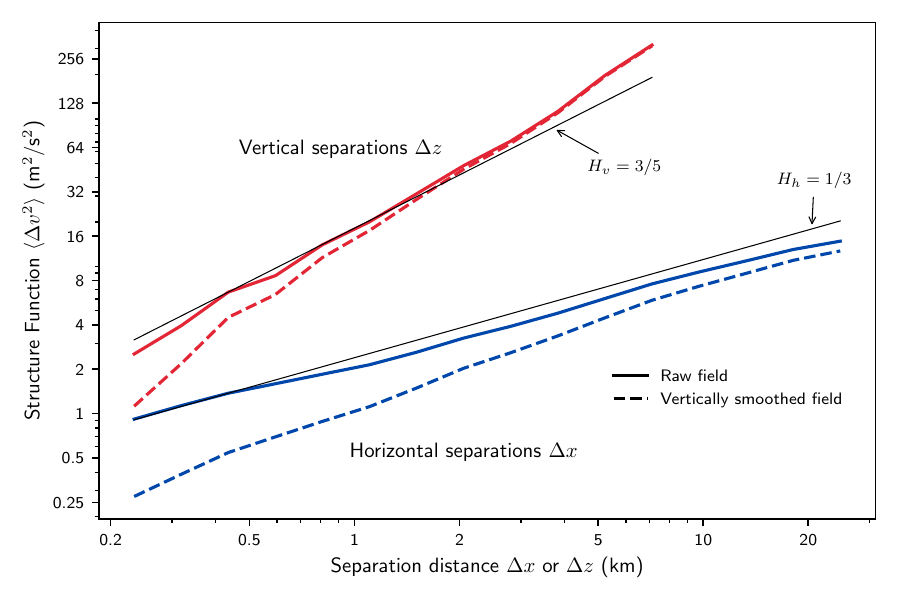}
    \centering \caption{Vertical (red) and horizontal (blue) structure functions calculated for the SAM simulation. Structure functions are calculated both for the original wind field with no smoothing applied (solid) and after the wind field is smoothed along the vertical direction (dashed). The theoretical values $H_h=1/3$ and $H_v=3/5$ according to Lovejoy-Schertzer theory are shown for reference (thin black).}\label{fig:SAM spectra}
\end{figure}

Vertical and horizontal structure functions calculated from the SAM wind field lie close to the Lovejoy-Schertzer scaling (Fig. \ref{fig:SAM spectra}). Omitting separation distances equal to one grid point, which may be more affected by numerical artifacts, calculated Hurst exponents for the raw wind field are $H_h = 0.305 \pm 0.008$ and $H_v=0.69 \pm 0.02$. As anticipated, when vertical smoothing is applied using a vertical Gaussian convolution with a standard deviation of two grid points or $200\,\mathrm{m}$, the horizontal Hurst exponent increases to $H_h=0.42 \pm 0.01$. The vertical Hurst exponent is also increased to a value of $H_v=0.79 \pm 0.03$. Notably, smoothing is only performed along the vertical direction but the horizontal Hurst exponent is nonetheless strongly affected.

Because the domain size is smaller than the horizontal measurements from IGRA, the result that vertical smoothing increases the horizontal Hurst  exponent should be viewed as qualitative rather than quantitative. More rigorous testing, perhaps with large-scale multifractal simulations with known Hurst exponents, would be necessary to determine whether the magnitude of the discrepancy between the IGRA-derived horizontal Hurst exponents and theory could be explained by vertical smoothing. Even so, Fig. \ref{fig:SAM spectra} suggests that vertical smoothing may plausibly explain why observations of $H_h$ are a little higher than expected by Lovejoy-Schertzer turbulence -- without needing to invoke any new set of physics such as quasi-geostrophic turbulence.

We should note that applying a vertical smoothing directly to Eqn. \ref{eq:2D structure function} can only decrease, rather than increase, the calculated horizontal Hurst exponent $H_h$. The reason is that Eqn. \ref{eq:2D structure function} applies for mean values of $\Delta v^2$ that are averaged over many realizations of the flow. This implies that the full distribution of values for $\Delta v^2$, rather than simply the mean $\langle \Delta v^2 \rangle$, must be considered to fully explain why $H_h$ is higher due to vertical smoothing. This is in contrast to the isobaric mechanism explanation for bias in measurements of $H_h$ suggested by \citet{lovejoy2009}, where the increase of $H_h$ may be derived directly from the mean statistics for $\Delta v^2$ represented by Eqn. \ref{eq:2D structure function} using the argument shown above. 

That the sounding data includes vertical smoothing may also explain \citet{lovejoy2007}'s prior finding of an increase in $H_v$ with altitude. Given that dropsondes such as those used by \citet{lovejoy2007} have a faster fall speed in the upper troposphere \citep[e.g.][]{vomel2023}, a given dropsonde-wind shear adjustment timescale would imply upper-tropospheric measurements are effectively smoothed over a larger vertical spatial scale than  lower-tropospheric measurements. If Hurst exponent calculations include scales affected by smoothing, a spurious altitude dependence of $H_v$ could result. \citet{lovejoy2007} included separations down to $5\,\mathrm{m}$ in their analyses, which is much smaller than our $200\,\mathrm{m}$ threshold and plausibly introduced a spurious dependence of $H_v$ on altitude that is not seen in Fig. \ref{fig:exponents with height}.

\section{Conclusions}

It is widely assumed that the dynamics of the atmosphere are controlled by a hierarchy of distinct dynamical mechanisms, each restricted to some limited range of spatial scales. The distribution of kinetic energy is thought to be determined by a quasi-two-dimensional enstrophy cascade at the largest scales \citep{charney1971,nastrom1984}, gravity waves at the mesoscale \citep{dewan1997,lindborg2006}, and a three-dimensional turbulent energy cascade at the smallest scales \citep{tennekes1972}. Such a hierarchy would imply clear transitions in the Hurst exponents for kinetic energy structure functions when calculated along either the horizontal or vertical directions, from $H=1$ at large scales to $H = 1/3$ at small scales. 

Here, we use high-resolution dropsonde and radiosonde measurements to calculate structure functions for horizontal wind separated both horizontally and vertically. We find a Hurst exponent close to $H_v\approx 0.6$ for vertical separations between $200\,\mathrm{m}$ and $8\,\mathrm{km}$, which is inconsistent with both small-scale isotropic turbulence and mesoscale gravity waves. Along the horizontal direction, large scale structure functions show a Hurst exponent with value $H_h\approx 0.4$ %
for separation scales ranging from  $200\,\mathrm{km}$ to $2000\,\mathrm{km}$, which is inconsistent with a large-scale enstrophy cascade.  We argue that these measured structure functions  are closely consistent with a lesser known theory of ``Lovejoy-Schertzer'' turbulence with $H_h = 1/3$ and $H_v = 3/5$ at all scales \citep{schertzer1985}. We show that the small difference between the observed and Lovejoy-Schertzer value of $H_h$ is plausibly due to vertical smoothing of radiosonde data. 

Thus, we find that the canonical ``transition'' paradigm has little empirical support. Instead, it appears that the dynamics of the troposphere and most of the stratosphere are controlled by a single wide-ranging anisotropic turbulent cascade rather than a hierarchy of independent dynamical mechanisms.  Looking forward, more measurements of structure functions or spectra as a function of separation direction will be necessary to confirm Lovejoy-Schertzer scaling. Scales smaller than $200\,\mathrm{m}$, which were not resolved here, are of particular interest for determining whether small-scale turbulence is isotropic with $H_h = H_v$. Simultaneous wind observations separated in both the horizontal and vertical direction will be necessary for such an analysis. The extent to which vertical smoothing affects calculations of horizontally-separated structure functions could also be quantified, perhaps by considering multifractal simulations with known exponents \citep{lovejoy2010}.

\setlength{\bibhang}{0in}
\bibliographystyle{ametsocV6}
\bibliography{sources}

\newcommand{\steamfigroot}{}

\chapter{Toward non-hydrodynamic simulation of moist convection} \label{sec:steam}

The cloud feedback remains the most uncertain component of Earth's climate sensitivity, partly due to the fact that important physical processes span a very wide range of spatial scales. The default assumption is that increasing model resolution will reduce uncertainty \citep{shukla2009, slingo2022}, although ever-larger simulations built over the past 40 years have yet to reduce model spread in estimates of either the cloud radiative feedback \citep{ceppi2017} or the overall climate sensitivity \citep{ipcc6}. %

It is argued that the next generation of kilometer-scale models will break the trend because they can resolve individual clouds \citep{slingo2022}. The implicit assumption is that clouds are typically kilometer-scale such that they may be resolved by kilometer-scale grid spacings. The ubiquity of this assumption is exemplified by the equivalent terminology ``Global Cloud-Resolving Models'' (GCRMs). However, the assumption that cloud dynamics are kilometer scale appears contradicted by the property of scale invariance in cloud shapes and sizes, which has been observed to apply globally from subkilometer scales up to thousands of kilometers \citep{wood2011,guillaume2018,dewitt2024,rees2024,dewitt2024b,dewitt2026}. Scale invariance implies that kilometer scales are no more fundamental than any other scale within the invariant regime: kilometer-scale phenomena are simply a rescaled version of 100-kilometer-scale phenomena without being qualitatively different \citep{lovejoy2013}. The implication would be that a physically accurate cloud parameterization should, in principle, work equally well in a 100-kilometer simulation as a 1-kilometer simulation. Conceivably, then, the reason the cloud feedback is uncertain is not that the models have insufficient resolution but rather that our parameterizations for clouds are insufficient.

On a fundamental level, any parameterization is equivalent to a theory for cloud dynamics, as both attempt to represent a more complex phenomenon through simpler means \citep{arakawa2004}. This could be viewed as analogous to how the Navier-Stokes equations themselves represent a substantial simplification over a particle-by-particle accounting of the same system. This is a phenomenon referred to as ``emergence'', where some simplified set of laws can still describe many important features of a complex system without needing to explicitly simulate every smaller-scale detail. Just as the Navier-Stokes ``emerge'' out of a sufficiently complex collection of particles, a hypothetical theory for cloud dynamics might ``emerge'' out of a sufficiently complex turbulent flow. That parameterizations work at all suggests that such an emergent theory may exist, even if it is not yet understood.

At the root of the problem of the cloud radiative effect, then, is a theoretical problem: how can the extreme complexity of a cloud field be represented in some simplified state space, using some set of emergent laws? How can the dynamics at one scale, such as those resolved in a GCM, be related to the dynamics at other scales, such as those being parameterized? Although these questions may be quite difficult to answer in any sufficiently general way, we have no choice but to attempt an answer. Parameterizations must be used whether or not they have a plausible physical foundation.

There is a long history of analytical formulations of clouds and cloud fields in terms of a large number of constituent elements, even if what serves as the basic element is not yet agreed upon. An early example is \cite{arakawa1974}, who derived a convective parameterization by modeling individual clouds as convective "plumes" using similarity theory. However, given that plumes require a constant source of buoyancy, their applicability to moist convection has been called into question. Instead, thermals or ``bubbles'' have been proposed as a better starting point for a similarity theory-based cloud model \citep{yano2016}, suggesting cloud fields may be thought of as a large number of superimposed thermals.
Other work has attempted to take clouds themselves as the basic constituent element in order to derive overall statistics of fields containing large numbers of clouds \citep{craig2006, cohen2006, garrett2018}. Unfortunately, even individual clouds are highly complex structures, and cloud theories therefore tend to make major simplifications such as ignoring interactions between clouds \citep{craig2006} or making clouds binary objects \citep{garrett2018}. 

Although it is less commonly applied to clouds, a much older tradition exists of conceptualizing fluid flows as being made up of constituent parts: a turbulent cascade. As first articulated by \citet{richardson1922}, here a turbulent flow is envisioned to be made up of large numbers of ``eddies'' that recursively break up into smaller and smaller eddies through a ``cascade'' while conserving some relevant physical quantity such as the kinetic energy dissipation rate. 

Despite being widespread in turbulence theory, the cascade is typically only invoked as a conceptual argument used to motivate scaling arguments such as the Kolmogorov -5/3 energy spectrum, where kinetic energy $E$ is related to wavenumber through $E\propto k^{-5/3}$. It is rarely taken as a literal description of the dynamics that could be used to make simulations. Multiplicative cascades are a partial exception, but even here eddies are difficult to interpret as physical constituents of the flow, and the most realistic simulation methods substitute a cascade for more physically opaque mathematical methods \citep{lovejoy1990,lovejoy2010FIF}.

Conceptually, eddies and bubbles bear some resemblance insofar as both form the constituent elements from which the overall field is built. However, turbulence and the dynamics shaping clouds are typically thought to be distinct physical processes operating at different spatial scales, primarily because buoyancy is central to cloud dynamics but is omitted from Kolmogorov's theory of turbulence. It is therefore understandable that bubbles and eddies have thus far been considered separately. What is less widely recognized is that buoyancy may be incorporated into the theory of turbulence itself, in which case the concepts of eddies and bubbles might be unified to construct a theory of cloud dynamics. 

As suggested by \citet{schertzer1985}, when buoyancy is present in a turbulent flow, it is not kinetic energy flux that is conserved through the cascade, but a different flux related to buoyancy. The reason is that kinetic energy may be converted into potential energy in a gravitational field and is therefore not conserved. The main empirical consequence of ``Lovejoy-Schertzer'' turbulence is that Kolmogorov's -5/3 energy cascade only applies when the power spectrum is computed along the horizontal direction, where gravity has no effect: $E\propto k_x^{-5/3}$. When computed along the vertical, a different relationship was proposed, namely $E\propto k_z^{-11/5}$. This directional dependence of the kinetic energy spectrum was recently found to be closely consistent with sonde observations and LES output from scales of 200 m to 2000 km. These scales include nearly the full range of motions that are typically associated with cloud dynamics, suggesting that ``bubbles'' may indeed be equivalent to ``eddies'' -- so long as the theory of turbulence considered is that proposed by Lovejoy and Schertzer, not Kolmogorov.

\begin{sidewaysfigure}
\centering
\includegraphics[height=0.93\portraitwidth]{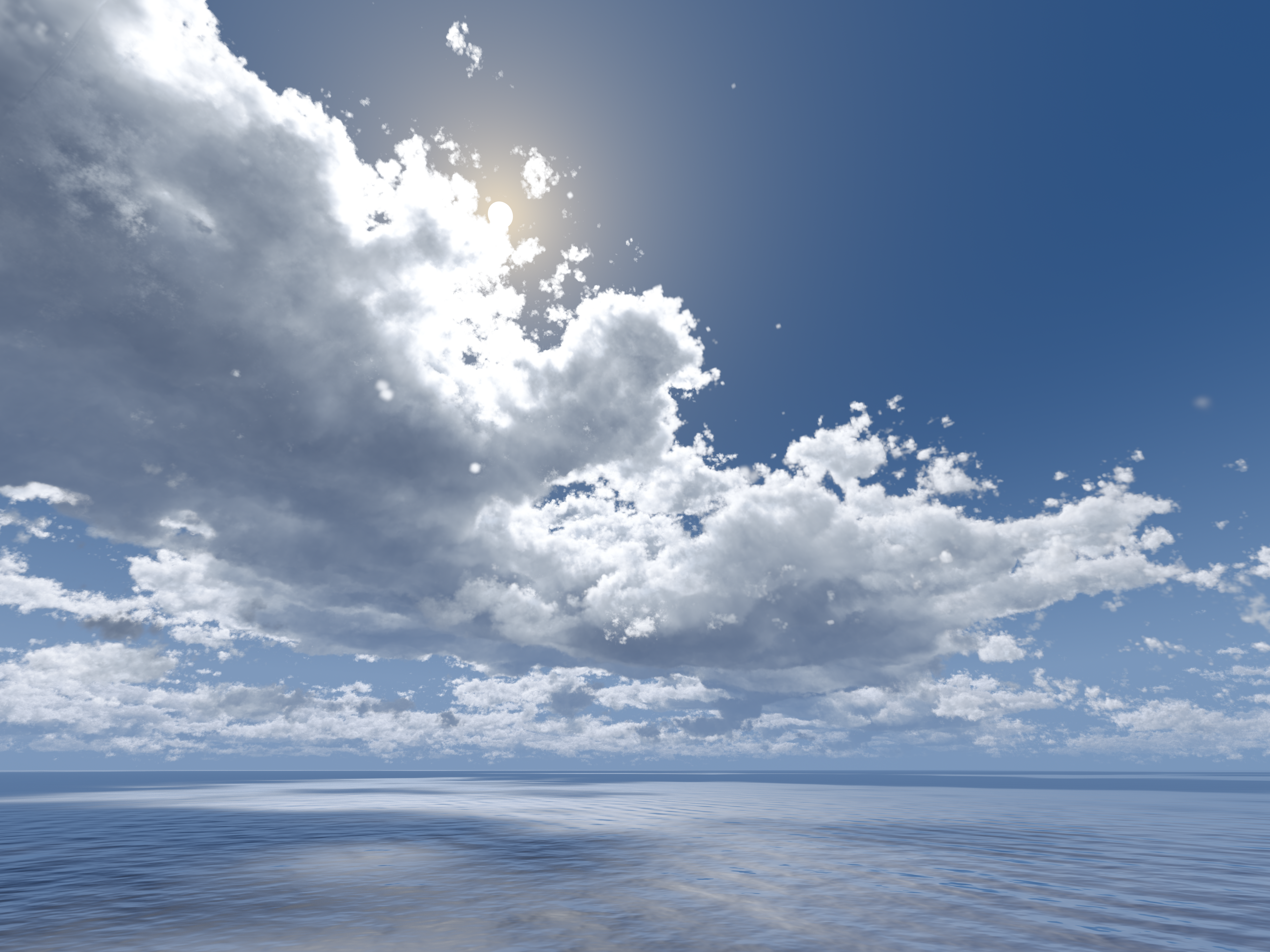}
\caption{Simulated rendering of a STEAM-simulated cumulus field.} \label{fig:intro rendering}
\end{sidewaysfigure}

Here, we propose a model for clouds by formalizing the conceptual model of a Lovejoy-Schertzer turbulent cascade in a physically realistic manner. The model is based on the idea that a cloudy turbulent flow can be thought of as being composed of a large number of discrete constituents, which we refer to as ``turbulons''. A turbulon is like an ``eddy'' or ``bubble'' but has a specific mathematical form and applies to an arbitrary vector or scalar field. We propose an algorithm that may be used to simulate three-dimensional volumes of water vapor and condensate, temperature, and moist static energy using turbulons, called the Superposition of Turbulons and Eddies Atmospheric Model (STEAM). Renderings of the generated cloud fields using a ray-tracing algorithm display striking realism (Fig. \ref{fig:intro rendering}), suggesting the model may pass the ``Palmer-Turing test'' for visual fidelity \citep{palmer2016,christensen2021}. We show quantitatively in Section \ref{sect:field comparison} that many aspects of STEAM-simulated fields are comparable to state-of-the-art high-resolution hydrodynamic models, although not in every case. Some differences between STEAM and the hydrodynamic benchmark suggest concrete steps for future work, while others are more mysterious. Conceivably, some discrepancies may even indicate STEAM is more realistic than the hydrodynamic comparison: using a comparison to MODIS satellite observations in Section \ref{sect:satellite comparison}, we show that STEAM reproduces several statistics describing cloud geometry more accurately than the hydrodynamic baseline. STEAM is also remarkably computationally efficient: in Appendix~\ref{app:steam supplement} we estimate that STEAM simulations run between $10^5$ and $10^6$ times faster than a comparable hydrodynamic simulation per independent snapshot simulated, mainly because STEAM does not require any spin-up period.

\section{Theoretical basis for STEAM} \label{sect:theoretical basis}

In this section, we provide a high-level overview of the theoretical considerations used to formulate the Superposition of Turbulons and Eddies Atmospheric Model (STEAM). A complete description of the STEAM algorithm is provided in Appendices \ref{sect:steam algorithm details} and \ref{ssect:steam algorithm spec}.

\subsection{A new basic element of turbulent flow: the turbulon}

\begin{figure}[t]
\centering
\includegraphics[width=\textwidth]{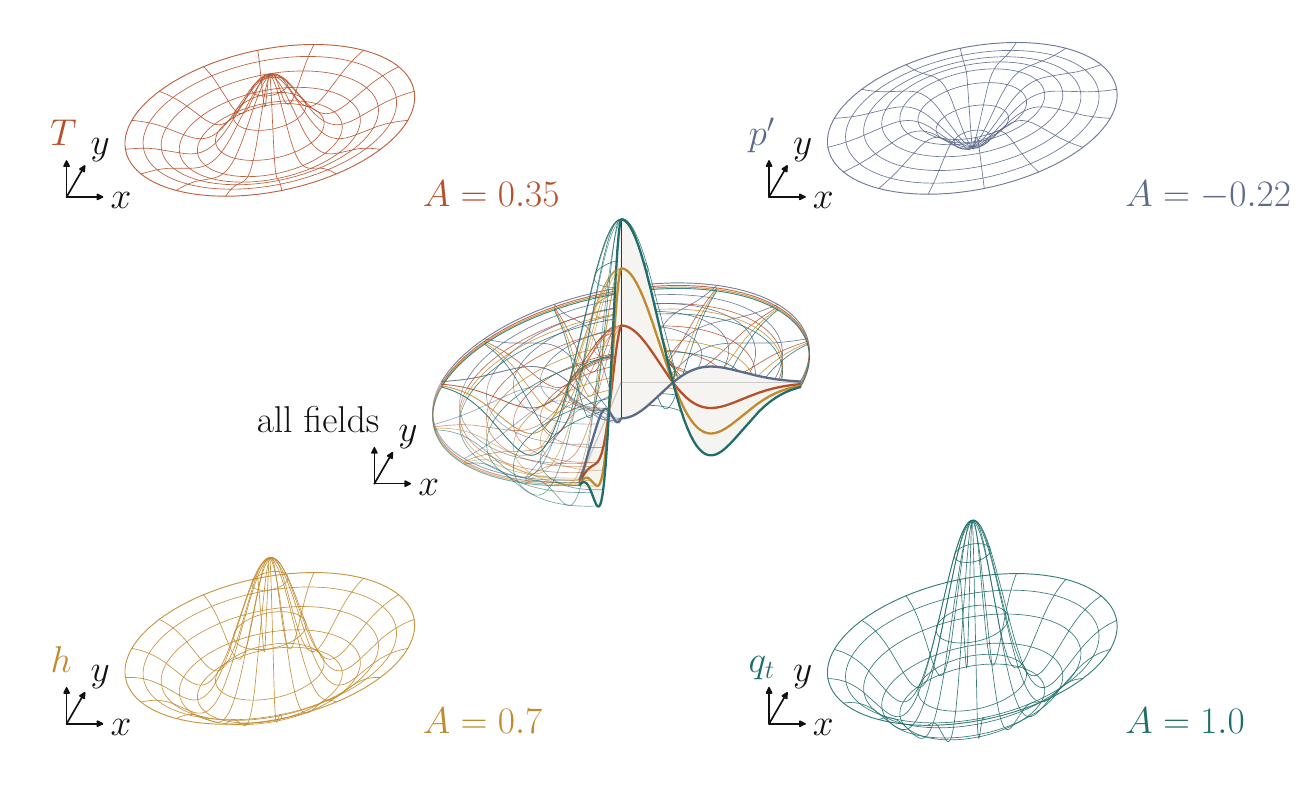}
\caption{A single two-dimensional ``turbulon'': a set of localized perturbations in each field from which a turbulent field is built. Each perturbation (corners) associated with a single turbulon $\mathfrak{t}$ (center) has a varying amplitude $A$ but a common shape, defined by the envelope function $\mathfrak{T}$, here a two-dimensional Mexican Hat analogous to that used for STEAM (Eqn. \ref{apxeq:turbulon shape}).} \label{fig:turbulon concept}
\end{figure}

The basic idea of a turbulent cascade is to consider the fluid volume as being made up of a discrete set of constituent elements called ``eddies'', i.e. circulations in the wind field. In order to formalize this idea, we must first specify exactly what we mean by a turbulent eddy and extend the concept to fields other than the wind field. In place of the term ``eddy'', we introduce the term ``turbulon'' to refer to an individual constituent element of a turbulent field (Fig. \ref{fig:turbulon concept}). A turbulon $\mathfrak{t}$ is defined as a set of localized perturbations in all fields that make up a turbulent flow. Perturbations in each field have an identical shape but varying amplitude. For example, a given turbulon might be associated with a strong temperature perturbation but a weak pressure perturbation. The turbulon shape is defined by some ``envelope'' function $\mathfrak{T} (\ell, \mathbf{r})$ with size parameter $\ell$. Any individual turbulon is made up of components that are a scaled and translated version of the envelope function with a specified size parameter $\mathfrak{t}_i = a(\mathbf{r}, \ell)\mathfrak{T}(\ell, \mathbf{r} - \mathbf{r}')$, where $a$ is the perturbation amplitude. Conceptually, turbulons are broader than the concept of ``eddies'' and more precise. Eddies are rarely defined mathematically, but if they were, an eddy would only represent the component of a turbulon that is associated with a perturbation in the wind field.

\begin{figure}[t]
\centering
\includegraphics[width=\textwidth]{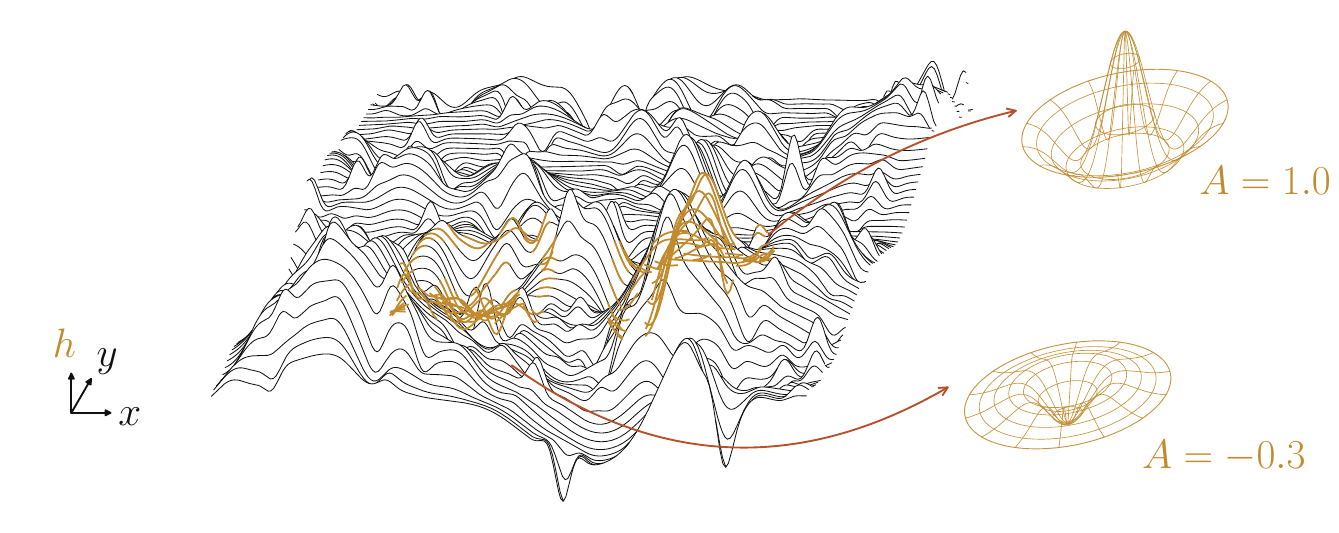}
\caption{A two-dimensional turbulent field of moist static energy $h$ (left) is built from a superposition of a large number of individual turbulons with varying amplitudes, positions, and size parameters (Eqn. \ref{eq:basic field construction with turbulons}). Shown on the right are the $h$-component of two such turbulons.} \label{fig:field superposition}
\end{figure}

The overall field $F$ defining a component of the atmospheric state, perhaps representing temperature or moist static energy, is represented as a superposition of a large number of turbulon components as shown in Fig. \ref{fig:field superposition}:
\begin{align}
  F(\mathbf{r}) = \sum_{\ell, \mathbf{r}'}^{} a(\mathbf{r}, \ell)\mathfrak{T}(\ell, \mathbf{r} - \mathbf{r}'). \label{eq:basic field construction with turbulons}
\end{align}

Mathematically, the approach is equivalent to a wavelet transform, a method already common in turbulence theory and discussed further below. However, almost exclusively wavelets are used as an analysis tool to determine how much variance a given signal has as a function of location and frequency. We are proposing to invert this interpretation. Given sufficient theoretical constraints on the statistics of the turbulons, the wavelet formalism may not only be used to analyze atmospheric volumes but also to simulate them. More fundamentally, we also propose an ontological re-evaluation of wavelets as more than just a mathematical tool used to evaluate frequency content. In this conception, a cloud is made up of constituent elements called turbulons, in the same way that an ideal gas is made up of particles. This conceptual shift can provide a new and complementary perspective on atmospheric turbulence and cloud dynamics.

The wavelet equivalence implies that any possible field $F$ may be simulated given the correct turbulon statistics using Eqn. \ref{eq:basic field construction with turbulons}. Therefore, any given question of atmospheric statistics or dynamics posed in real space has an equivalent formulation in the ``turbulon space'' represented by the turbulon amplitudes and locations $a_i(\mathbf{x},\ell)$. As clouds and turbulence are inherently multiscale phenomena, turbulon space has the benefit of explicit separation as a function of the scale parameter $\ell$. And as we will now show, turbulence theory provides several important constraints on the statistics of the turbulons as a function of the size parameter $\ell$.

\subsection[Constraints on turbulon statistics from turbulence theory]{Constraints on turbulon\\* statistics from turbulence theory}

Most commonly, theories of turbulence provide a constraint on the kinetic energy spectrum of some component of the wind field $v$. This constraint on the frequency content may be represented in either wavenumber space, using the spectrum, or real space, using wavelets or, more commonly, structure functions. The structure function is defined as
\begin{align}
  \Delta v^q (r)\equiv \langle \big|v(\mathbf{x})-v(\mathbf{x}+\mathbf{r})\big|^q\rangle_\mathbf{x}; \qquad \ell \equiv |\mathbf{r}|, \label{eq:structure function}
\end{align}
where the brackets $\langle \rangle_\mathbf{x}$ indicate averaging over many spatial locations $\mathbf{x}$. The order parameter $q$ is typically set to 2 so that the difference in Eqn. \ref{eq:structure function} may be interpreted as isolating the mean kinetic energy that is associated with motion at scale $\ell$.

Turbulence theories typically constrain how the structure function scales with the separation distance $r$.
If the kinetic energy spectrum follows a power law $E\propto k^{-\beta}$, the second-order structure function with $q=2$ scales as \begin{align}
  \langle\Delta v^2 (r)\rangle\propto r^{\xi(2)}\qquad \xi(2) = \beta - 1. \label{eq:isotropic KE structure function}
\end{align}
For the homogeneous isotropic turbulence considered by Kolmogorov, $\beta =5/3$.
One advantage of the structure function is that, unlike the spectrum, structure functions may be defined for arbitrary orders $q$. This property becomes useful when the flow becomes ``intermittent'' and the strongest fluid gradients become clustered spatially rather than evenly distributed. The strength and character of turbulent intermittency can be quantified via the shape of the function $\xi(q)$ \citep{lovejoy2013}. 

Atmospheric turbulence does not in general follow the isotropic law (Eqn. \ref{eq:isotropic KE structure function}) due to the effects of buoyancy. A more accurate, but less well understood, description of atmospheric flow is ``Lovejoy-Schertzer'' turbulence where structure functions for horizontal wind components are a function of separation direction as well as distance. The hypothesized reason for the directional separation is that only vertical motions are affected by gravity, and so the quantity controlling the turbulent cascade along the vertical direction is related to buoyancy. In contrast, the horizontal component of the cascade is not affected by buoyancy and therefore structure functions follow the Kolmogorov law when calculated along a purely horizontal direction. \citet{lovejoy1985} proposed that structure functions should follow
\begin{align}
  \begin{cases}
    \langle\Delta v (\Delta x)\rangle = \varepsilon^{1/3} \Delta x^{H_h}; \qquad H_h=1/3; \qquad \Delta x \equiv |\mathbf{r}(\Delta x, \Delta z=0)|\\
      \langle\Delta v (\Delta z)\rangle = \phi^{1/5} \Delta z^{H_v}; \qquad H_v=3/5;\qquad \Delta z \equiv |\mathbf{r}(\Delta x=0, \Delta z)| 
\end{cases}\label{eq:LS directional structure functions}
\end{align} 
where $H_h$ and $H_v$ are called Hurst exponents for horizontal and vertical separations, respectively. The proportionality constant $\varepsilon$ represents Kolmogorov's ``kinetic energy flux'', i.e. the rate of kinetic energy transfer from large to small scales $\partial v^2/\partial t$. The analogous quantity $\phi=\partial f^2/\partial t$, called ``buoyancy variance flux'', is also a down-cascade variance flux analogous to kinetic energy but for buoyancy $f$ and was originally proposed by \citet{bolgiano1959} and \citet{obukhov1959} as being relevant for atmospheric turbulence. For both cases in Eqn. \ref{eq:LS directional structure functions}, the exponent values are determined simply by dimensional consistency after assuming $\langle \Delta v(\Delta x)\rangle$ is only a function of $\varepsilon$, which has units $\mathrm{m^3\,s^{-2}}$, while $\langle \Delta v(\Delta z)\rangle$ is only a function of $\phi$, which has units $\mathrm{m^5\,s^{-2}}$.

A key feature of Eqns. \ref{eq:LS directional structure functions} is the fact that the Hurst exponent differs based on the direction considered. As we will show, this directional dependence determines turbulon heights as a function of their widths. Considering isolines of constant $\langle \Delta v \rangle$ from Eqn. \ref{eq:LS directional structure functions}, the height and width of isolines may be related through
\begin{align}
  \Delta z = \ell_s^{4/9} \Delta x^{H_z}\qquad H_z=5/9;\qquad \ell_s = \frac{\varepsilon^{5/4}}{\phi^{3/4}}.\label{eq:spheroscale definition}
\end{align}
In Section \ref{sect:distribution of turbulon amplitudes across scale}, Eqn. \ref{eq:spheroscale definition} is reinterpreted as specifying the aspect ratio of the turbulons, with $\Delta z$ being proportional to turbulon height and $\Delta x$ being proportional to turbulon width. The constant $\ell_s$, termed the ``spheroscale'', determines the scale at which turbulon aspect ratios are equal to unity and the turbulons are isotropic. The exponent $H_z$ determines the rate at which turbulons become elongated horizontally as they grow larger, which is referred to as being ``stratified'' in LS turbulence \citep{lovejoy2013}. At large scales where $\Delta x\gg \ell_s$, atmospheric cross sections attain a highly layered appearance that would be consistent with, for example, a strong inversion that is also thermally stratified. Though it may be that these two notions of ``stratification'' are related or identical, it is not yet understood how thermal stratification affects the constants $\ell_s$ and $H_z$.

Given that the wind field transports numerous other quantities, the value of the Hurst exponent for wind affects the Hurst exponents for other variables, including any conserved advected scalar $\Phi$ following
\begin{align}
  \frac{D\Phi}{Dt} = 0.\label{eq:advection conservation}
\end{align} 
For simulations, below we will take $\Phi$ to be equal to moist static energy and total water content, but other conserved scalars could be simulated as well. 
In general, Hurst exponents for $\Phi$ are thought to be equal to the Hurst exponents for the wind field \citep{lovejoy2013}. In classical isotropic turbulence theory, the Hurst exponent is $1/3$, a result known as the Corrsin-Obukhov law of scalar advection. For Lovejoy-Schertzer turbulence, the analogous law is 
\begin{align}
  \begin{cases}
  \langle\Delta \Phi (x)\rangle\propto \Delta x^{H_h}; \quad & H_h=1/3, \\
  \langle\Delta \Phi (z)\rangle \propto \Delta z^{H_v}; \quad & H_v=3/5,
  \end{cases}\label{eq:LS scalar directional structure functions}
\end{align}
where $H_h$ and $H_v$ are equal to their values for the horizontal component of the wind field (Eqn. \ref{eq:LS directional structure functions}). 

\subsubsection[Observational evidence for anisotropic turbulence]{Observational evidence for\\* anisotropic turbulence} \label{sec:observational LS turb}

Lovejoy-Schertzer turbulence is not yet fully understood. The values of the horizontal and vertical Hurst exponents used in Eqn. \ref{eq:LS directional structure functions} were first proposed in two separate isotropic theories that assumed $H_h=H_v$ but took a different flux ($\varepsilon$ for \citet{kolmogorov1941}, $\phi$ for \citet{bolgiano1959,obukhov1959}) as being physically primary. It is not obvious a priori that these isotropic values would appear unmodified in Eqn. \ref{eq:LS directional structure functions}. Nonetheless, there is some empirical evidence that Eqn. \ref{eq:LS directional structure functions} does apply to atmospheric flow. 

In \citet{dewitt2025preprint} empirical structure functions were computed for horizontal wind components measured using global dropsonde and radiosonde datasets. When computed along the vertical direction for separations between $200\,\mathrm{m}$ and $7\,\mathrm{km}$, calculated exponents were near $H_v\approx 0.7$, just above the expected value of $3/5$. This value is consistent with the range for $H_v$ found by \citet{lovejoy2007}, also using dropsonde data, which was between roughly 0.6 and 0.75 depending on altitude. Along the horizontal direction, and for separations between $200\,\mathrm{m}$ and $2000\,\mathrm{km}$, \citet{dewitt2025preprint} obtained values for $H_h$ that were between roughly 0.4 and 0.5, which is slightly above the value $H_h=1/3$ in Eqn. \ref{eq:LS directional structure functions} but nonetheless closer to Lovejoy-Schertzer value than that predicted by quasi-geostrophic turbulence ($H_h=1$). Likewise, using aircraft data \citet{pinel2012} found values of $H_h$ between 0.37 and 0.35 for horizontal separations near $100\,\mathrm{km}$ and values for $H_v$ between 0.68 and 0.63 for vertical separations near $10\,\mathrm{m}$.

There is also evidence for anisotropic Lovejoy-Schertzer scaling in other atmospheric fields. \citet{lilley2004} found Hurst exponents for the aerosol backscatter ratio of $H_h=0.33\pm0.03$ and $H_v=0.60\pm0.04$, in excellent agreement with Eqn. \ref{eq:LS scalar directional structure functions}. For clouds, \citet{guillaume2018} found that the aspect ratio of cloud cross-sections scale almost exactly as prescribed by Eqn. \ref{eq:spheroscale definition}, although their reported results require some algebraic manipulation as we show in Appendix \ref{sec:guillame}.

Thus, for the wind field, for conserved advected scalars fields such as aerosols, and even for cloud condensate fields that are not conserved, it appears that atmospheric perturbations generally scale anisotropically according to Eqn. \ref{eq:spheroscale definition}. Values for $H_h$ and $H_v$ are often found to be slightly larger than the theoretical values $1/3$ and $3/5$, respectively, but nonetheless closer to LS theory than the alternatives. The broad aspect ratio scaling supports the turbulon picture, where perturbations such as circulations in the wind field or clouds are associated with turbulons that follow the same aspect ratio scaling law as the perturbations themselves.

\subsubsection{Structure functions to turbulons} \label{sect:structure functions to turbulons}

\begin{figure}[t]
\centering
\includegraphics[width=\textwidth]{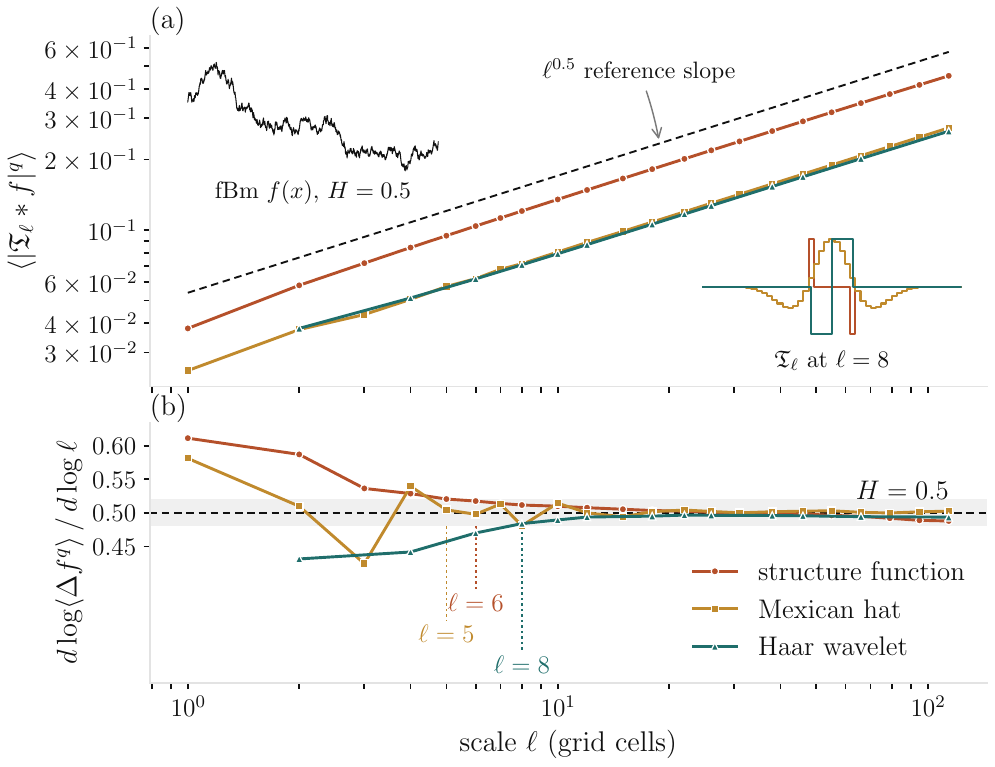}
\caption{First-order ``fluctuation functions'' (Eqn. \ref{eq:mean wavelet transform}), computed for a random walk (panel a, top left) using $q=1$. Shown are fluctuation functions computed using three choices of the envelope function $\mathfrak{T}$: the structure function (sienna), a Mexican hat (ochre), and a Haar wavelet (teal), alongside a $\ell^{0.5}$ reference slope (black dashed) representing the asymptotic power-law dependence in the limit of a perfectly well-resolved random walk. Local power-law exponents of the curves in panel (a), computed as $H = d\log \langle\Delta f\rangle_r/ d \log \ell $, depart from the theoretical value $H=0.5$ at small $\ell$ (panel b) due to discretization.} \label{fig:wavelet discretization}
\end{figure}

The structure functions considered above are typically used to analyze observed fields. But using the more general scale decomposition framework of wavelets, the process may be reversed to construct simulations. To see how structure functions and wavelets are related, consider that the differences in Eqn. \ref{eq:LS directional structure functions} may be replaced by a turbulon envelope function $\mathfrak{T}$ defined, in one dimension, by
\begin{align}
  \mathfrak{T}_\ell(x) = \delta\left(x \right) - \delta\left(x - \ell\right), \label{eq:delta pair envelope}
\end{align}
where $\delta$ is the Dirac delta function. The convolution $(\mathfrak{T}_\ell * f)(x) = f(x) - f(x - \ell)$ recovers the field differences at separation $\ell$ as in Eqn. \ref{eq:structure function}, and the structure function may be equivalently defined by
\begin{align}
\langle \Delta f^q\rangle(\ell) = \langle\big|\mathfrak{T}(\ell) * f\big|^q\rangle_r. \label{eq:mean wavelet transform}
\end{align}
If, for each $q$, the structure function Eqn.~\ref{eq:mean wavelet transform} follows a power law in $\ell$, there are other choices of the turbulon envelope function $\mathfrak{T}$ that result in an identical power-law exponent, at least over the range where discretization effects can be neglected \citep{lovejoy2023}. Eqn. \ref{eq:mean wavelet transform} therefore represents a generalization of the structure function to which we refer to as a ``fluctuation function.'' Where discretization effects dominate, different choices of $\mathfrak{T}$ do affect the power-law exponent. For example, Fig. \ref{fig:wavelet discretization} shows the mean ($q=1$) wavelet transform for a random walk, for three common choices of $\mathfrak{T}$ defined through a structure function, a Mexican hat, and a Haar wavelet. For each function, a local power law exponent is computed as $d\log \langle \Delta f^q\rangle/d\log \ell$ and shown in Panel (b). This local power-law exponent represents the Hurst exponent at each scale $\ell$, and it approaches its theoretical value of 0.5 only above a value of $\ell$ ranging from about 5 to 8. The consequence is that the choice of the turbulon envelope function does not affect the resulting exponent of the fluctuation function, but only at scales much larger than the grid scale, and that such discretization effects are not a specific limitation of the model but a general feature of discretization. The only limitation on the shape of $\mathfrak{T}$ is the admissibility criterion for discretized wavelets: namely, that they have zero mean.

Discretization effects aside, the wavelet-based definition of the fluctuation function in Eqn. \ref{eq:mean wavelet transform} allows us to interpret the Lovejoy-Schertzer law of scalar advection (Eqn. \ref{eq:LS scalar directional structure functions}) as specifying the mean amplitude $\langle a(\ell)\rangle_\mathbf{r}$ of turbulons $\mathfrak{t}$ as a function of their size $\ell$. This is because the convolution $\mathfrak{T}*F$ may be written as in Eqn. \ref{eq:basic field construction with turbulons}. Because a wavelet transform is invertible, the convolutions in Eqns. \ref{eq:mean wavelet transform} and \ref{eq:basic field construction with turbulons} may be used either to analyze existing fields, as is more common, or to construct new fields in simulation, as we turn to now.

\subsection{Adaptation to STEAM simulations}

The Lovejoy-Schertzer law (Eqn. \ref{eq:LS scalar directional structure functions}) provides the first important constraint on the statistics of turbulons, but more is required. We may separate the question into two parts: first, how turbulon amplitudes are distributed across scale $\ell$, averaging across spatial locations, and second, how turbulon amplitudes are distributed in space at a given scale.

\subsubsection[Distribution of turbulon amplitudes across scale]{Distribution of turbulon\\* amplitudes across scale} \label{sect:distribution of turbulon amplitudes across scale}

\begin{figure}[t]
\centering
\includegraphics[width=\textwidth]{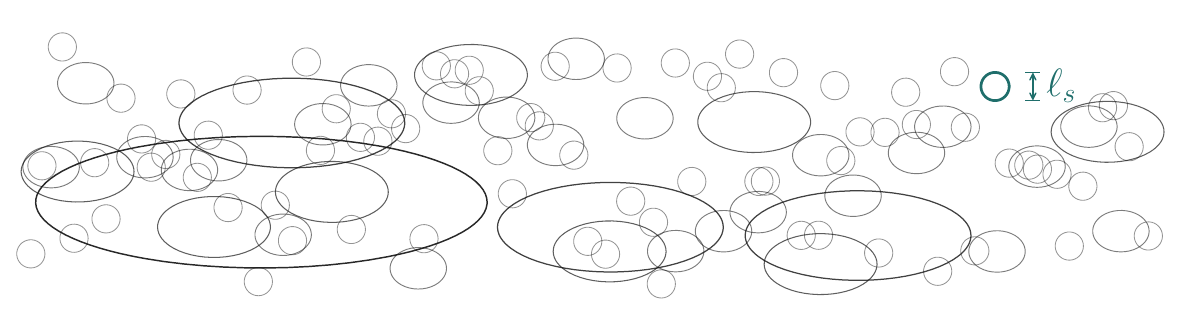}
\caption{Vertical slice along an ($x$--$z$) plane of an anisotropic cascade. Ellipses represent the support of individual turbulons of varying size $\ell$ and location. Turbulon aspect ratios follow Eqn. \ref{eq:turbulon aspect ratio scaling} such that large turbulons are horizontally elongated whereas turbulons near in size to the spheroscale $\ell_s$ have unit aspect ratio. } \label{fig:cascade slice}
\end{figure}

The Lovejoy-Schertzer law (Eqn. \ref{eq:LS scalar directional structure functions}) requires that mean turbulon amplitudes follow 
\begin{align}
  \langle |\mathfrak{t}|\rangle \propto \ell^{H_h},\label{eq:turbulon amplitude vs scale}
\end{align} 
with aspect ratios given by Eqn. \ref{eq:spheroscale definition} but with horizontal and vertical size parameters of the turbulons $\ell$ and $\ell_z$, respectively, in place of $\Delta x$ and $\Delta z$:
\begin{align}
  \frac{\ell_z}{\ell} = \ell_s^{1-H_z} \ell^{H_z-1},\label{eq:turbulon aspect ratio scaling}
\end{align}
Resulting turbulon cross-sections are pictured in Fig. \ref{fig:cascade slice}. Since Eqn. \ref{eq:turbulon aspect ratio scaling} implies $\ell_z\propto\ell^{H_z}$, then from Eqn. \ref{eq:turbulon amplitude vs scale} we have $\langle |\mathfrak{t}|\rangle\propto \ell_z^{H_h/H_z}$, where $H_h/H_z=3/5$ if $H_h=1/3$ and $H_z=5/9$, consistent with Eqn. \ref{eq:LS scalar directional structure functions} but with $\ell$ and $\ell_z$ replacing $\Delta x$ and $\Delta z$, respectively.

Thus the first input parameters for STEAM simulations are the horizontal Hurst exponent $H_h$, the anisotropy exponent $H_z$, and the spheroscale $\ell_s$. For the simulations considered here, we use $H_h=0.45$, which is slightly above the theoretical value $1/3$ and more consistent with observations (Section \ref{sec:observational LS turb}), $H_z=5/9$, and consider several different cases for $\ell_s$. Different values of $H_h$ and $H_z$ could also be used to simulate different turbulent flows, for example isotropic turbulence where $H_h=H_v=1/3, H_z=1$. 

The constant of proportionality in Eqn. \ref{eq:turbulon amplitude vs scale} must also be constrained. Consider that,
at its core, the basic action of a turbulent cascade is to transfer variability from large scales to small scales. In such a downscale cascade, the overall magnitude of the fluctuations at each scale are determined by the amplitude of the fluctuations at the largest scales. For STEAM, this idea is operationalized by computing the large scale fluctuation magnitudes using the ensemble mean profile. For simplicity, we consider a horizontally uniform domain mean, and so the large scale fluctuation magnitude is determined from the ensemble mean profile using differences at the scale of the largest simulated turbulon,  as described in Appendix \ref{sect:steam algorithm details}. This is done individually for each simulated scalar field. The resulting amplitudes at each height are interpreted as the mean absolute amplitude of the turbulon components at the outer scale. 
Finally, using Eqn. \ref{eq:turbulon amplitude vs scale}, outer scale amplitudes are used to set the proportionality constants for the mean turbulon amplitudes at every other scale. 

Because the gradient of the mean profile typically varies with height, the proportionality constants in Eqn. \ref{eq:turbulon amplitude vs scale} will also vary with height. This has two consequences. First, fields which have horizontally and vertically uniform mean profiles will not produce any variability in the simulated scalars, as should be expected. Convective activity should not be expected to modify local concentrations of an already well-mixed gas like argon. Second, any level within the domain that has a stronger mean gradient will produce more small-scale variance, as might also be expected intuitively. Stratocumulus fields near a strong inversion, for example, display much greater moisture variance within the cloud layer than in the mixed layer below because the proximity to the strong inversion allows for very dry pockets of air to intrude from the free troposphere above. However, as we argue in Section \ref{sect:field comparison}, it may be simplistic to set the constant for Eqn. \ref{eq:turbulon amplitude vs scale} using only the gradient magnitude as done here. In the future the sign of the gradient could also be considered.
 
\subsubsection{Distribution of turbulon amplitudes in space} \label{sect:distribution of turbulon amplitudes in space}

Although Eqn. \ref{eq:LS scalar directional structure functions} constrains the mean absolute turbulon amplitude as a function of scale $\ell$ globally, at any given location a turbulon amplitude should also depend on the local structure of the field. There are two particular aspects to consider: the amount of structure in the scalar field that is available to be advected, and the strength of the eddies that cause such advection to occur. Because advection represents a rearrangement of the field, there must be some nonzero gradient spanning the region in which a given eddy operates if that eddy is to introduce any perturbation in a scalar field. If the scalar field was initially uniform, an eddy might transport molecules from one location to another, but the scalar concentration would not change. If the scalar field does have a nonzero gradient, then it might be expected that the size of the perturbation caused by the eddy would be proportional to the magnitude of the gradient. 

As a first approximation, consider a rearrangement of a linearized gradient in some scalar field $\Phi$. The amplitude of the perturbation due to the rearrangement will be proportional to the gradient magnitude and the length of the eddy that causes this rearrangement. Since the eddies are highly anisotropic, they have very different vertical and horizontal lengths, so we can model advection by introducing a perturbation that is proportional to 
\begin{align}
  \delta \Phi \propto \ell\left|\nabla_h \Phi\right| + \ell_z\left|\frac{\partial \Phi}{\partial z}\right|, \label{eq:anisotropic gradient weighing}
\end{align}
where $\nabla_h$ represents the horizontal component of the gradient operator.

The second aspect of local field structure to consider is the strength of the eddy that causes a perturbation. In a turbulent cascade, eddy strength is determined by the down-cascade turbulent flux, i.e. the dynamically-relevant quantity that is conserved as it is passed from large scales to small. In isotropic Kolmogorov turbulence this quantity is the kinetic energy dissipation rate $\varepsilon$. In Lovejoy-Schertzer turbulence, there are two such fluxes $\varepsilon_h$ and $\phi$ (Eqn. \ref{eq:LS directional structure functions}), with the relevant flux dependent on the direction of the flow. To model eddy strength, STEAM simulations include a normalized flux quantity $\mathcal{F}$ as a prognostic variable alongside $h$ and $q_t$. The main difference between the turbulent flux and the scalar fields is that the turbulent flux is conserved at each scale through the cascade, implying that amplitudes of the flux component of the turbulons do not depend on scale. Instead, at each scale $\ell$ the mean absolute perturbation $\langle |\mathcal{F}_\ell |\rangle$ is equal to a constant that is independent of $\ell$, specified by the value of the nondimensional parameter $\varpi$. Because we use a normalized and dimensionless quantity for $\mathcal{F}$, the exact physical nature of the flux is not important and might depend on the flow being simulated. The important property for a turbulent flux is that it is conserved across scale. For each turbulon, we assume that the magnitude of the conserved scalar field component is proportional to the amplitude of the flux component.

The flux $\mathcal{F}$ is not spatially uniform. Local regions within a convective cell will have higher eddy dissipation rates than adjacent clear air, for example. We hypothesize that perturbations in the flux field at scale $\ell$ form with a mean absolute amplitude that is proportional to the local value of $\mathcal{F}_{>\ell}$, that is, the flux field associated with turbulons larger than scale $\ell$. As shown in Appendix \ref{sect:steam algorithm details}, this hypothesis naturally creates multiplicative intermittency in the flux field $\mathcal{F}$, with the strength of the intermittency determined by the mean absolute fluctuation perturbation $\langle |\mathcal{F}_\ell |\rangle$, or equivalently $\varpi$.

Thus, the flux controls the strength of the eddies while the scalar gradient controls the concentration that is available to be advected. Together, in STEAM these physical effects are modeled as follows, described in further detail in Appendix \ref{sect:steam algorithm details}. The simulation proceeds from large scales to small. At each scale, the first step is to compute perturbations in $\mathcal{F}_\ell$ that are associated with each turbulon of size $\ell$. This is done by drawing a random variable with zero mean and unit amplitude from the distribution described in Appendix \ref{sect:steam algorithm details}, and then scaling the perturbation by the local field $\mathcal{F}_{>\ell}$ while enforcing that $\langle | \mathcal{F}_\ell | \rangle$ is a constant, independent of scale, with its value determined by the input parameter $\varpi$. Perturbations in $h_\ell$ and $q_{t,\ell}$ are then computed such that the local value is proportional to, from Eqns. \ref{eq:anisotropic gradient weighing} and \ref{eq:turbulon aspect ratio scaling},
\begin{align}
  \delta h \propto \mathcal{F}_\ell\left(\ell|\nabla_h h| + \ell_s^{1-H_z}\ell^{H_z}|\frac{\partial h}{\partial z}|\right) \\
  \delta q_t \propto \mathcal{F}_\ell\left(\ell|\nabla_h q_t| + \ell_s^{1-H_z}\ell^{H_z}|\frac{\partial q_t}{\partial z}|\right).
\end{align}
The proportionality constant is set by enforcing that the ensemble mean $\langle |h_\ell|\rangle$ or $\langle |q_{t,\ell}|\rangle$ follows Eqn. \ref{eq:turbulon amplitude vs scale}. Finally, for all three fields, a perturbation with shape defined by the turbulon envelope function and amplitude scaled as described above are added to the parent fields, and the simulation proceeds to the next smaller scale.

\subsubsection{Diagnostic variables} \label{sect:diagnostic variables}

Once the cascade simulation has proceeded to the smallest resolvable scale for the three prognostic variables $h$, $q_t$, and $\mathcal{F}$, we then diagnose additional cloud-relevant variables as described in Appendix \ref{sect:steam algorithm details}. In short, we use a saturation adjustment scheme where any saturated grid cell is assigned a relative humidity of exactly 100\%, and the remaining water is converted to condensate. Temperatures and pressures that are required for saturation calculations are solved iteratively from the definition of moist static energy, a saturation vapor pressure parameterization \citep{bolton1980}, and an assumed surface pressure. Condensate is then partitioned into liquid and ice according to a simplistic linear partitioning based on the temperature, ramping from purely liquid at 0°C to purely ice at -38°C, matching the partitioning used in the Global Atmospheric Research Program Atlantic Tropical Experiment (GATE) SAM simulation described below. Overall, the diagnostics we use are simplistic and are intended only as a first proof-of-concept that the cascade simulation method can produce plausible simulated atmospheric volumes. A particularly egregious simplification is that condensate does not convert to precipitation, which can lead to unphysically high condensate values in dense convective cores (Section \ref{sect:field comparison}). In the future, precipitation could be added through an autoconversion process analogous to LES, where larger, saturated turbulons produce precipitation over the course of their lifetime. Additional microphysical processes, perhaps including aerosol-limited condensation that produces positive supersaturation, should also be added.

The full simulation process in STEAM is summarized in Fig. \ref{fig:steam pipeline} and described in more detail in Appendices \ref{sect:steam algorithm details} and \ref{ssect:steam algorithm spec}. Beginning at the largest scales of the simulation, mean $q_t$ and $h$ field gradients are used to set turbulon amplitudes at every scale according to the Lovejoy-Schertzer law of scalar advection. The cascade is then simulated from large scales to small, with the amplitude of each turbulon set by the larger-scale field gradient, the turbulent flux, and a random number. Once the cascade is run to the smallest scale, diagnostic variables are computed using a saturation adjustment scheme.

\begin{sidewaysfigure}
\centering
\includegraphics[width=\textwidth]{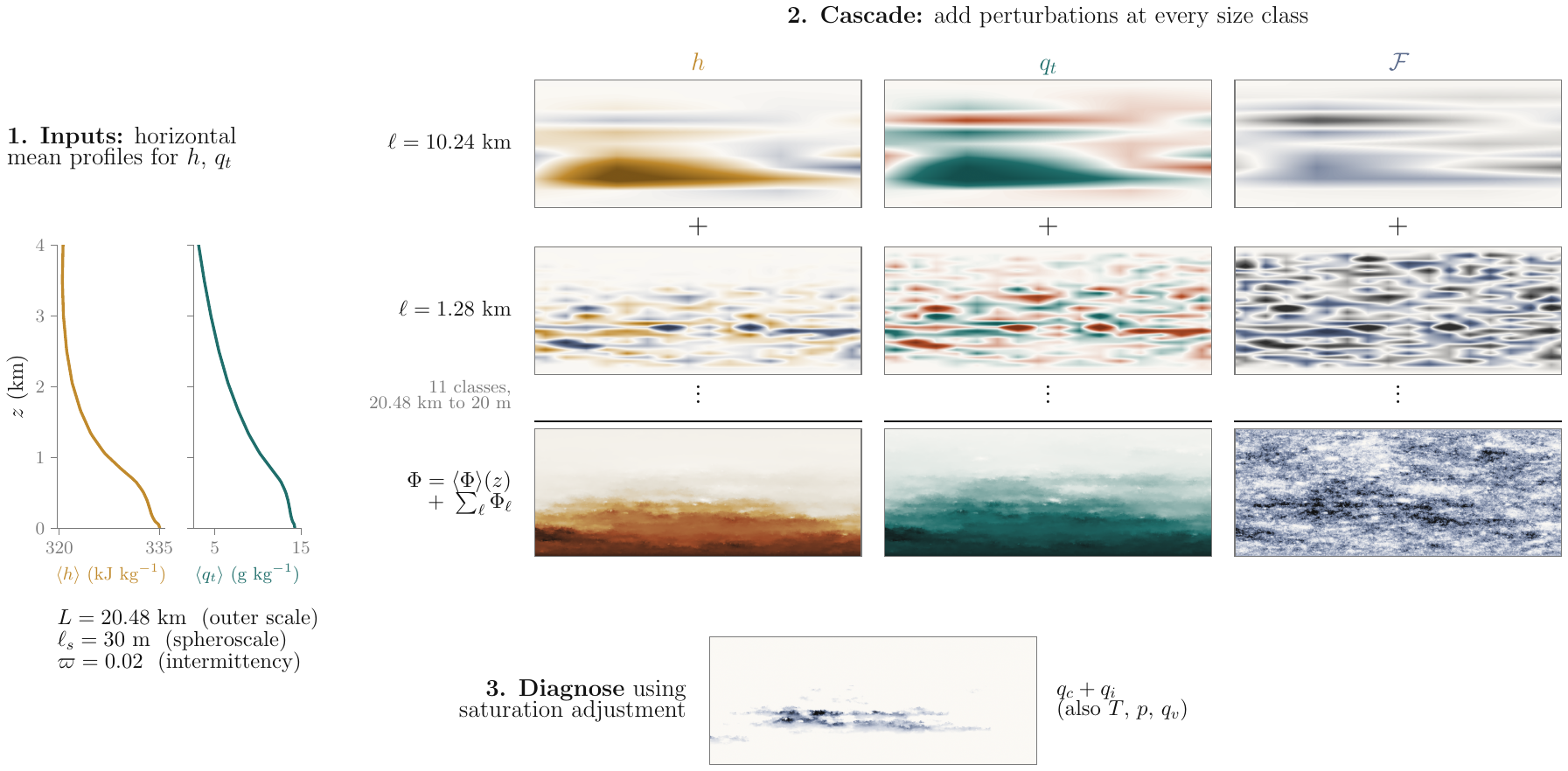}
\caption{The three steps of a STEAM simulation, illustrated with an $x$--$z$ slice through a 10.24~km by 4~km simulated region. (1) Inputs are horizontal mean profiles of $h$ and $q_t$ together with three constants. (2) Perturbations are added sequentially for each size class $\ell$; two of the eleven classes are shown. Each prognostic field $\Phi$ is its mean profile plus the sum of the perturbations across classes. (3) Condensate and the remaining diagnostic variables are computed by saturation adjustment (Section \ref{sect:diagnostic variables}).} \label{fig:steam pipeline}
\end{sidewaysfigure}

\section{Evaluation of STEAM simulation output} \label{sect:evaluation}

As a comparison for STEAM outputs, we consider output from two complementary sets of hydrodynamic simulations as well as satellite imagery derived from MODIS. The primary hydrodynamic comparison is to observationally constrained large-eddy simulations of deep convection, referred to as ``giga-LES'' simulations because their numerical grids contain over a billion points. The second hydrodynamic comparison is to a multimodel ensemble of coarser cloud-resolving simulations from the Radiative-Convective Equilibrium Model Intercomparison Project \citep[RCEMIP;][]{wing2018, wing2020}. Using STEAM and the giga-LES simulations, we also construct simulated cloud masks that are analogous to satellite imagery and compare their statistics to real MODIS imagery.

\subsection[Hydrodynamic comparisons and configuration of STEAM simulations]{Hydrodynamic comparisons and\\* configuration of STEAM simulations} \label{sect:hydrodynamic comparisons}

The two giga-LES cases were performed with the System for Atmospheric Modeling \citep[SAM;][]{khairoutdinov2003} with identical grid configurations spanning $204.8\,\mathrm{km}\times204.8\,\mathrm{km}$, using a doubly periodic domain with $100\,\mathrm{m}$ horizontal grid spacing and a vertical grid spacing of $100\,\mathrm{m}$ through the mid-troposphere.
For the first case, boundary conditions were taken from conditions observed during the Tropical Warm Pool -- International Cloud Experiment (TWP-ICE), which sampled monsoonal deep convection near Darwin, Australia in early 2006. The simulation was forced by large-scale thermodynamic tendencies derived from the observed conditions, following the specifications of the TWP-ICE cloud-resolving model intercomparison \citep{dazlichGigaLarge2013, fridlind2010}.
Because only two hours of giga-LES output were preserved, we use a single snapshot taken from the statistically steady portion of the simulation. The simulated wind field for this case was previously shown to closely follow the directional structure function predictions of Lovejoy-Schertzer turbulence (Eqn. \ref{eq:LS directional structure functions}) \citep{dewitt2025preprint}.

The second giga-LES case is forced by idealized mean conditions observed during Phase III of the Global Atmospheric Research Program Atlantic Tropical Experiment (GATE), which sampled oceanic deep convection over the tropical east Atlantic in 1974 \citep{khairoutdinov2009}. In contrast to the time-varying TWP-ICE forcing, the prescribed large-scale tendencies are constant in time. Over the 24-hour simulation, convection develops from shallow cumulus into deep convection within the first six hours and is quasi-steady beyond approximately hour 12 \citep{khairoutdinov2009}. We use a single snapshot taken from hour 23.

Both giga-LES simulations are constrained by observations. Large-scale advective tendencies of temperature and moisture are prescribed as horizontally uniform profiles, and the horizontal mean wind profile is relaxed toward the observed mean profile \citep{khairoutdinov2009, fridlind2010}. The TWP-ICE case additionally has its horizontal mean temperature and humidity profiles relaxed toward observed profiles above ${\sim}15\,\mathrm{km}$, where the observationally derived forcing tendencies are poorly constrained \citep{fridlind2010}. STEAM's use of specified mean profiles for $h$ and $q_t$ is analogous to the tendencies prescribed in SAM in the sense that only the horizontal mean is prescribed. In both the giga-LES and in STEAM, all local variability arises from the intrinsic dynamics of the model.

As a complementary form of evaluation, we also compare STEAM simulations to hydrodynamic model output from the Radiative-Convective Model Intercomparison Project (RCEMIP). RCEMIP simulations are run in radiative-convective equilibrium, an idealization of the tropical atmosphere in which convection interacts with interactive radiation and surface fluxes over a fixed, horizontally uniform sea surface temperature, with spatially uniform insolation, no rotation, and no diurnal cycle \citep{wing2018}. Radiative-convective equilibrium is a less realistic boundary condition than the forcings prescribed for the giga-LES simulation as it does not include latitudinal variability in forcing. It is included here as a complementary evaluation that establishes the magnitude of the variability present between different hydrodynamic models even under identical forcings. 

We use the \texttt{RCE\_large300} configuration for our comparisons: an elongated, doubly periodic channel of approximately $6000\times400\,\mathrm{km}^2$ with $3\,\mathrm{km}$ horizontal grid spacing and a sea surface temperature of 300 K, integrated for 100 days. In this configuration, convection produces moist convecting regions and dry, nearly cloud-free regions whose number, spatial scale, and orientation differ among models \citep{wing2020}. 

From the RCEMIP archive we use the nine models that provide usable three-dimensional snapshot output for the 300 K channel: SAM, CM1, SCALE, UCLA-CRM, three configurations of the Met Office Unified Model, and two configurations of ICON. We use three snapshots separated by ten days near the end of the integration, chosen so that each snapshot is equilibrated and approximately statistically independent. Two archived models are excluded for data integrity reasons. A complete specification of the comparison data and analysis is provided in Appendix~\ref{app:steam supplement}.
Notably, SAM appears in both comparison sets, at $100\,\mathrm{m}$ resolution for the giga-LES cases and at $3\,\mathrm{km}$ resolution in RCEMIP. The hydrodynamic ensemble therefore contains a shared underlying model spanning very different domain and boundary condition configurations (SAM), while also containing a set of shared domain and boundary condition configurations that span very different underlying models (RCEMIP).

In any hydrodynamic simulation, statistics computed at the grid scale are often strongly affected by numerical approximations such as hyperdiffusion or filtering that tend to damp grid-scale variability. A model's ``effective'' resolution is often considered to be a few times larger than the grid spacing due to these approximations. As we will show, such small-scale artifacts can be easily identified using fluctuation functions, most commonly manifesting as a reduction in small-scale variability relative to what scale invariance would predict.
In contrast, STEAM's effective resolution is equal to the grid spacing because its fields are scale invariant to the grid scale by construction. 
To reduce these small-scale artifacts in the hydrodynamic models, we first coarsen output by averaging consecutive $2\times 2 \times 2$ grid points before computing any one-point statistics such as standard deviations. Analysis functions that measure variability across scales, such as fluctuation functions and cloud fractal metrics, are computed at native output resolution in order to show numerical artifacts directly.

A distinct but closely related challenge is that one-point statistics such as standard deviation depend strongly on the resolution of the data they are computed from, especially for scale invariant fields \citep{lovejoy2013}. Due to this effect, standard deviations for different model output can only be expected to agree if the resolutions are the same.
For any comparisons made using one-point statistics, STEAM is configured such that both the horizontal and vertical resolutions are nearly matched to the coarsened hydrodynamic output being compared (see Appendix~\ref{app:steam supplement}). 

Because STEAM requires horizontal mean profiles of the prognostic variables $q_t$ and $h$ as inputs, we run an ensemble of STEAM simulations for each individual hydrodynamic case. For example, for the CM1 channel case we consider three timesteps. From these timesteps, we compute horizontal mean profiles for $q_t$ and $h$ and then run the STEAM ensemble using these profiles as inputs. The hydrodynamic model configuration from which profiles are sourced (here, the CM1 channel case) is referred to as the ``host'' model. The grid resolution of the STEAM ensemble is matched to the coarsened host, and a full list of horizontal and vertical resolutions is shown in Appendix \ref{ssect:steam config}.

STEAM also requires as inputs a profile for the spheroscale, a constant $\varpi$ specifying the amount of intermittency in the flux, and a constant for the outer length scale $L$.
The outer scale $L$, which represents the size of the largest turbulon, is most simply set to the domain length such that the largest turbulon is the largest that could fit inside the domain. For the channel configurations, it is ambiguous which domain length constrains the size of the largest turbulon, so we consider two comparisons for the channel cases: the outer scale being set to either the channel width or length. For the spheroscale profile, we use a uniform value of $10\,\mathrm{m}$ which is consistent with observations \citep{dewitt2025preprint}. In the real atmosphere the spheroscale is likely to display large spatial variability, so the constant used here is a key approximation that we return to in Section \ref{sect:visual renderings}. It is less clear what the value for the flux perturbation scaling factor $\varpi$ should be, and it is more difficult to constrain observationally. We consider three cases: $\varpi\in\{0.02, 0.05, 0.17\}$.
For each of the 11 host configurations and three values for $\varpi$, 10 STEAM simulations are run so that the total number of STEAM simulations is $11\times 10\times 3 = 330$.

\subsection[Comparison between hydrodynamic- and STEAM-simulated fields]{Comparison between\\* hydrodynamic- and STEAM-simulated fields} \label{sect:field comparison}

We begin our comparison using profile statistics and single-level probability density functions for fields simulated by STEAM and the hydrodynamic models. These comparisons are made between coarsened hydrodynamic models and STEAM simulations with domain geometry and mean profiles for $h$ and $q_t$ matched to each hydrodynamic case.

\begin{figure}
\centering
  \includegraphics[width=\textwidth]{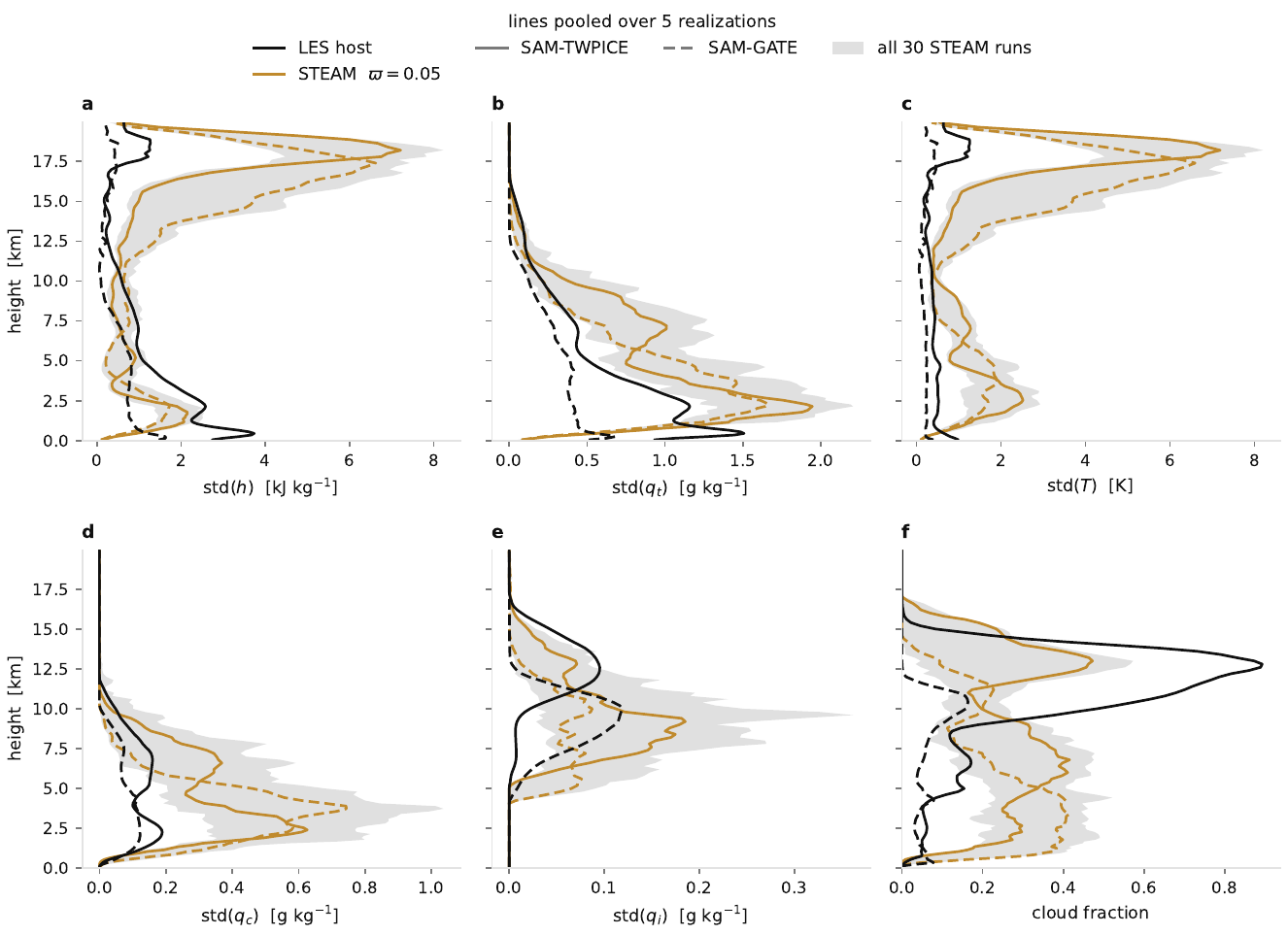}
  \caption{
Standard deviation profiles of moist static energy $h$, total water mixing ratio $q_t$, cloud liquid and ice mixing ratios $q_c$ and $q_i$, temperature $T$, and cloud fraction. For STEAM, profiles (ochre lines) are computed from five statistically independent fields for the moderate value $\varpi=0.05$, using the TWPICE or GATE profiles for $h$ and $q_t$ as inputs. Profiles for other values of $\varpi$ are shown in Appendix~\ref{sect:additional comparison figures} and not substantially different from the $\varpi=0.05$ case. Standard deviation profiles for the hydrodynamic simulations (black lines) are instead computed using only a single statistically independent snapshot. The grey shading represents snapshot-to-snapshot variability in STEAM, showing the minimum and maximum at every height for profiles computed for each individual STEAM field, spanning the five ensemble members and three values for $\varpi$.} \label{fig:profile stats}
\end{figure}

STEAM shows some agreement with several notable departures from the host models (Fig. \ref{fig:profile stats}).
In both GATE and TWPICE, the upper-level peak in cloud fraction occurs at nearly the same altitude as it does in the host simulations, but the magnitude only matches for the GATE case, and in all cases STEAM produces roughly two to five times larger cloud fraction at lower levels. Per-level standard deviations for liquid and ice condensate are generally higher in STEAM, with the exception of the GATE case where STEAM standard deviation for $q_i$ is more comparable to its host. For variability in $h$, departures between STEAM and its host are of a similar magnitude to departures between the two hydrodynamic cases below approximately 12~km. At upper levels STEAM has far higher variability in $h$ than either SAM output, which also drives a peak in temperature variability there. At lower levels, STEAM also has two to four times more variability in temperature $T$ than either hydrodynamic simulation, but at least some snapshots are closer to their hydrodynamic hosts in the mid troposphere between about 5~km and 12~km.

The high altitude peak in $h$ and $T$ variability, along with higher variability in $q_c$ and, to a lesser extent higher low-level cloud fraction, are the clearest departures between STEAM and the host simulations. A less obvious departure is $h$ and $q_t$ variability at the surface, where variability drops in all models but only reaches zero in STEAM. Elsewhere, departures between STEAM and the host model are generally of a comparable magnitude to departures between the two hydrodynamic models themselves. Intriguingly, the value of the flux perturbation scaling parameter $\varpi$ appears not to substantially affect profiles for standard deviation, full plots for which are shown in Appendix~\ref{sect:additional comparison figures}.

\begin{sidewaysfigure}
\centering
\includegraphics[width=\textwidth,height=0.88\portraitwidth,keepaspectratio]{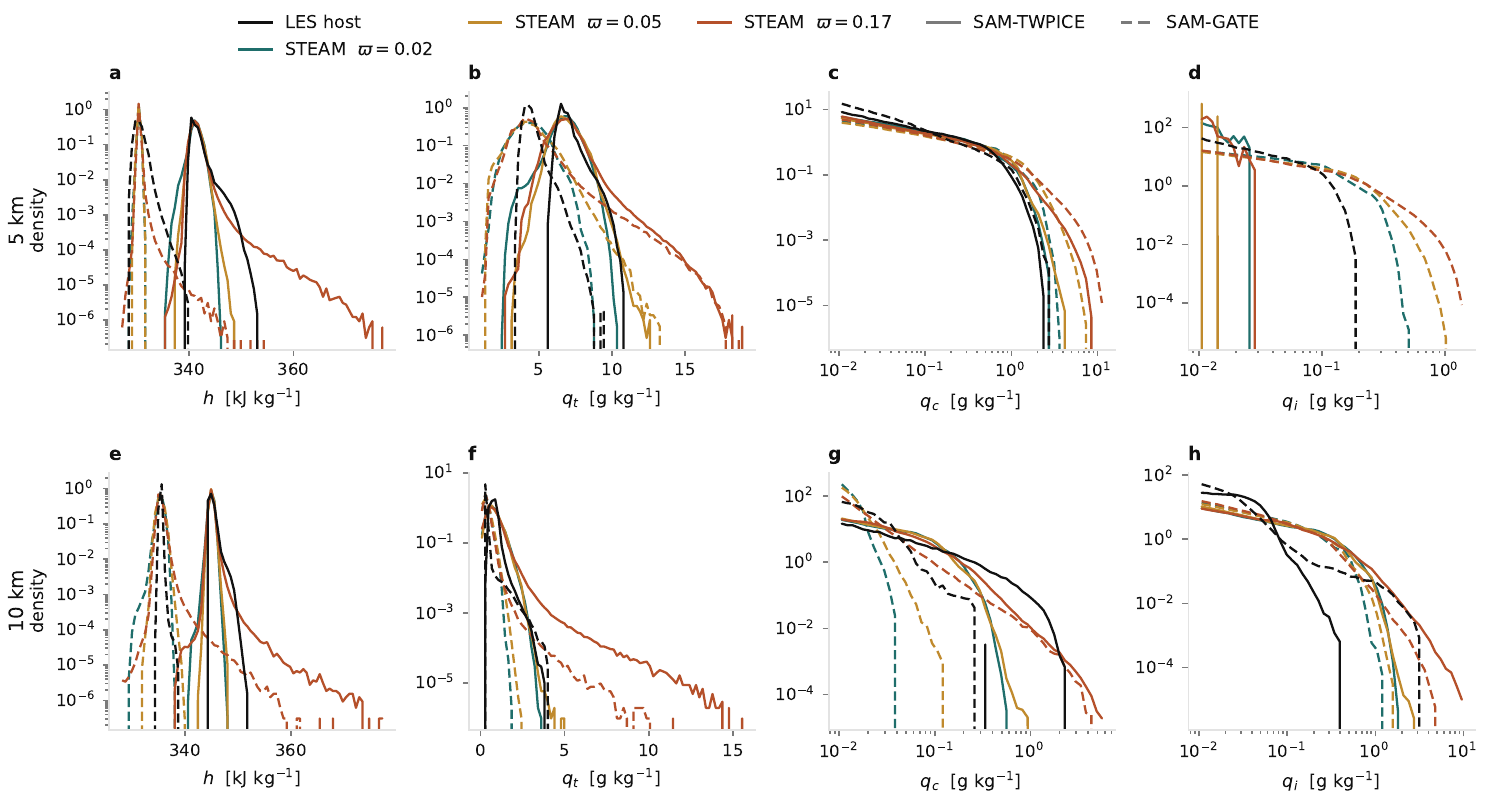}
\caption{Probability density functions for moist thermodynamic variables at the model levels nearest 5~km and 10~km altitude for the giga-LES host and matched STEAM simulations, for values of $\varpi$ ranging from $0.02$ (teal), $0.05$ (ochre), and $0.17$ (sienna).} \label{fig:level pdfs}
\end{sidewaysfigure}

The probability density functions for 5~km and 10~km altitudes shown in Fig. \ref{fig:level pdfs} show general agreement between the host and STEAM outputs, and in this case the effect of changing $\varpi$ is more noticeable.
In nearly all cases the host PDF is bracketed by the STEAM outputs across values for $\varpi$, with larger values for $\varpi$ showing fatter tails in the distributions.

\begin{sidewaysfigure}
\centering
\includegraphics[width=\textwidth,height=0.88\portraitwidth,keepaspectratio]{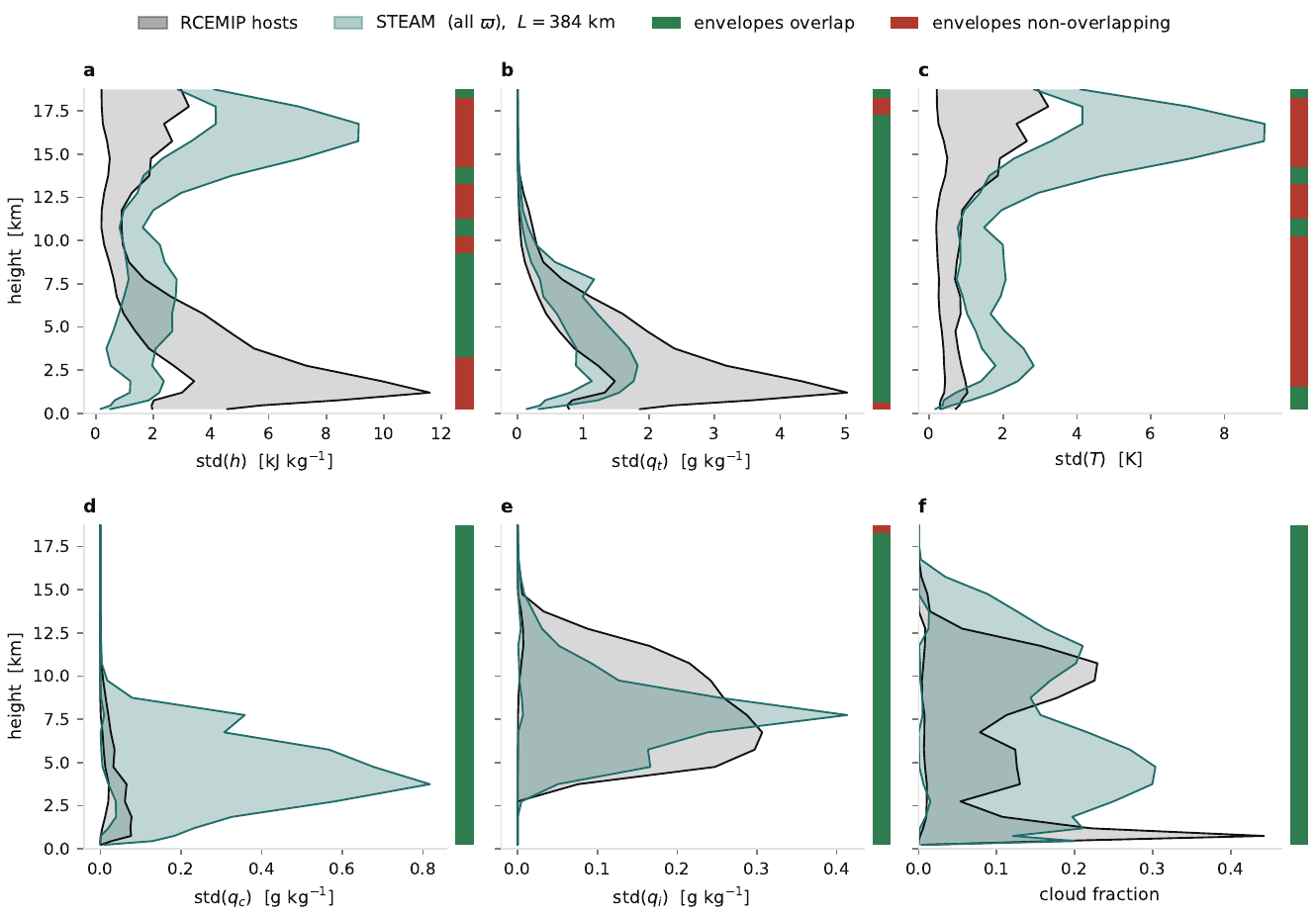}
\caption{As in Fig. \ref{fig:profile stats} but for the nine RCEMIP channel hosts and the matched STEAM ensemble. For both the hydrodynamic and STEAM ensembles, only the profile spread is shown for visual clarity.} \label{fig:rcemip profiles}
\end{sidewaysfigure}

Similar themes are present in the RCEMIP comparison, at least for the configuration where the outer scale is set to the channel width (Fig. \ref{fig:rcemip profiles}).  For these cases, at least some STEAM profiles are overlapping with at least some RCEMIP profiles for most variables and heights. The most notable exception is again high-elevation moist static energy and temperature variability, which is substantially higher for STEAM. Surface-level variability in $h$ and $q_t$ is also lower for STEAM. For liquid condensate, most STEAM members show vastly higher variability, although the driest STEAM members overlap the wettest RCEMIP members at all levels. For ice condensate, the STEAM and RCEMIP ensembles are very similar. Profiles and single-level distributions for the alternate configuration with larger outer scale are shown in Appendix~\ref{sect:additional comparison figures}, where variability is generally increased relative to Fig. \ref{fig:rcemip profiles}.

Increased variability of liquid condensate in STEAM would be consistent with values for $q_c$ being artificially inflated because precipitation is not present in the model. The high-altitude peak in $h$ and $T$ variability likely arises because the mean profile for $h$ warms sharply at the tropopause, where STEAM's gradient weighting mechanism (Eqn. \ref{eq:anisotropic gradient weighing}) reads a substantial large-scale gradient magnitude and cascades that large-scale variability down to smaller scales regardless of its sign. In the real atmosphere, the stratospheric inversion instead suppresses motion, suggesting that in the future, the sign of the large scale vertical gradient should influence STEAM's dynamics. Both aspects suggest future work is needed to address these limitations in STEAM. 

The broad agreement otherwise is promising and perhaps somewhat remarkable. In these simulations, STEAM reads only the large-scale vertical gradient in the mean profile, translates this into a horizontal perturbation, and then extrapolates that perturbation over approximately a factor of 1000 in horizontal scale. That the resulting variability at each horizontal level would match that of a hydrodynamic model even to within an order of magnitude is not obvious a priori.

\begin{sidewaysfigure}
\centering
\includegraphics[width=\textwidth,height=0.88\portraitwidth,keepaspectratio]{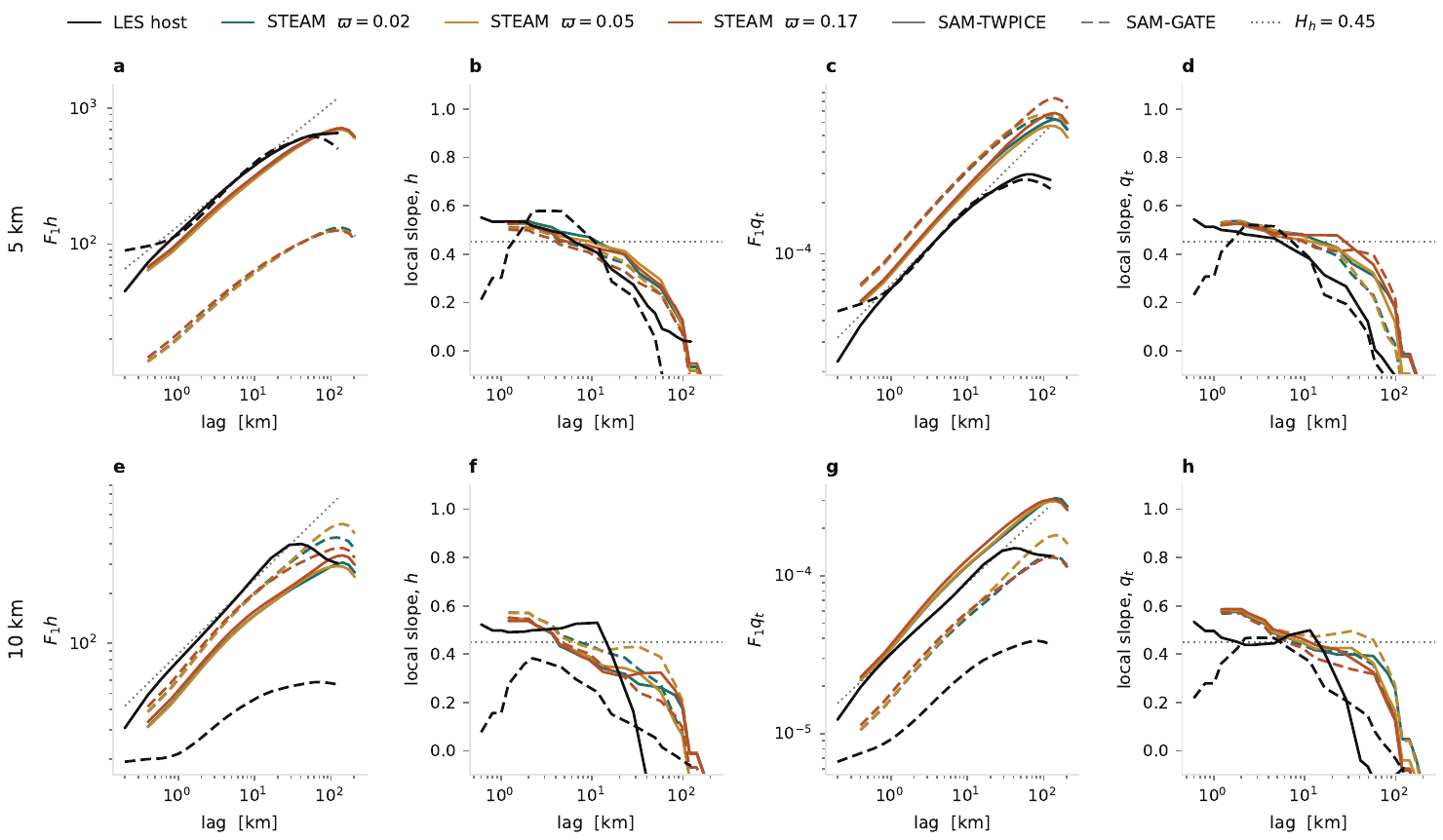}
\caption{Horizontal fluctuation functions of $h$ and $q_t$ at the levels nearest 5~km and 10~km for the giga-LES hosts and matched STEAM simulations, with local power-law exponents computed as described in Section \ref{sect:structure functions to turbulons}.} \label{fig:fluctuation functions}
\end{sidewaysfigure}

Moving beyond one-point statistics, we now consider fluctuation functions as defined in Eqn. \ref{eq:mean wavelet transform}, computed for 5~km and 10~km levels in the horizontal direction using a Mexican Hat wavelet. Computations are performed using the Python package \verb|scaleinvariance| \citep{dewitt2026scaleinvariance}. To estimate power-law exponents, we use a linear regression to the logarithm of the fluctuation function vs. the logarithm of the scale parameter. To better examine the scaling behavior, we define a ``local exponent'' for each scale $r$, which represents the power-law exponent fit to scales within the half order-of-magnitude surrounding $r$. This function is plotted in Fig. \ref{fig:fluctuation functions} next to the fluctuation functions themselves for moist static energy $h$ and total water content $q_t$ for the giga-LES comparison. For a scale-invariant power-law function with a constant exponent, the local exponent will be constant with scale, as it is for STEAM and the TWPICE SAM case at mid-range scales near 10~km. As the scale is increased toward the outer scale, in all cases the fluctuation function rolls off and local slopes decrease toward zero. The outer scale is plausibly 2 to 4 times smaller in the TWPICE hydrodynamic simulation than in STEAM. Local slope plots for the SAM GATE simulation are much less constant, indicating that simulation is substantially less scale invariant than either STEAM or the other hydrodynamic case. Also notable is that the magnitude of the fluctuation functions agrees with the TWPICE case over much of the observed range, while for GATE fluctuation function magnitudes are much lower, particularly for the 10~km altitude slice.

\subsection[Comparison of simulated and observationally-derived satellite imagery]{Comparison of simulated and\\* observationally-derived satellite imagery} \label{sect:satellite comparison}

A complementary way of evaluating STEAM simulation output is to consider the geometry of cloud objects as if they were viewed from space. Because atmospheric clouds are fractal, it is most natural to quantify their geometry by considering scale invariant metrics \citep{dewitt2026}. Such metrics are also well suited for comparisons between models and observations with differing resolutions, such as the giga-LES and satellite observations we consider here. This is because they represent statistical relationships across many different scales rather than at any single scale, and so their value is not affected by resolution unlike the metrics considered in Section \ref{sect:field comparison} \citep{dewitt2026}.

Most generally, fractal metrics quantify how some measure of cloud structure scales with some measure of cloud size. We consider four specific metrics relating cloud structure to cloud size that are defined below: the individual and ensemble fractal dimensions $D_i$ and $D_e$ and size distribution exponents for cloud area $\tau_\mathrm{area}$ and perimeter $\tau_{\mathrm{per.}}$. 
The first metric, the individual fractal dimension, measures how convoluted individual cloud perimeters become as the cloud under consideration grows larger. It is defined by measuring the rate at which individual cloud perimeters increase as the cloud's bounding box area grows, with higher values indicating a more convoluted perimeter. 
The second dimension is the ``ensemble'' fractal dimension, which quantifies both how convoluted individual cloud perimeters are in addition to the relative frequency of small and large clouds. The dimension is obtained via a linear regression to a logarithmically-transformed plot of the count of cloud-edge pixels that are separated less than a distance $r$, a function known as the ``correlation integral''.  The computation of both fractal dimensions follows the recommendations of \citet{dewitt2026}.
The third and fourth metrics, the size distribution exponents $\tau_\mathrm{area}$ and $\tau_{\mathrm{per.}}$, are computed using a linear regression to a logarithmically-transformed histogram of cloud area or perimeter. 

An extensive theoretical, empirical, and methodological investigation of all four metrics was described in \citet{dewitt2024,dewitt2024b} and \citet{dewitt2026}, with recommendations implemented in the Python package \verb|objscale| \citep{dewitt2026objscale}. We defer further methodological details and choices to those papers and use \verb|objscale| functions with defaults for the computation of the four metrics. When a metric is flagged as not being fit over a sufficiently wide range of scales by \verb|objscale|, we do not report it. For RCEMIP, the narrow channel geometry precludes a sufficiently wide range of cloud areas or perimeters to be resolved for any metric, and so we exclude RCEMIP entirely.

We also compute the fractal metrics for a selection of 72 satellite images, collected using MODIS during January 2021. These particular granules were filtered by \citet{dewitt2026} to remove granules containing bright noncloud objects such as sun glint or land, allowing a simple reflectivity-based threshold to be used to define cloud boundaries. 

STEAM simulations are constructed with a domain geometry that is broadly similar to MODIS granules, using a $1\,\mathrm{km}$ horizontal grid spacing in a square domain spanning $2048\,\mathrm{km}\times 2048\,\mathrm{km}$. Both are comparable to the resolution and extent of the MODIS imagery, which has a nadir resolution of approximately $1\,\mathrm{km}$ and a domain size of roughly $1950\times2030\,\mathrm{km}$. For each value of the intermittency parameter $\varpi$, we simulate ten independent realizations.

For both STEAM and the giga-LES volumes, simulated albedo fields are constructed by first computing a column optical depth field $\tau_c$ from the liquid and ice water paths, as implemented in the \verb|cloudyview| Python package as
\begin{align}
  \tau_c = \frac{3}{2}\frac{\mathrm{LWP}}{\rho_w\, r_{e,\mathrm{liq}}} + \mathrm{IWP}\left(a + \frac{b}{r_{e,\mathrm{ice}}}\right), \label{eq:bulk optics}
\end{align}
where the liquid and ice water paths $\mathrm{LWP}$ and $\mathrm{IWP}$ are obtained by vertically integrating the condensate fields within each column, $\rho_w$ is the density of liquid water, and the assumed effective radii are $r_{e,\mathrm{liq}} = 10\,\mathrm{\mu m}$ and $r_{e,\mathrm{ice}} = 30\,\mathrm{\mu m}$. The liquid term is the geometric optics relationship from \citet[Eqn. 7.86 of][]{petty2006} and the ice term is the visible-band ice parameterization of \citet{ebert1992}, with $a = 3.448\times10^{-3}\,\mathrm{m^2\,g^{-1}}$ and $b = 2.431\,\mathrm{\mu m\, m^2\,g^{-1}}$ for ice water paths in $\mathrm{g\,m^{-2}}$. 

From the column optical depth, visual albedo is computed using the two-stream solution for the reflectivity of a nonabsorbing cloud layer \citep[Eqn. 5.51 of][]{bohren2008}:
\begin{align}
  A = \frac{(1-g)\,\tau_c}{(1-g)\,\tau_c + 2}, \label{eq:two stream albedo}
\end{align}
with asymmetry parameter $g = 0.85$. With the solar zenith adjustment described in Appendix~\ref{app:steam supplement}, MODIS reflectivity and model albedo are comparable for clouds. For both, cloud masks are defined from albedo $A$ or reflectivity $R$ using thresholds of $\{0.1, 0.2, 0.3\}$, the range used by \citet{dewitt2026}. Because $A$ increases monotonically with $\tau_c$, each albedo threshold is equivalent to a threshold in column optical depth, with the three values of $A$ corresponding to $\tau_c \approx 1.5$, $3.3$, and $5.7$.

With the exception of the TWPICE giga-LES case at R=0.3, all of the geometric scaling functions defining the four fractal metrics are well represented by a power law (not shown), indicating that the geometry of observed and most simulated clouds is scale invariant.
The calculated values of the fractal metrics are shown in Table \ref{tab:cloud geometry}.
\begin{table}[t]
\centering
\caption{Cloud geometric exponents as a function of reflectance threshold $R$ for STEAM, the giga-LES hydrodynamic model SAM, and MODIS observations. $D_e$ is the ensemble fractal dimension, $D_i$ the individual fractal dimension ($D_i$ in DeWitt et al., 2026), and $\tau_{\mathrm{area}}$, $\tau_{\mathrm{per}}$ the size distribution exponents for individual cloud area and perimeter. ``CF'' is cloud fraction. Bolded model-derived estimates represent those that are closest (or tied for closest) to the MODIS value. Estimates are omitted when they fail the most stringent statistical robustness checks described in Appendix~\ref{app:steam supplement} as recommended by \citet{dewitt2024b, dewitt2026,dewitt2026objscale}.}
\label{tab:cloud geometry}
\begin{tabular}{lccccc}
\hline
 & CF & $D_e$ & $D_i$ & $\tau_{\mathrm{area}}$ & $\tau_{\mathrm{per}}$ \\
\hline
\multicolumn{6}{l}{\textbf{$R > 0.1$}} \\
MODIS & 0.54 & 1.77 & 1.37 & 0.85 & 1.27 \\
\cline{1-6}
SAM-GATE & 0.31 & 1.68 & 1.50 & - & - \\
SAM-TWPICE & 0.95 & 1.47 & -- & - & - \\
STEAM $\varpi = 0.02$ & 0.52 & 1.60 & 1.26 & 0.68 & 1.06 \\
STEAM $\varpi = 0.05$ & 0.52 & 1.64 & 1.27 & 0.70 & 1.13 \\
STEAM $\varpi = 0.17$ & {0.54} & \textbf{1.73} & \textbf{1.31} & \textbf{0.80} & \textbf{1.23} \\
\hline
\multicolumn{6}{l}{\textbf{$R > 0.2$}} \\
MODIS & 0.41 & 1.75 & 1.40 & 0.88 & 1.28 \\
\cline{1-6}
SAM-GATE & 0.23 & 1.62 & \textbf{1.28} & - & - \\
SAM-TWPICE & 0.78 & 1.57 & 1.19 & - & - \\
STEAM $\varpi = 0.02$ & {0.48} & 1.60 & 1.26 & 0.64 & 1.05 \\
STEAM $\varpi = 0.05$ & {0.49} & 1.63 & 1.26 & 0.67 & 1.06 \\
STEAM $\varpi = 0.17$ & 0.50 & \textbf{1.73} & \textbf{1.27} & \textbf{0.80} & \textbf{1.21} \\
\hline
\multicolumn{6}{l}{\textbf{$R > 0.3$}} \\
MODIS & 0.32 & 1.73 & 1.38 & 0.89 & 1.29 \\
\cline{1-6}
SAM-GATE & {0.19} & 1.58 & 1.24 & - & - \\
SAM-TWPICE & 0.58 & 1.61 & 1.21 & - & - \\
STEAM $\varpi = 0.02$ & {0.45} & 1.59 & \textbf{1.27} & 0.62 & 1.02 \\
STEAM $\varpi = 0.05$ & {0.45} & 1.62 & 1.26 & 0.66 & 1.05 \\
STEAM $\varpi = 0.17$ & 0.47 & \textbf{1.72} & \textbf{1.28} & \textbf{0.73} & \textbf{1.18} \\
\hline
\end{tabular}

\end{table}
The only parameter that can be reliably estimated in all datasets is the ensemble fractal dimension $D_e$, with the individual fractal dimension also estimated for some of the SAM cases. Size distribution exponents could not be estimated reliably for any of the SAM cases. In all cases, metrics computed for the STEAM simulation with $\varpi=0.17$ are equal or closer to MODIS observations than any of the alternatives. 

\subsection{Visual renderings of STEAM clouds} \label{sect:visual renderings}

As a final analysis of STEAM outputs, we turn to renderings of simulated clouds. \citet{palmer2016} proposed a ``Palmer-Turing test''\footnote{Referred to there as a ``climatic Turing test''; we use the terminology put forward by \citet{christensen2021}.} as a standard for model fidelity: at present, outputs from climate simulations may be quite easily distinguished from observational data. Clearly, if a given simulation was fully realistic, its outputs should not be distinguishable from real imagery if the simulation and observation resolutions are matched. Visual realism can therefore be seen as a necessary but not sufficient condition for a model to be considered realistic. Visual renderings are also useful because they can guide intuition about model behavior, such as how the spheroscale $\ell_s$ or the flux intermittency parameter $\varpi$ affect model outputs as we will show.

We use a ray-marching algorithm from the cloud visualization engine \verb|cloudyview| \citep{dewitt2026cloudyview}.\footnote{A more physically-based Monte Carlo method, also from \texttt{cloudyview}, that would produce physically accurate pixel radiances was attempted but found computationally intractable due to the slow convergence of multiple scattering in even moderately thick clouds.} For each pixel in the rendered image, a ray is traced outward from the camera location through the 3D model output. When a ray intersects a grid point containing liquid or ice condensate, an additional ray march from the grid point toward the solar light source determines that grid point's illumination and how much light is scattered toward the camera. Phase functions and multiple scattering are approximated by visually tuned heuristics so that lighting and opacity of rendered clouds are tuned while the three-dimensional geometry and perspective from the camera's location are physically correct. The algorithm is computationally inexpensive enough that a consumer GPU can render views in real time as if in a video game. An interactive demo of STEAM-simulated clouds, alongside hydrodynamic model output for comparison, is available for this purpose at \texttt{thomasddewitt.com/soar}.

We focus here on renderings for a small domain periodic channel simulation spanning 20.48~km in the longer dimension, 10.24~km in the shorter, and 4~km in the vertical. The horizontal grid spacing is set to 10~m for a grid size in the horizontal of 2048$\times$1024. Inside this ``parent'' simulation, we also simulated a ``nested'' domain, which is produced by continuing the cascade over a small subregion of the parent, as described in Appendix \ref{sect:steam algorithm details}. Specifically, within a 1.28 by 2.56~km region of the parent, the cascade is continued to a final horizontal grid spacing of 2.5~m, corresponding to a grid size of 512 by 1024 for the nested domain. Profiles for $h$ and $q_t$ are used from the CM1 channel simulation, which were found to result in cloud fractions that are conducive to visualization. 

\begin{sidewaysfigure}
\centering
\includegraphics[width=\linewidth]{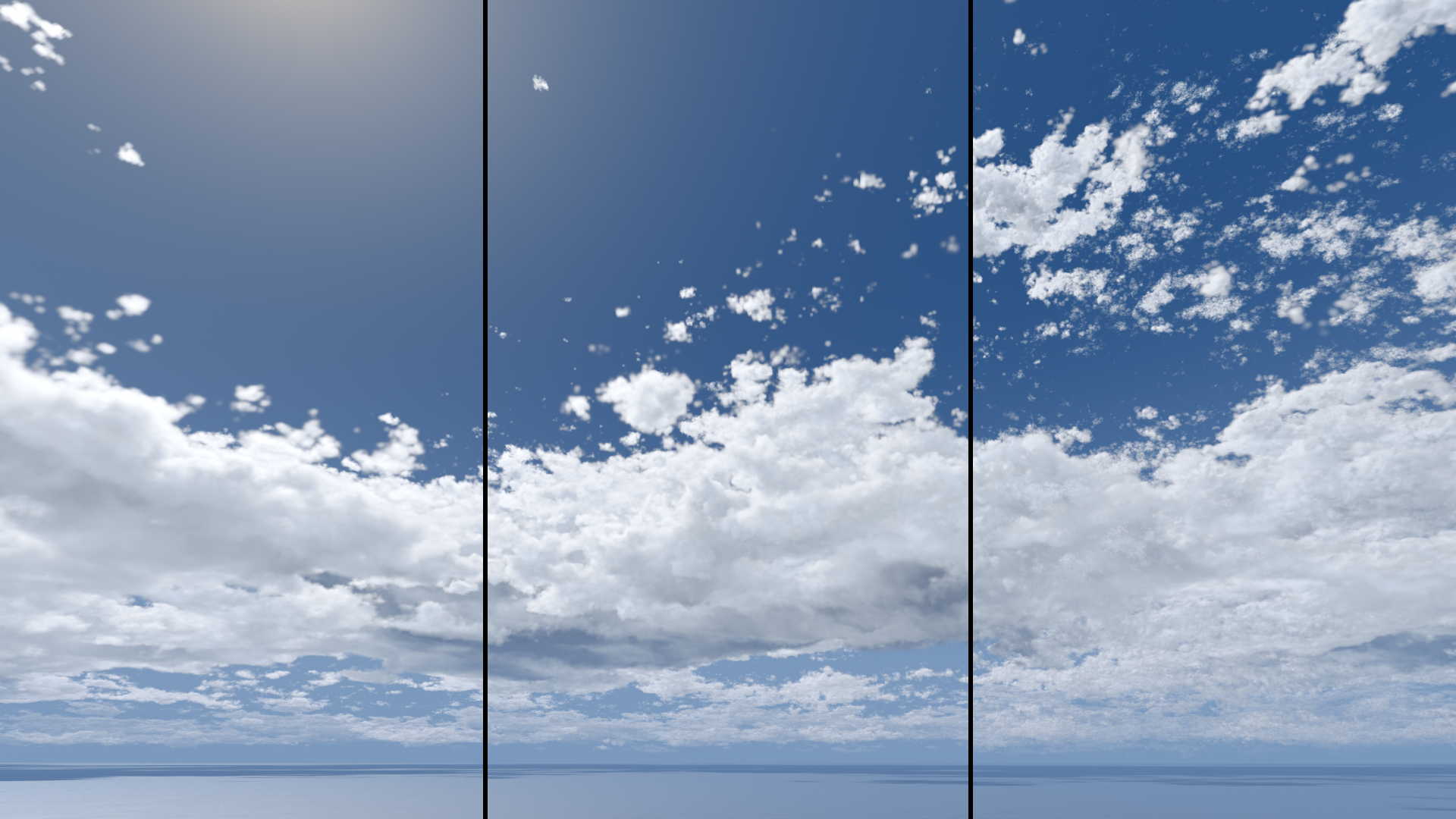}
\caption{Renderings of three small-domain simulations sharing a random seed but differing in the flux perturbation scaling parameter, with values $\varpi = 0.02$ (left), $\varpi=0.05$ (middle), and $\varpi = 0.17$ (right).} \label{fig:renderings intermittency}
\end{sidewaysfigure}

The first visual comparison, in Fig. \ref{fig:renderings intermittency}, is between cases with higher and lower intermittency, controlled by the flux perturbation scaling parameter $\varpi$. A composite image is shown for three small-domain simulations with the same random seed but different values of $\varpi$, so that exactly the same cloud is simulated in all three cases but with varying degrees of intermittency. The lower value $\varpi=0.02$ has a smaller cloud fraction in the rendered region and is dominated by larger clouds, as is consistent with the smaller size distribution exponents obtained for this value (Table \ref{tab:cloud geometry}). The higher value $\varpi=0.17$ fills clear regions with more small and dense clouds. 

\begin{figure}[t]
\centering
\includegraphics[width=\linewidth]{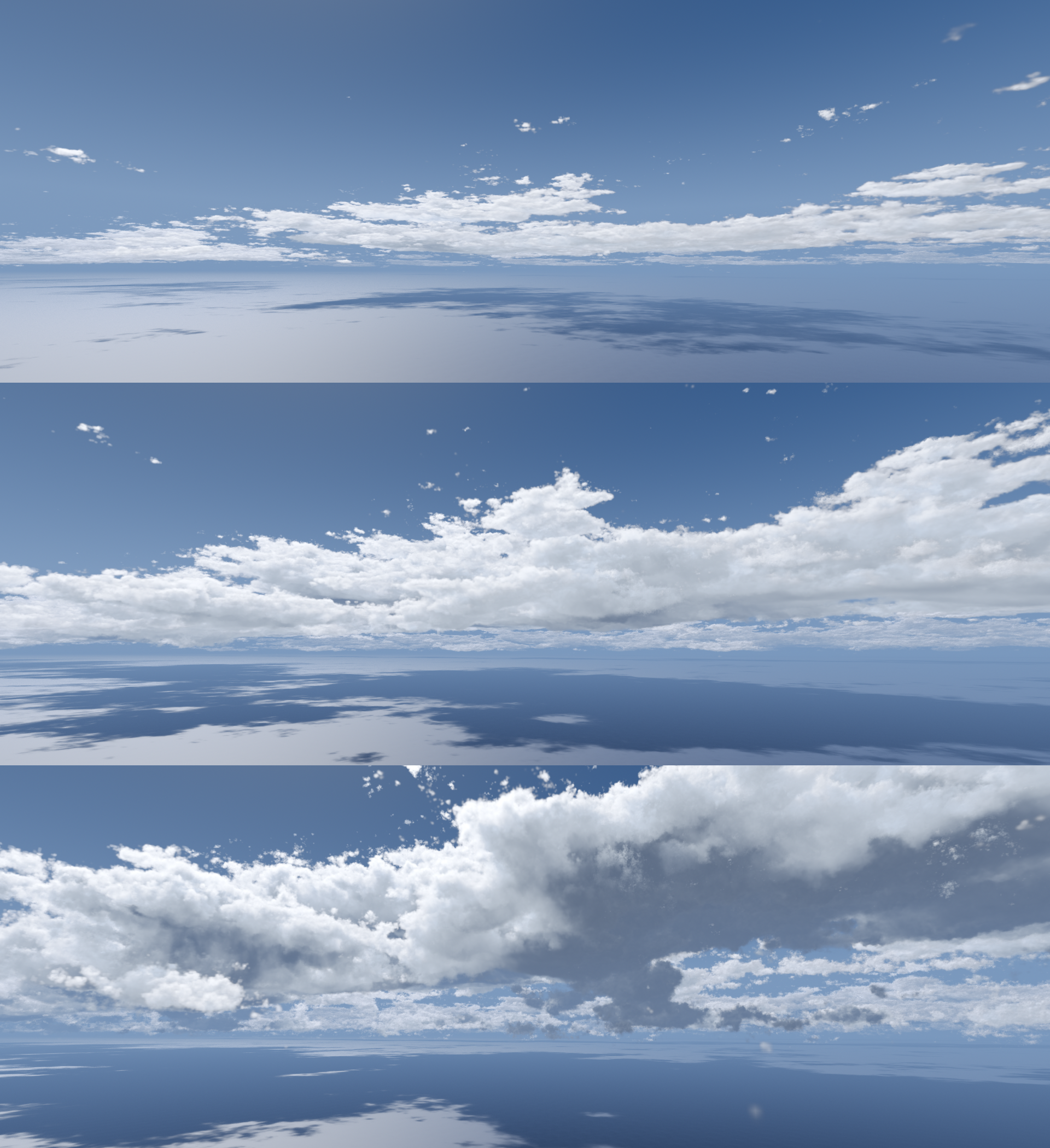}
\caption{Renderings of small domain simulations sharing a random seed but differing in the spheroscale, with values $\ell_s = 10$ (top), $30$ (middle), and $100\,\mathrm{m}$ (bottom).} \label{fig:renderings spheroscale}
\end{figure}

The other particularly useful visual comparison is the effect of changing the spheroscale. For simplicity, we only used a single constant value of $\ell_s=10\,\mathrm{m}$ in the quantitative comparison above, given that this parameter has at least been estimated in past work \citep{dewitt2025preprint}, unlike $\varpi$. Nonetheless, a constant value is unlikely to be realistic: consider the renderings shown in Fig. \ref{fig:renderings spheroscale} for spheroscales 10~m, 30~m, and 100~m. Each case is visually realistic but displays a different type of cloud, where smaller values create thin, scattered cumulus, while larger spheroscales cause cumulus congestus to form. It seems likely that, in the real atmosphere, the spheroscale in deep convection could grow much larger, perhaps to a few kilometers or more. 

Given that the simulated clouds are made up of turbulons, the cloud aspect ratios, measured as cloud width divided by cloud height, will on average follow turbulon aspect ratios from Eqn. \ref{eq:spheroscale definition}. Because all simulations here have a spheroscale that is prescribed to be a constant throughout the domain, this implies that all clouds of a given width will have nearly the same height. This is unlikely to be realistic. 
In the real atmosphere, a cloud of width 10~km could be several kilometers tall if it is a deep cumulus congestus or cumulonimbus cloud. But other clouds such as altostratus or stratocumulus might be only a few hundred meters thick while still spanning 10~km horizontally. Conceivably, this could be the result of a spatially varying spheroscale. From Eqn. \ref{eq:spheroscale definition}, and as can be seen in Fig. \ref{fig:renderings spheroscale}, a larger spheroscale reduces the aspect ratio of clouds of fixed width by making them taller.

A potential future improvement to STEAM, then, would be to incorporate the spheroscale into the dynamics of the model rather than being prescribed such that it may vary spatially. Its value could, for example, depend on the local stratification in some way: locally unstable regions create more convective clouds while more stable regions create stratocumulus. Such a dependence may be partially anticipated by the definition of the spheroscale (Eqn. \ref{eq:spheroscale definition}), where its value depends on the local turbulent fluxes that vary spatially. However, more work remains to determine exactly how such a modification should be implemented.

The figure displayed in the introduction (Fig. \ref{fig:intro rendering}) was rendered for a simulation using the small domain configuration with $\varpi=0.02$ and $\ell_s=100\,\mathrm{m}$. At least for cumulus, we tentatively suggest that STEAM passes the Palmer-Turing test for visual fidelity. Without identification, it might be difficult to determine whether the image was of a simulation or a real cloud field.

\section{Conclusions}

Even setting aside the challenging question of microphysics, fully resolving the dynamics of moist convection would require a domain spanning hundreds of kilometers but with grid spacings near millimetric scales. Sufficient computers can only be found in science fiction. 
In practice, unresolved small-scale motions are instead represented through parameterizations. Despite increasingly sophisticated parameterizations and ever finer resolution, model-to-model spread in climate sensitivity has not narrowed over the past 40 years \citep{ipcc6}. What is missing is not detail but theory: an emergent description of atmospheric dynamics that accounts for the collective effect of small-scale interactions without needing to explicitly simulate them.

Several nascent attempts at such a theory exist, but none are yet developed enough to make plausible simulations of a moist convective atmosphere. 
Here, we develop a generalization of turbulence theory that applies to stratified atmospheres, originally proposed by \citet{schertzer1985}. Our modifications are sufficient to produce numerical simulations of three-dimensional moist thermodynamic quantities that are broadly comparable to existing state-of-the-art hydrodynamic simulations.
The central idea is to formalize and generalize the concept of a ``turbulent eddy'', and to replace it with a mathematical object we term a ``turbulon''. A convective cloud field is then nothing more than a superposition of a large number of turbulons. 

Cloud fields simulated with STEAM are plausible but do not match those coming from existing state-of-the-art hydrodynamic models in every way. As we show, some departures are expected a priori, while others are unexpected but suggest clear future research directions. Despite its limitations, STEAM is promising as a method of simulating Earth's atmosphere without using the equations of hydrodynamic motion at all, analogous to how fluid mechanical simulations themselves simulate fluid flow without simulating individual particle trajectories. The model requires a few physical constants and ensemble mean profiles for moist static energy and total mixing ratio, and outputs three-dimensional fields for moist static energy, water vapor, cloud liquid and ice condensate, temperature, and pressure. Horizontal and vertical variability of the generated fields is broadly similar to that computed for state-of-the-art hydrodynamic simulations.
Comparing simulated albedo fields from STEAM clouds and hydrodynamic models to MODIS imagery suggests that STEAM clouds are, surprisingly, more aligned with MODIS observations than the hydrodynamic models, raising the possibility that STEAM's discrepancies with the hydrodynamic models could, in some cases, indicate a deficiency in the hydrodynamic model rather than STEAM.

Using a cloud rendering algorithm, we suggest that STEAM tentatively passes the ``Palmer-Turing test,'' where outputs are visually indistinguishable to the human eye, at least for some cloud types. These visualizations suggest future research directions such as allowing the spheroscale to vary spatially, investigating the role of the intermittency parameter, and improving STEAM's simplistic treatment of thermal inversions.

Perhaps most surprising is STEAM's computational advantage, estimated rigorously in Appendix~\ref{app:steam supplement}. Primarily because the model is not time dependent and therefore does not require computation of intermediate states, STEAM simulations are between $10^5$ and $10^6$ times faster per independent snapshot than the hydrodynamic model CM1 on the same hardware, as shown in Appendix~\ref{app:steam supplement}. If the limitations we discuss can be addressed, perhaps fully resolving a convective system is not science fiction after all. The real question is whether it would be necessary, or if an emergent physical theory for atmospheric dynamics is all that is needed.

\setlength{\bibhang}{0in}
\bibliographystyle{ametsocV6}
\bibliography{sources}

\chapter{Conclusions} \label{sec:conclusion}

Our attempts to model and understand Earth's atmosphere face a conundrum: the only reasonably complete physical laws that describe atmospheric motion imply such an enormous state space that, for a single moment in time,  the troposphere's temperature field alone would require $\sim 10^{29}$ bytes to fully resolve digitally --- approximately one billion times larger than humanity's globally-summed digital storage capability in 2007 \citep{hilbert2011}.
Despite this fundamental intractability, the typical approach is to simulate as much as is computationally possible using a simplified set of equations and to hope that the error incurred is bounded and decreasing with model generation \citep{shukla2009}, despite a track record that indicates model-to-model parameter spread has remained steady over the past 40 years for key metrics such as the climate sensitivity \citep{ipcc6}.

What is sorely needed is a set of emergent laws that extract only those aspects of the atmospheric state that are important for climate without requiring the full state space to be simulated. It is likely that such laws will not only consist of some new mathematical equation but a wholly new set of concepts for the ``ontology'' of the atmosphere, or what the atmosphere \emph{is}. Here, I argue that the most promising candidates are organized around a symmetry principle known as scale invariance, but at present these laws are limited in scope and lack any clear physical interpretation. Broadening their scope and establishing an ontological foundation may enable a profoundly different way of simulating and understanding Earth's atmosphere. But first, it is worth establishing why scale invariance should be expected to be so fundamental.

Some of the first empirical evidence for wide-ranging scale invariance came from cloud geometry \citep{lovejoy1982}. But soon thereafter large discrepancies in the literature arose: cloud area distributions were generally found to follow a power law, but estimates for the value of its exponent and on the scale at which the power law terminates did not agree. Chapter \ref{sec:finite-domains} showed that much of this disagreement can be reproduced simply by varying the domain size and the treatment of clouds truncated by the boundary of the measurement domain, even without any change to the underlying cloud population. Removing such truncated clouds from a distribution introduces a spurious cutoff scale, but retaining them produces a spurious pileup near the domain scale. Without careful treatment, measured power law exponents can be biased by 20\,\% to 30\,\% or more. The remedy is fortunately simple: fits should be restricted to size bins in which fewer than half of the objects are truncated, a criterion that applies regardless of the underlying distribution and to any objects measured within a finite domain, whether clouds, percolation clusters, or predator-prey ecosystem sizes \citep{dewitt2024b}. What might appear at first to be a mundane methodological detail may have obscured a key organizing principle of the atmosphere, scale invariance, from our view.

Chapter \ref{sec:metrics} moved from cloud sizes to cloud shapes, and to the fractal dimensions that quantify them. A distinction that has not been widely appreciated turns out to be quantitatively large: the fractal dimension $D_i$ of an individual cloud edge ($D_i \approx 1.4$) is not the fractal dimension $D_e$ of the ensemble of edges in a cloud field ($D_e \approx 1.7$). The two are related through the exponent $\beta$ of the cloud perimeter distribution via $D_e = \beta D_i$. Here too, measurement subtleties obscure the underlying symmetry: the common technique of fitting individual cloud perimeters against areas to obtain $D_i$ is corrupted by cloud holes. We showed that filling the holes discards a geometrically meaningful property of the field while leaving them violates the definition of the dimension $D_i$ being measured. We argued that the ensemble dimension $D_e$, calculated as a correlation integral, bypasses these difficulties, and that it provides a more objective basis for comparing simulations against observations than visual classification schemes, being grounded in a scaling symmetry rather than in subjectively defined and discrete morphological categories \citep{dewitt2026}.

Chapter \ref{sec:sondes} turned from clouds to the wind field, and to the common view that the atmosphere is governed by a hierarchy of distinct dynamical regimes. Quasi-two-dimensional turbulence is presumed to govern large scale motion, gravity waves the mesoscale, and three-dimensional isotropic turbulence is expected at subkilometer scales. Even though these theories make clear quantitative predictions for atmospheric kinetic spectra when computed both along the horizontal and the vertical directions, past empirical tests have neglected the vertical direction. Using three carefully quality-controlled dropsonde and radiosonde datasets, we calculated structure functions for horizontal wind along both vertical and horizontal separations. Neither exponent agrees with any member of the proposed hierarchy: vertical separations between 200\,m and 8\,km yield $H_v \approx 0.6$, inconsistent with both gravity waves ($H_v = 1$) and isotropic turbulence ($H_v = 1/3$), while horizontal separations between 200\,km and 1800\,km yield $H_h \approx 0.4$, inconsistent with quasi-geostrophic turbulence ($H_h = 1$). No transition appears where the hierarchy requires one. Instead, the observations are closely consistent with the anisotropic cascade hypothesis proposed by Lovejoy and Schertzer, in which a single dynamical regime with $H_v = 3/5$ and $H_h = 1/3$ spans all scales, and the modest discrepancy in $H_h$ is plausibly an artifact of vertical smoothing applied to operational sonde data \citep{dewitt2025preprint,schertzer1985,lovejoy1985}.

A recurring pattern across this Dissertation is that apparent breaks in atmospheric scaling can often be explained via measurement artifacts rather than novel dynamical regimes. The influence of a finite observational domain, the treatment of holes in a binary cloud mask, or smoothing that is applied during data processing is often neglected. None of these mechanisms is exotic. 
Perhaps what has allowed them to pass unexamined is that a scale break is expected \emph{a priori}. The prevailing ontological move within atmospheric science is to divide phenomena into distinct dynamical objects and to study them separately. This tendency surfaces when we divide clouds based on their appearance into cirrus, cumulus, and stratus, or cloud fields into sugar, gravel, fish, and flowers. It surfaces when we assume different physics control the micro, meso, and synoptic scales, and when we partition dynamics into a stratified, time-invariant background and a dynamic perturbation field. This tendency for division even appears in the fact that we identify "clouds" as discrete objects.

Division is not forced by nature, but it is a latent assumption that circumscribes the space of possible emergent theories that can be considered. The concept of a continuous field is itself required as a prerequisite to even writing down the Navier-Stokes equations, and this concept is entirely different than a discrete set of infinitesimal particles. It would not be possible to discover the Navier-Stokes equations without first inventing the mathematical concept of a continuous field.
Likewise, the most successful emergent laws for atmospheric dynamics are unlikely to be based on concepts handed down to us by our ancestors such as cloud entities. In Chapter \ref{sec:steam} we propose an alternative ontological foundation, inspired by turbulence theory: a ``quantum'' of the atmosphere that we term a ``turbulon''. As we show, turbulons and their associated dynamical laws can be used to simulate three-dimensional fields of moist thermodynamic variables that broadly reproduce the structure and variability of fields simulated using the equations of hydrodynamic motion, and at a computational cost that is smaller by a factor of up to a million.

The model in its present form also has clear limitations, detailed in Chapter \ref{sec:steam}. STEAM does not predict its own mean state: horizontally uniform ensemble mean profiles of moist static energy and total water must be supplied as inputs, here taken from a host hydrodynamic simulation. Of these profiles, only the magnitude of the vertical gradient influences the simulated variability, not its sign, so a stable inversion increases local variability rather than suppressing motion as it would in the real atmosphere. This shortcoming is the likely origin of the largest departure from the hydrodynamic comparisons, namely a spurious peak in moist static energy and temperature variability near the tropopause.

From a visual perspective, a substantial limitation is that the strength of the stratification, controlled by the value of the spheroscale $\ell_s$, is prescribed to be a constant throughout a given simulation. This means that the model can simulate deep convection and thin cirrus equally well, but never within the same volume. A principled way of allowing $\ell_s$ to vary spatially could be a promising future direction. In particular, the definition of $\ell_s$ (Eqn. \ref{eq:spheroscale}) suggests that $\ell_s$ should be a function of the two local turbulent fluxes, namely kinetic energy flux and buoyancy variance flux. 
This might be implemented in STEAM by, at each cascade step, computing the vertical size of the next smaller size class of turbulons using a location-dependent $\ell_s(\mathbf{r})$, whose value is proportional to the local flux raised to some power $\mathcal{F}^\psi$, where $\psi$ is a new parameter. 

It would also be worth investigating the physical nature of the flux in STEAM on a theoretical level, possibly to determine how the mean flux depends on the stability at each height. This may also bear on the inversion question. If the model was modified such that the mean flux was suppressed proportionally to the strength of an inversion, then the spheroscale would be reduced at stable levels, which would also reduce the vertical outer scale at those levels. This would in turn reduce the magnitude of the cascade through the $C_{\Phi,L}$ constants. This might substantially reduce or eliminate the stratospheric peak in moist static energy and temperature variability, and also cause cirrus occurring at stable levels to be naturally more stratified than deep convection below.

Finally, the methods used to infer diagnostic variables may be simplistic. Condensate is obtained by saturation adjustment and partitioned linearly between liquid and ice, leaving room for more realistic microphysical parameterizations. More egregious is the lack of precipitation in STEAM, which might be implemented by considering how the lifetime of a turbulon increases with its size. As a first simplified implementation, diagnostic variables could be computed not for only the final field but also for each individual size class of turbulon. Saturated turbulons could be assigned to be precipitating above a certain condensate threshold (analogous to a Kessler scheme; \citet{kessler1969}), and precipitation allowed to fall throughout the lifetime of the turbulon so that taller rainshafts would be instantiated for larger and longer-lived turbulons. The resulting field would then be a superposition of precipitating turbulons, although the nonlinear relationship between condensation and temperature may require some care.

None of these limitations appears fundamental to the turbulon picture itself. In fact, the turbulon picture might plausibly make these questions easier to answer, or implement, than they would in a Fractionally Integrated Flux, for example. It is probably more intuitive to ask why a given location might shear and stretch turbulons along different axes, rather than to ask why the generator of the local scale transformation might have a nonzero off-diagonal element. This is true even if both approaches are mathematically equivalent.

Even if the approach taken for STEAM does not turn out to be fruitful, some analogous approach is needed if we wish to understand the dynamics underpinning Earth's atmosphere. One would never claim to understand the motion of a fluid using only a description based in particles, and  a purely fluid-based approach will likewise be forever incomplete for understanding atmospheric motion. This is already well known on an intuitive level, which is why meteorologists think in terms of ``parcels'', ``clouds'', and ``thunderstorms''. These are emergent concepts in their own right. What is less widely questioned, and should be, is whether these concepts represent the best possible way to carve up the atmosphere into simpler constituent elements, or if they only remain due to historical inertia.

\setlength{\bibhang}{0in}
\bibliographystyle{ametsocV6}
\bibliography{sources}

\numberofappendices = 1   %
\appendix       %

\setcounter{tocdepth}{3}

\chapter{Supporting Material} \label{app:supporting}
\addtocontents{toc}{\protect\setcounter{tocdepth}{2}}

\section{Supporting material for Chapter~\ref{sec:sondes}} \label{app:sondes}

\subsection[Vertical Hurst exponents for individual meteorological stations]{Vertical Hurst exponents for\\* individual meteorological stations}  \label{sec:individual stations}  %

Reported vertical Hurst exponents from the IGRA data were computed after the structure function was averaged over all available stations to obtain a global mean value of $H_v$. Here, we calculate the exponents for individual stations by identifying which stations contained at least 730 complete soundings, corresponding to at least two years. Then, mean structure functions and Hurst exponents were calculated for individual $2\,\mathrm{km}$-thick vertical layers as in Section \ref{sec:results} and Fig. \ref{fig:exponents with height}. Finally, as visual inspection of plotted structure functions was not possible, we filtered the calculated exponents by requiring the 95\% confidence interval of the least-squares fit to be less than 0.05. This ensures that we are not assuming a power-law form for structure functions that are poorly described by a power-law fit. This filtering process was only used in this section and resulted in a total of 184 stations for consideration.

\begin{figure}
    \includegraphics{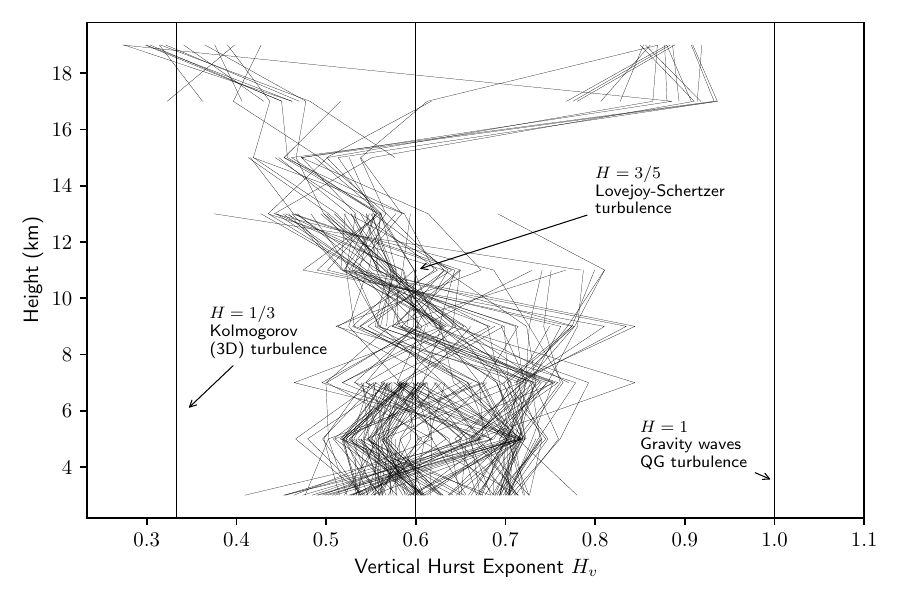}
    \centering \caption{As in Fig. \ref{fig:exponents with height}, but for individual IGRA stations (thin black lines).  }\label{fig:exponents with height each station}
\end{figure}

Fig. \ref{fig:exponents with height each station} shows the vertical Hurst exponents as a function of height for each IGRA station.
Overall, a large majority of values are much closer to the Bolgiano-Obukhov value of 3/5 than the values of 1/3 or 1 that correspond to three-dimensional or two-dimensional turbulence, respectively. There also appears to be a bimodal distribution below $8\,\mathrm{km}$, with a cluster of values near 0.6 and a cluster closer to 0.75. At first glance, this might indicate evidence of two separate turbulence regimes, possibly depending on latitude or some other climatological factor. However, we also observe a nearly perfect split between the two regimes based on sounding nationality. For example, for the layer between $4\,\mathrm{km}$ and $6\,\mathrm{km}$, 82 out of 84 U.S.A.-based soundings had a Hurst exponent smaller than 0.63, whereas 79 out of 84 Hurst exponents measured from China-based soundings were larger than 0.63. Recall that the value of the exponent is sensitive to data processing methods, such as the period of the vertical smoothing applied to the raw data as shown in Section \ref{sec:discussion}. Since it is likely that the methods are uniform within a country but differ between different countries, the bimodal distribution in Fig. \ref{fig:exponents with height each station} is likely an indication of data processing artifacts rather than some dependence on local meteorology.

\section{Supporting material for Chapter~\ref{sec:steam}} \label{app:steam}

\subsection{STEAM algorithm} \label{sect:steam algorithm details}

Here we provide an explicit description of the STEAM algorithm for modeling moist static energy $h$, total water mixing ratio $q_t$, and the conserved turbulent flux $\mathcal{F}$.
As a point of notation, there are three distinct types of averages requiring disambiguation. Consider a large ensemble of statistically independent turbulent flows. Any variable may be averaged along the spatial directions, which we denote $\langle \rangle_\mathbf{x}$, or alternatively $\langle \rangle_\mathbf{x}(z)$ when the average is taken over the horizontal dimensions only. The result may in principle still have time dependence and differ between ensemble members. Alternatively, if we average only along the time direction, denoted by $\langle \rangle_t$, in principle the result may differ spatially and among ensemble members. Finally, we might average across ensemble members, which we denote $\langle \rangle_i$, which in principle could result in an ensemble mean that still varies in time and space. When ergodicity applies, such as in an RCE simulation, $\langle \rangle_t=\langle \rangle_i$. Unit vectors are denoted as $\hat{x}, \hat{y}, \hat{z}$, and we denote the horizontal gradient magnitude of some arbitrary field $f$
\begin{align}
  |\nabla_h f| = \sqrt{\left(\frac{\partial f}{\partial x}\right)^2 + \left(\frac{\partial f}{\partial y}\right)^2}.
\end{align}
For an arbitrary variable $\Phi$, we refer to the component of $\Phi$ that is associated with perturbations of a given size class $\ell$ as $\Phi_\ell$. The component of $\Phi$ associated with perturbations of size larger than $\ell$ is written $\Phi_{>\ell}$, so that 
$$\Phi = \langle \Phi \rangle_t + \sum_{\ell'=\ell_\text{min}}^{\ell'=L} \Phi_{\ell'},$$
$$\Phi_{>\ell} = \langle \Phi \rangle_t + \sum_{\ell'>\ell}^{L} \Phi_{\ell'} ,$$
$$\Phi_{\ge\ell} = \langle \Phi \rangle_t + \sum_{\ell'=\ell}^{L} \Phi_{\ell'} .$$

We assume the ensemble means $\langle h \rangle_t$ and $\langle q_t \rangle_t$ are known and horizontally uniform, in which case they may be simply represented by a 1D profile along the vertical direction. We also assume the outer scale $L$ and the spheroscale $\ell_s$ are known, along with physical bounds $[\Phi_{\min}, \Phi_{\max}]$ for each advected scalar, and we consider logarithmically-spaced size classes for the turbulons such as $L, L/2, L/4, \dots, 2dx$, implying the outer scale must be a power of two factor of the horizontal grid resolution. The smallest size class is $2dx$ rather than $dx$ so that the finest turbulons are Nyquist-sampled by the output grid. The cascade density may be increased to an integer $n_c\ge 1$ size classes per octave, so that adjacent classes are separated by the scale ratio $2^{1/n_c}$. When $n_c>1$, an adjustment factor described below is included to compensate for the increased variance due to more size classes. The simulations here use dyadic classes, $n_c = 1$.

First, define the three-dimensional turbulon envelope shape $\mathfrak{T}_\ell$ as a function of size class $\ell$. For simplicity we will use the three-dimensional Mexican hat (the negative Laplacian of a Gaussian) with width parameter $\sigma = \ell/\pi$, chosen so that the envelope's power spectral density peaks near wavelength $\ell$:
\begin{align}
  \mathfrak{T}_\ell(||r||) = \left(\varkappa -\frac{\pi^2||r||^2}{\ell^2}\right)\exp \left(-\frac{\pi^2||r||^2}{2\ell^2}\right) \label{apxeq:turbulon shape}
\end{align}
Here, $||r||$ is a vertically-stretched distance metric defined by
\begin{align}
  ||r|| = \left(x^{2} +y^{2} + \left(z\frac{\ell}{\ell_z}\right)^{2}\right)^{1/2}. \label{apxeq:stretched distance}
\end{align}
The familiar one-dimensional Mexican hat would have $\varkappa = 1$, but $\varkappa$ must be modified in order to keep $\mathfrak{T}_\ell(||r||)$ zero mean, both because the wavelet is three-dimensional and because it is resolved on a coarse grid. We use $\varkappa=2.9678$, a constant computed numerically for default ``sparsity factors'' defined below.

The turbulon aspect ratio $\ell/\ell_z$ is defined by
\begin{align}
  \begin{cases}
    \ell_z>\ell_s; \quad & \ell_z = \ell_s \left(\frac{\ell}{\ell_s}\right)^{H_z}, \\
    \ell_z \le \ell_s; \quad & \ell_z = \ell,
  \end{cases}
   \label{apxeq:vertical turbulon size class}
\end{align}
which recovers Eqn. \ref{eq:turbulon aspect ratio scaling} at scales larger than $\ell_s$ and is isotropic at smaller scales. This piecewise aspect ratio scaling function was found to produce more visually realistic subspheroscale clouds (not shown). However, all simulations that are analyzed quantitatively here are fully within the anisotropic regime due to the small spheroscales considered. The visualizations in Section \ref{sect:visual renderings} are the only simulations containing subspheroscale turbulons.

In practice the stretched metric is realized by constructing an envelope that is isotropic in grid-index space on a working grid whose vertical spacing is compressed in proportion to $\ell_z/\ell$ (see Appendix~\ref{app:steam supplement}), which is equivalent to Eqn. \ref{apxeq:stretched distance} when the sparsity factors defined below are equal.

On a basic level, the three scalar fields $\mathcal{F}$, $h$ and $q_t$ are simulated by constructing a hierarchy of 3D arrays for each turbulon size class $\ell$, called $\mathcal{F}_\ell$, $h_\ell$ and $q_{t,\ell}$, each representing the contribution of turbulons of a given size to the resulting scalar fields $\mathcal{F}$, $h$ and $q_t$. The simulation proceeds in a sequential manner, beginning with $\ell=L$ and progressively stepping down until the smallest $\ell=2dx$ is reached. The final fields are the sum of the ensemble mean field and the perturbation due to each turbulon size class:
\begin{align}
  h = \langle h \rangle_t + \sum_\ell h_\ell, \\
  q_t = \langle q_t \rangle_t + \sum_\ell q_{t,\ell}, \\
  \mathcal{F} = \langle \mathcal{F} \rangle_t + \sum_\ell \mathcal{F}_\ell.
\end{align} 
Note that $\mathcal{F}_\ell$, $h_\ell$, and $q_{t,\ell}$ only represent transient fluctuations, i.e. $\langle\mathcal{F}_\ell\rangle_t = 0$, $\langle h_\ell\rangle_t = 0$, and $\langle q_{t,\ell}\rangle_t = 0$.

At each size class, a computationally efficient means of superimposing turbulons is use a convolution with the turbulon envelope function
\begin{align}
  h_\ell &=  \mathfrak{T}_\ell * A_h, \\
q_{t,\ell} &=  \mathfrak{T}_\ell * A_{q_t}, \\
\mathcal{F}_\ell &=  \mathfrak{T}_\ell * A_\mathcal{F}
\end{align}
where $A_\Phi$ represents turbulon amplitudes as a function of $\mathbf{x}$ and $\ell$.
For a given location $\mathbf{x}$ and size class $\ell$, the turbulon amplitude fields are defined in a similar way for the advected scalar fields $h$ and $q_t$, but differently for the turbulent flux due to their different physical roles.

The flux is represented by a dimensionless field $\mathcal{F}$, initialized to one everywhere at the outer scale, i.e. $\langle \mathcal{F}\rangle_t=1$. Due to scale-by-scale flux conservation, the mean absolute amplitude of flux perturbations is independent of scale. But for a given location within a given realization, perturbation amplitudes are scaled by the local large-scale flux. Specifically, at a given scale $\ell$ flux perturbations $\widetilde{\mathcal{F}_{\ell}}$ are randomly generated according to 
\begin{align}
  \mathcal{F}_{\ell} = \mathfrak{T}_{\ell} * \left(\left(\exp\left(\varpi\, n_c^{-1/\alpha}\gamma\right) - 1\right)\mathcal{F}_{>\ell}\right) \label{eq:flux at scale}
\end{align}
where $\gamma$ is a sparse field of independent random variables drawn at the turbulon centers and the input parameter $\varpi$ sets the amplitude of the noise. The factor $n_c^{-1/\alpha}$ converts $\varpi$, an amplitude per octave of scale, to the noise scale of a single size class. The factor $\mathcal{F}_{>\ell}$ represents the component of the flux that is due to turbulons of a scale larger than $\ell$. This factor implements the hypothesis of Section \ref{sect:distribution of turbulon amplitudes in space} whereby the absolute magnitude of flux increments is, on average, proportional to the local large-scale flux. 

The random variables $\gamma$ are independent, identically distributed, and drawn from an extremal L\'evy $\alpha$-stable distribution with $\alpha = 1.8$ and skewness parameter -1.\footnote{This value for the skewness parameter, which is responsible for the ``extremal'' label, is forced because any other value for the skewness parameter makes all moments for $e^\gamma$ infinite.} This particular distribution choice is made because it causes the resulting flux $\mathcal{F}$ to be a ``universal multifractal'' as defined by \citet{lovejoy2013}. For our purposes, it can be simply seen as a convenient distribution that is relatively general but requires only one parameter $\alpha$ to specify its form. In the limit $\alpha=2$, a Gaussian is obtained. The center of the distribution for $\gamma$ is chosen such that $\langle \exp(\varpi\, n_c^{-1/\alpha}\gamma)\rangle = 1$, so that the mean perturbation $\langle \mathcal{F}_\ell\rangle=0$. We use the value $\alpha=1.8$ because it approximately matches empirical values for multifractal exponents as shown in \citet{lovejoy2013}. 

By $\alpha$-stability, the factor $n_c^{-1/\alpha}$ makes the noise added per octave of scale invariant in distribution under the cascade density $n_c$, so the intermittency of the flux is set by $\varpi$ alone. The only residual dependence on $n_c$ enters through the clipping step described below. All simulations here use $n_c=1$.

At each cascade step, after the perturbation field is generated according to Eqn. \ref{eq:flux at scale} and the result added to the running flux to obtain $\mathcal{F}_{\ge \ell}$, the flux is clipped at zero and renormalized to preserve its volume mean $\langle\mathcal{F}_{>\ell}\rangle_\mathbf{x}$ to ensure positivity of $\mathcal{F}$.\footnote{Perhaps counterintuitively, in practice clipping occurs most frequently when large positive factors for $e^\gamma$ are generated in regions with small large-scale flux. This occurs because $e^\gamma$ is not bounded above and so the negative lobes of the envelope function push $\mathcal{F}$ negative.} This clipping step is only necessary because the scale separation between cascade steps is a finite number. In the limit of a dense cascade with steps separated by an infinitesimal scale ratio, the per-class noise scale $\varpi\, n_c^{-1/\alpha}$ goes to zero and perturbations large enough to cause negative $\mathcal{F}$ happen with probability zero. Note, however, that convergence to a dense cascade can be very slow.

Scalar turbulon amplitudes are then computed following Eqn. \ref{eq:anisotropic gradient weighing}. For each scalar, the local advective weight $W_{\Phi,\ell}$, the bound factor $g_\Phi$, and the flux weighting factor $S_\ell$ are combined into an amplitude pattern
\begin{align}
  P_{\Phi,\ell} = g_\Phi\, W_{\Phi,\ell}\, S_\ell, \label{apxeq:amplitude pattern}
\end{align}
which sets the spatial structure of the turbulon amplitudes but not their magnitude. The amplitude field is computed as
\begin{align}
  A_\Phi = C_{\Phi, \ell}\, \frac{P_{\Phi,\ell}}{\left\langle \left|P_{\Phi,\ell}\right| \right\rangle_\mathbf{x}(z)},
  \label{apxeq:amplitude propto gradient normalized}
\end{align}
where the mean is taken over the turbulon centers holding a nonzero $P_{\Phi, \ell}$ at each level. The normalization of $P_{\Phi, \ell}$ ensures that mean absolute amplitudes are controlled by $C_{\Phi,\ell}$ alone, as specified by the Lovejoy-Schertzer law of scalar advection (Eqn. \ref{eq:LS scalar directional structure functions}) such that
\begin{align}
    C_{\Phi, \ell} = n_c^{-1/\alpha}\, C_{\Phi, L}\left(\frac{\ell}{L}\right)^{H_h}.\label{apxeq:mean turbulon amplitude}
\end{align}
The factor $n_c^{-1/\alpha}$ compensates for the cascade density in the same way as the scaling factor for the flux noise. For the flux, the exponent follows exactly from $\alpha$-stability as described above. For the scalars, we assume the same compensation applies because $S_\ell$ is built from the same noise. A fuller theoretical investigation of STEAM may be required to confirm or reject this assumption. In any case, the simulations presented here use $n_c = 1$ for which the factor is unity.

The overall normalization factor $C_L$
represents the mean absolute amplitude of the largest size class of turbulons. Physically, this represents the large-scale fluctuation magnitude that is available to cascade to smaller scales, computed from the vertical structure of the ensemble mean profile as described in Section \ref{sect:distribution of turbulon amplitudes across scale}. An overturning eddy of depth $\ell_{z,L}$ centered at height $z$ exchanges fluid across its depth, so anomaly magnitudes are set by the difference of the mean profile across the eddy. We therefore take the characteristic available anomaly to be the Haar fluctuation of the mean profile at the local vertical outer scale, assuming horizontally uniform mean profiles $\langle \Phi\rangle_t(\mathbf{x}) = \langle \Phi\rangle_t(z)$:
\begin{align}
  C_{\Phi, L}(z) = \vartheta \left| \overline{\langle \Phi \rangle_t}^{\,\mathrm{up}} - \overline{\langle \Phi \rangle_t}^{\,\mathrm{low}} \right|(z), \label{apxeq:norm factor computation}
\end{align}
where the overbars denote averages over the upper and lower halves of a window of width $\ell_{z,L}(z) = \ell_s(z)\left(L/\ell_s(z)\right)^{H_z}$, the local vertical outer scale. A first-difference (odd) measure such as the Haar fluctuation is required here to ensure a constant vertical gradient would produce a nonzero $C_{\Phi,L}$. An even zero-mean kernel, such as the turbulon envelope itself, would unphysically produce $C_{\Phi,L}=0$ for a constant gradient. The constant $\vartheta$ converts the measured Haar fluctuation to the turbulon amplitude convention of Eqn. \ref{apxeq:turbulon shape}. It is fixed by an iterative process described in Appendix~\ref{app:steam supplement}. The target for iteration is that vertical Haar fluctuations, computed for a simulated field and fitted to a power law with exponent $H_v$, intersect the vertical fluctuation function of a linear input profile at the vertical outer scale. The constant $\vartheta$ compensates for constant factors for spectral power that are introduced by our choice of Haar and Mexican Hat conventions. 

The Haar response (Eqn. \ref{apxeq:norm factor computation}) is evaluated level by level using a height-dependent spheroscale, with the profile edge-padded (extended beyond the domain edge by repeating its boundary values) to produce $C_{\Phi,L}(z)$. The resulting profile is interpolated during the simulation to each size class's height levels. 

The next factor in Eqn. \ref{apxeq:amplitude pattern}, $S_\ell$, is the amplitude of the flux component of a turbulon, equal to
\begin{align}
S_\ell = \left(e^{\gamma} - 1\right)\mathcal{F}_{>\ell}, \label{apxeq:scalar flux factor}
\end{align}
whose convolution generates the flux perturbation in Eqn. \ref{eq:flux at scale}. This term implements the hypothesis discussed in Section \ref{sect:distribution of turbulon amplitudes in space}, namely that the amplitude of a turbulon's scalar component is proportional to the amplitude of its flux component, both being set by the strength of the eddy's overturning motion. The factor is shared between $h$ and $q_t$ because a single eddy transports both scalars, which inherit the skewness of the flux perturbations. Note that $S_\ell$ depends only on size classes larger than $\ell$, so each cascade step is fully determined by the classes preceding it.

The weight $W$ follows the anisotropic advective form of Eqn. \ref{eq:anisotropic gradient weighing} in Section \ref{sect:distribution of turbulon amplitudes in space}:
\begin{align}
  W_{\Phi, \ell} =  | \nabla_h \Phi_{>\ell}| + \frac{\ell_z}{\ell}\left|\frac{\partial \Phi_{>\ell}}{\partial z}\right|, \label{apxeq:advective weight}
\end{align}
where $\ell_z/\ell$ is the turbulon aspect ratio of Eqn. \ref{apxeq:vertical turbulon size class}. 

The bound factor $g_\Phi$ accounts for the physical bounds on the scalar fields. Because advection simply rearranges the fluid, the value of a conserved scalar cannot exceed the minimum and maximum values set by the initial conditions. This is approximated in STEAM by forcing fluctuations to vanish as a bound is approached, using a linear taper for advective weight within a buffer distance $b_{\Phi,\ell}$ of either bound, computed as
\begin{align}
  g_\Phi = \max\left[0,\ \min\left(1,\ \frac{\Phi_{>\ell} - \Phi_{\min}}{b_{\Phi,\ell}},\ \frac{\Phi_{\max} - \Phi_{>\ell}}{b_{\Phi,\ell}}\right)\right], \label{apxeq:bound factor}
\end{align}
where $\Phi_{>\ell}$ is the same running field used for the gradients. Note that when field values lie outside the buffer region, $g_\Phi = 1$ and the bound does not affect the cascade. The buffer width $b_{\Phi,\ell}$ is set by the expected value of fluctuations that has yet to be added by the current and all smaller size classes. Because the amplitudes of the smaller classes form a geometric series (Eqn. \ref{apxeq:mean turbulon amplitude}), the sum takes a closed form,
\begin{align}
  b_{\Phi,\ell}(z) = n_b \sum_{\ell' \le \ell} C_{\Phi,\ell'}(z) = n_b\, \frac{C_{\Phi,\ell}(z)}{1 - 2^{-H_h/n_c}}, \label{apxeq:bound buffer}
\end{align}
with $n_b = 3$ and the right-hand side represents the analytical sum over the infinite series $n_b \sum_{\ell' \le \ell} C_{\Phi,\ell'}(z)$ with $\ell' \to 0$. Note that this sum is performed over only mean perturbations, and that the asymptotic behavior of the full stochastic cascade may require a more in-depth analysis.

Even with the tapered weighting via the factor $g_\Phi$, occasional large increments still occasionally carry a scalar past its bounds. The increment cannot simply be clipped where this occurs, because clipping alone would change the level mean, and wherever the mean value is small relative to the fluctuations the accumulated drift can become substantial. Each class's increment is therefore added through a bounded projection with three conditions at every level: the added perturbation has zero horizontal mean; its mean absolute amplitude is as close as possible to that of the unbounded increment, so that the amplitude at each scale remains set by the normalization alone; and the updated field respects the bounds pointwise. The three conditions cannot always be met at once. Where a level mean lies close to a bound, the zero-mean and bound conditions hold exactly and the amplitude condition yields. The construction of the projection is described in more detail in Appendix~\ref{app:steam supplement}.

A second projection is applied exactly once, when the final output fields are composed from the cascade state. The composition includes corrections for the finite resolution of the working grids described below, and these corrections can carry the composed field past its bounds even though every class was added within them. Each scalar field is therefore projected onto the set of fields that respect the bounds and preserve the horizontal mean at every level. The projection takes the form
\begin{align}
  \Phi \rightarrow \min\left[\max\left(\Phi - \mu,\ \Phi_{\min}\right),\ \Phi_{\max}\right], \label{apxeq:bound projection}
\end{align}
where the constant $\mu$ is chosen at each level such that the horizontal mean is unchanged. Because each grid point's projected value can only decrease as $\mu$ increases, the projected level mean is a continuous, piecewise-linear, and monotonically nonincreasing function of $\mu$, strictly decreasing so long as any point lies within the bounds. The root is therefore unique and is found by bisection, or exactly by sorting the breakpoints $\Phi - \Phi_{\max}$ and $\Phi - \Phi_{\min}$ at which grid points enter or leave the clipped state. Among all fields satisfying the two constraints, the projected field is the closest to the unprojected field in the least-squares sense. The flux output is treated analogously: it is clipped at zero and rescaled to its volume mean.

\begin{figure}[t]
\centering
\includegraphics[width=12cm]{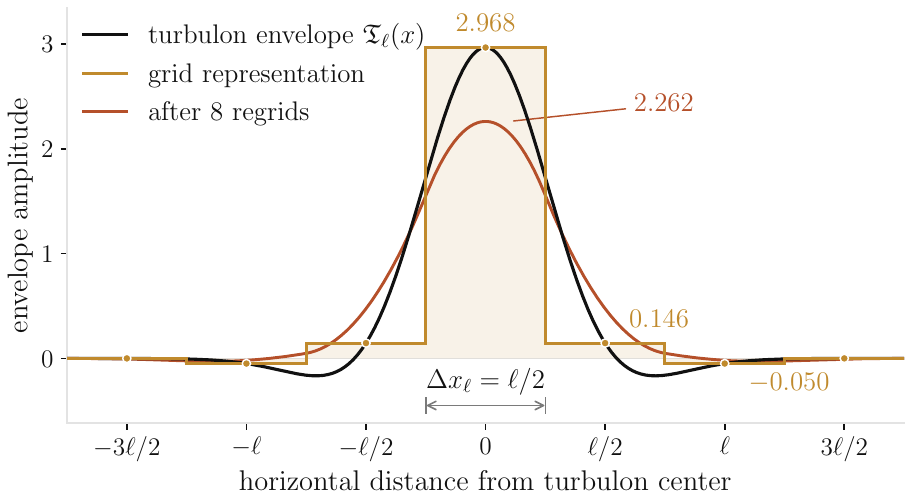}
\caption{One-dimensional transect of the turbulon envelope $\mathfrak{T}_\ell$ (Eqn. \ref{apxeq:turbulon shape}; black) as resolved on its own working grid at the most aggressive sparsity, $s=1$ (ochre), and after 8 regriddings (sienna). Note how repeated regridding via linear interpolation reduces the turbulon amplitude (sienna) relative to the theoretical shape (black). The leading constant $\varkappa$ is set to $\varkappa = 2.9678$ such that the discrete three-dimensional sum is zero. Note that the three-dimensional sum remains zero even as the sum of the one-dimensional transect shown here is positive. } \label{fig:discretized turbulon}
\end{figure}

For computational efficiency, the finest output grid is not used for all cascade steps. Because the field components $\mathcal{F}_\ell$, $h_\ell$, and $q_{t,\ell}$ do not contain variance at scales smaller than $\ell$, they may be computed on a coarsened grid without significant loss of accuracy. This grid is specified by defining positive integer ``sparsity factors'' $s_x, s_y, s_z$ such that the working resolution for size class $\ell$ is $\Delta x_\ell = \ell/(2s_x)$, $\Delta y_\ell = \ell/(2s_y)$, $\Delta z_\ell = \ell_z/(2s_z)$. When $s=1$, the grid spacing equals half the turbulon size, which Nyquist-samples the envelope's peak wavelength and is the most aggressive coarsening that can be achieved. When $s > 1$, each turbulon spans $2s$ grid cells along each dimension, providing better resolution of the turbulon shape at greater computational cost. On this coarsened grid, the random factors responsible for flux perturbations $\gamma$ are nonzero only at turbulon centers, placed every $s_x, s_y, s_z$ grid points along the respective dimensions, i.e. every $\ell/2$ in physical space. For the simulations here, we use the most aggressive sparsity factors $s_x=s_y=s_z=1$ and find that larger factors affect simulation output relatively little (not shown). The resulting discretized turbulon is shown in Fig. \ref{fig:discretized turbulon}. Because repeated linear interpolation causes the turbulon amplitude to be reduced, as shown by the sienna vs. black curves in Fig. \ref{fig:discretized turbulon}, empirical amplitude compensation factors are applied to the final per-class perturbation fields as described in Appendix~\ref{app:steam supplement}.

We use horizontally periodic boundary conditions, and along the vertical direction we use zero-padded convolutions while requiring that no turbulons are centered below the surface or above the domain top. Physically, this means that turbulons near the Earth's surface may be ``cut off'' by the boundary but are otherwise unaffected. However, the statistics of $h$ and $q_t$ near the surface remain affected in a nonobvious way. Without a domain boundary at the surface, there would be turbulons centered at negative values of $z$ that affect $h$ and $q_t$ at positive $z$ given that some portion of such turbulons might extend to positive $z$. Imposing a boundary at the surface removes the influence of such turbulons and therefore the overall fields $h$ and $q_t$ are affected. It remains to be determined whether such an effect is physically realistic.

\subsubsection[Recovering diagnostic variables from prognostic variables]{Recovering diagnostic variables\\* from prognostic variables}

Once three-dimensional fields of $h(\mathbf{r})$ and $q_t(\mathbf{r})$ are obtained, we recover temperature and water phase at each grid point. We assume the surface pressure $p_s$ is known and use the hypsometric equation to diagnose the pressure at each level; for the simulations described in the main text, $p_s$ is taken from the corresponding host model.

For each column $(x, y)$, we proceed upward from $z = z_{\min}$. At each level, we first test for saturation by assuming all water is in vapor phase, giving a tentative temperature
\begin{equation}
    T_{\mathrm{dry}} = \frac{h - L_v q_t - gz}{c_p}. \label{eq:Tdry}
\end{equation}
We then compute the saturation mixing ratio $q_{v,\mathrm{sat}}(T_{\mathrm{dry}}, p)$ using Bolton's formula \citep{bolton1980} for the saturation vapor pressure over liquid water,
\begin{equation}
    e_s(T) = 611.2 \exp\!\left(\frac{17.67\,(T - 273.15)}{T - 29.65}\right), \label{eq:bolton}
\end{equation}
and
\begin{equation}
    q_{v,\mathrm{sat}} = \frac{0.622\, e_s}{p - e_s}. \label{eq:qvsat}
\end{equation}
If $q_t \leq q_{v,\mathrm{sat}}(T_{\mathrm{dry}}, p)$, the grid cell is unsaturated: $T = T_{\mathrm{dry}}$, $q_v = q_t$, and $q_c = q_i = 0$.
If $q_t > q_{v,\mathrm{sat}}(T_{\mathrm{dry}}, p)$, condensation has occurred and the temperature must account for latent heat release. We solve
\begin{equation}
    h = c_p T + gz + L_v\, q_{v,\mathrm{sat}}(T, p) \label{eq:h_saturated}
\end{equation}
for $T$ using Newton's method. Defining $f(T) = c_p T + L_v\, q_{v,\mathrm{sat}}(T, p) + gz - h$, the derivative is
\begin{equation}
    f'(T) = c_p + L_v \frac{dq_{v,\mathrm{sat}}}{dT}. \label{eq:fprime}
\end{equation}
Differentiating Eqns.~\ref{eq:bolton}--\ref{eq:qvsat},
\begin{align}
    &\frac{de_s}{dT} = \frac{4302.6\, e_s(T)}{(T - 29.65)^2}, \label{eq:des_dT} \\
    &\frac{dq_{v,\mathrm{sat}}}{dT} = \frac{0.622\, p}{(p - e_s)^2} \frac{de_s}{dT}. \label{eq:dqvsat_dT}
\end{align}
Thus,
\begin{equation}
    f'(T) = c_p + \frac{0.622\, L_v\, p}{(p - e_s)^2} \cdot \frac{4302.6\, e_s(T)}{(T - 29.65)^2}. \label{eq:fprime_explicit}
\end{equation}

The unsaturated temperature $T_{\mathrm{dry}}$ serves as the initial guess, and a fixed five Newton iterations are performed, which is ample for convergence from this starting point. The vapor mixing ratio is then $q_v = q_{v,\mathrm{sat}}(T, p)$ and the total condensate is $q_t - q_v$, which is partitioned into liquid and ice as
\begin{equation}
    q_c = \omega\,(q_t - q_v), \quad q_i = (1 - \omega)\,(q_t - q_v), \label{eq:phase_partition}
\end{equation}
where
\begin{equation}
    \omega = \max\!\Big(0,\, \min\!\Big(1,\, \frac{T - 235.15}{273.15 - 235.15}\Big)\Big) \label{eq:liquid fraction}
\end{equation}
following \citet{khairoutdinov2003}. Finally, the pressure at the next level is computed using the hypsometric equation,
\begin{equation}
    p(z + \Delta z) = p(z) \exp\!\left(\frac{-g\,\Delta z}{R_d\, T_v}\right), \label{eq:hypsometric}
\end{equation}
where $T_v = T(1 + 0.608\, q_v)$ is the virtual temperature.

\subsection[Cloud aspect ratio scaling from \citet{guillaume2018}]{Cloud aspect ratio scaling\\* from \citet{guillaume2018}} \label{sec:guillame}

The study by \citet{guillaume2018} measured distributions of cloud chord lengths along the horizontal and vertical directions. Although they did not directly report cloud aspect ratios as a function of size, their reported size distribution exponents are algebraically equivalent. Note that a similar argument is provided by \citet{lovejoy2021}.

Specifically, we seek a function of the form 
\begin{align}
  h \propto w^{H_z}\label{fhdjkalbfuke}
\end{align}
where $h$ is cloud height, $w$ is cloud width, and $H_z$ is the unknown scaling exponent relating the two. Here, cloud widths and heights are understood as average values, and Eqn. \ref{fhdjkalbfuke} does not necessarily apply to every individual cloud. 

\citet{guillaume2018} instead reported size distribution exponents for the distributions for $h$ and $w$, which also represent statistical properties computed over many clouds. They found, over an intermediate scale invariant range, probability density functions followed
\begin{align}
  \frac{dn}{dw}\propto w^{-1.66} \\
  \frac{dn}{dh}\propto h^{-2.23},
\end{align}
which correspond to cumulative distribution functions
\begin{align}
  n(w)\propto w^{-0.66} \\
  n(h) \propto h^{-1.23}.
\end{align}
To relate these cumulative size distribution exponents to Eqn. \ref{fhdjkalbfuke}, first note that Eqn. \ref{fhdjkalbfuke} implies 
$$H_z=d\log h/d\log w.$$
Taking logarithms of the cumulative distribution functions gives
\begin{align}
  \log n = -0.66\log w +\text{const.}\\
  \log n = -1.23 \log h +\text{const.}.
\end{align}
Differentiating to eliminate the constant and then solving for $d\log h/d\log w$, we obtain
\begin{align}
  H_z = \frac{0.66}{1.23} = 0.54,
\end{align}
which is extremely close to the theoretical value $5/9=0.55\dots$ proposed by \citet{schertzer1985}.

\subsection{Additional comparison figures} \label{sect:additional comparison figures}
\begin{figure}[t]
\centering
\includegraphics[width=12cm]{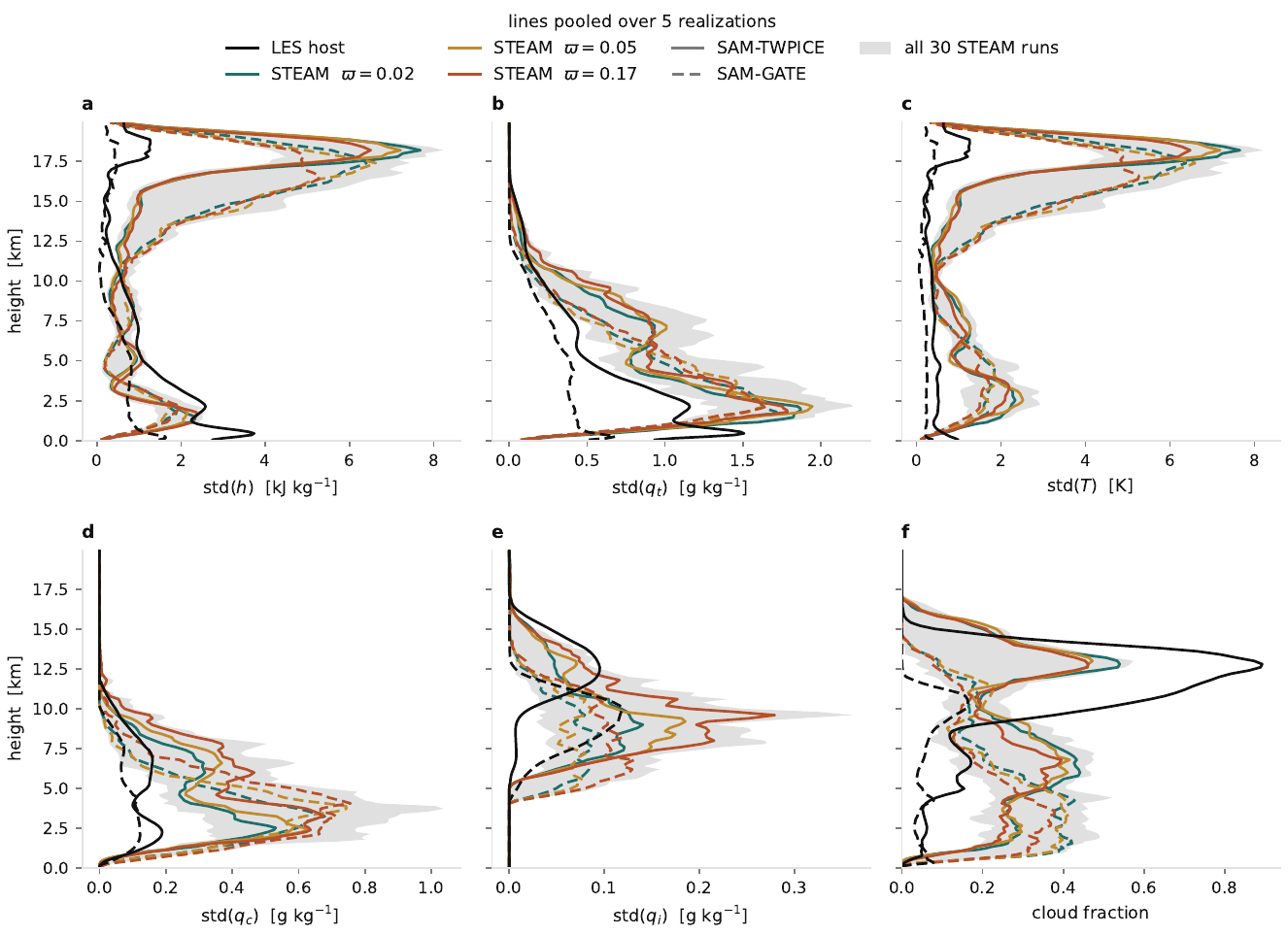}
\caption{Standard deviation profiles as in Fig. \ref{fig:profile stats}, shown for each value of $\varpi$.} \label{fig:profile stats per amplitude}
\end{figure}

\begin{figure}[t]
\centering
\includegraphics[width=12cm]{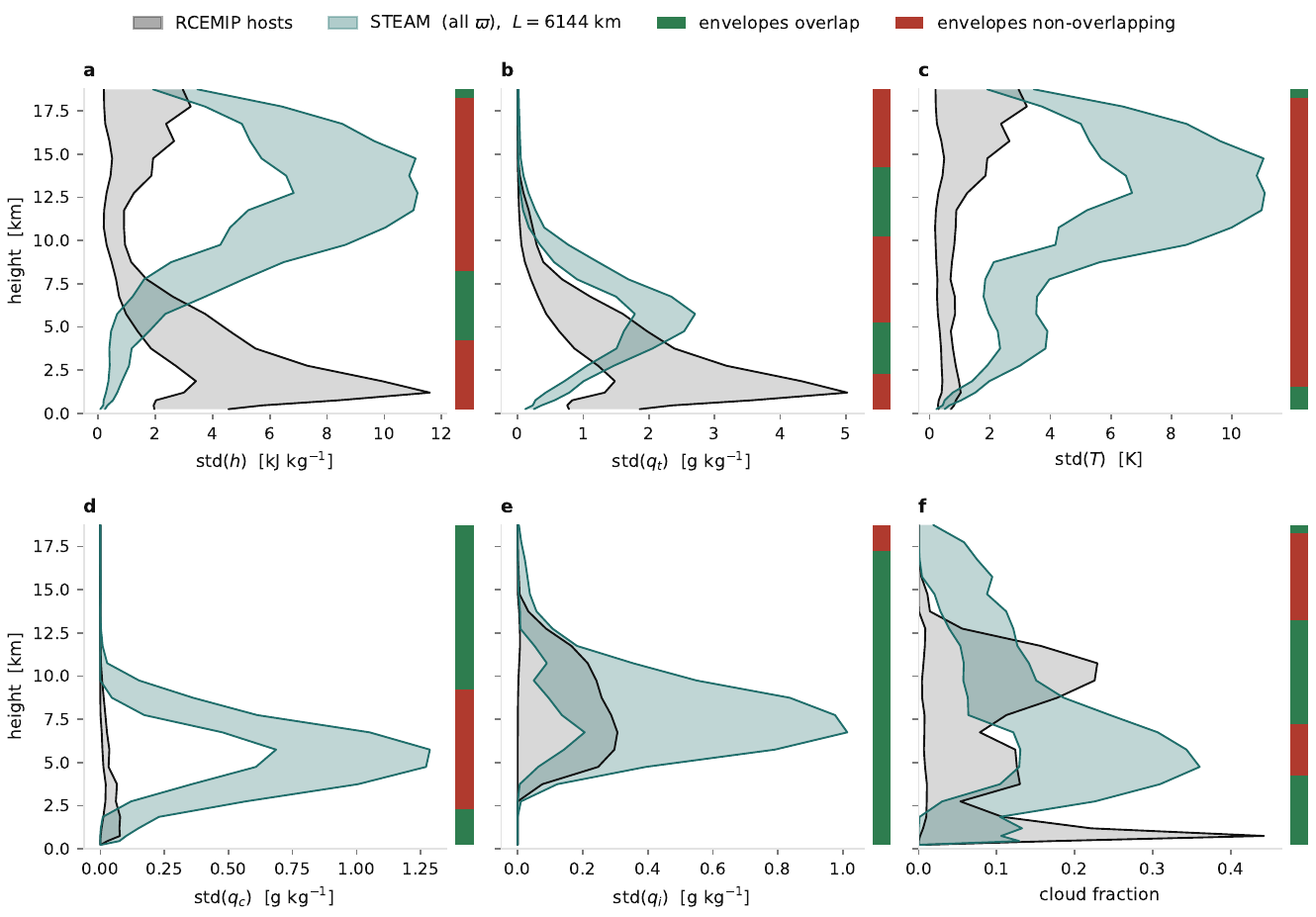}
\caption{Standard deviation profiles as in Fig. \ref{fig:rcemip profiles}, for the alternate configuration with the outer scale set to the channel length, $L = 6144$~km.} \label{fig:rcemip profiles Llong}
\end{figure}

\begin{figure}[t]
\centering
\includegraphics[width=12cm]{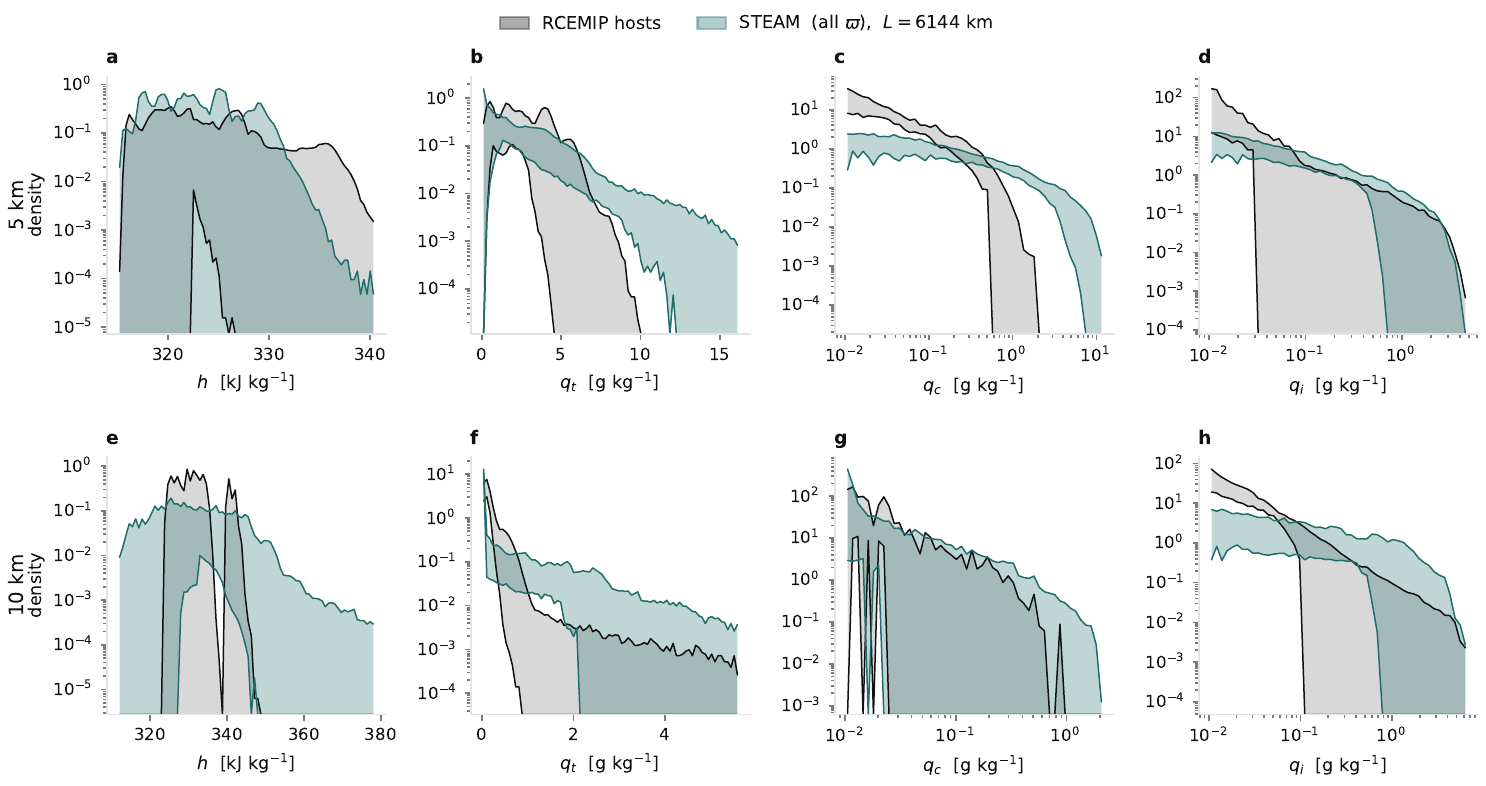}
\caption{Single-level probability density functions for the RCEMIP comparison, for the alternate configuration with the outer scale set to the channel length, $L = 6144$~km.} \label{fig:rcemip pdfs Llong}
\end{figure}

Figs. \ref{fig:profile stats per amplitude}, \ref{fig:rcemip profiles Llong}, and \ref{fig:rcemip pdfs Llong} are similar to Figs. \ref{fig:profile stats}, \ref{fig:rcemip profiles}, and \ref{fig:level pdfs} but contain different information.

\subsection{Supplementary material} \label{app:steam supplement}

\subsubsection[Complete specification of the STEAM algorithm]{Complete specification\\* of the STEAM algorithm}
\label{ssect:steam algorithm spec}

This section specifies the STEAM algorithm at the level of implementation
detail.
The physical construction and its motivation are described in the main
text and its appendices, while the intent here is that the simulation could be
reproduced exactly from this description together with the main-text
equations. The reference implementation is the \texttt{turbulon-model}
repository, and the full analysis for this paper is contained in the
companion \texttt{turbulon-analysis} repository.

\paragraph[Inputs and size classes]{Inputs and size classes.}

The inputs are listed in Table \ref{tab:steam inputs}.

\begin{table}[t]
\centering
\small
\begin{tabular}{lll}
\hline
Symbol & Input & Notes \\
\hline
$\langle h \rangle_t(z)$, $\langle q_t \rangle_t(z)$ & ensemble mean profiles & horizontally uniform, sampled at spacing $\Delta z_p$ \\
$\ell_s(z)$ & spheroscale profile & \\
$L$ & outer scale & conventionally a power-of-two multiple of $dx$ \\
$L_x$, $L_y$ & horizontal domain size & \\
$dx$, $dy$ & finest (output) grid spacing & \\
$L_z$ & domain height & \\
$[\Phi_{\min}, \Phi_{\max}]$ & physical bounds & one pair per scalar \\
$n_b$ & bound-buffer multiple & default 3 \\
$m$ & minimum center distance to ground & in units of $\ell_z(\ell)$, default 0 \\
$p_s$ & surface pressure & \\
--- & random seed & \\
$s_x$, $s_y$, $s_z$ & sparsity factors & positive integers, default 1 \\
$n_c$ & cascade density & size classes per octave, default 1 \\
\hline
\end{tabular}
\caption{Inputs to the STEAM algorithm.}\label{tab:steam inputs}
\end{table}

The implied vertical outer scale $\ell_{z,L}(z) = \ell_s(z)\left(L/\ell_s(z)\right)^{H_z}$ is enforced to be smaller than the domain height at every level.
Logarithmically spaced size classes are defined as $\ell_i = L\, 2^{-i/n_c}$
for $i = 0, 1, \dots, n$, proceeding from the outer scale down to the
class nearest $\ell = 2dx$, the smallest turbulon that is Nyquist-sampled
by the output grid such that $n = \mathrm{round}[n_c\log_2(L/(2dx))]$. When the
outer scale is a power-of-two multiple of $dx$, the conventional
configuration, the finest class is $2dx$ exactly. For each $\ell_i$ the vertical turbulon size is $\ell_{z,i}(z) =
\ell_s(z)(\ell_i / \ell_s(z))^{H_z}$, evaluated with the local spheroscale. The value of $H_z$ is set to $5/9$ for $\ell_i>\ell_s$ and 1 otherwise when the default ``piecewise isotropic'' scaling function is used, as in all simulations in this work. The alternative ``canonical'' scaling function keeps $H_z=5/9$ for all sizes.

The list of size classes does not depend on the
sparsity factors, and so the cascade always terminates at the class nearest
$\ell = 2dx$. The sparsity factors
instead set how finely each class is resolved on its own working grid
as described in the following section. At $s = 1$ the finest class works directly on the
output grid, while for $s > 1$ every class works on a
correspondingly finer grid and the output is written on the finest
working grid with spacing $dx/s_x$.

\paragraph[Normalization]{Normalization.}\label{ssect:normalization}

The outer-scale amplitude profile $C_{\Phi, L}(z)$ is computed for each
scalar $\Phi \in \{h, q_t\}$ from the mean profile as described in
Appendix A. At each profile level, the Haar fluctuation at the local
vertical outer scale is computed by taking the mean of the profile over the upper half of
a $\ell_{z,L}(z)$-wide window minus the mean over the lower half. The
profile is read as the piecewise-linear function its samples define,
and the two half-window means are its exact integrals over
$[z, z + \ell_{z,L}/2]$ and $[z - \ell_{z,L}/2, z]$. Beyond its ends the
profile is extended by its end values. The windows are therefore not
rounded to whole profile cells, which would otherwise bias the
response high wherever a half-window spans only a few cells. The result is multiplied by the calibration
constant $\vartheta$ and taken in absolute value.

$\vartheta$ is set by the consistency requirement of Appendix A: the
vertical Haar fluctuation of the simulated fields at the vertical outer
scale must equal that of the mean profile. It is calibrated by running
a full simulation on linear mean profiles for both scalars, on a domain
whose vertical extent contains one vertical outer scale. The
calibration runs place the scalar bounds far beyond any value the
fields can reach, so the bound projection described in Appendix A never engages and the value of $\vartheta$ does not depend on the bound configuration.
The mean
absolute vertical Haar fluctuation of the simulated columns is computed
at logarithmically spaced lags. A line of fixed slope $H_v = H_h/H_z$ is
anchored to the fluctuation at the second-smallest lag (the smallest,
at two grid cells, is more affected by discretization) and extrapolated to
$\ell_{z,L}$. The criterion is evaluated on this extrapolated line rather
than on the fluctuations measured at $\ell_{z,L}$ itself, because near the
outer scale the Haar response contains contributions from both the fluctuations and the mean profile itself. The ratio of
the profile's Haar fluctuation at $\ell_{z,L}$ to the extrapolated value,
averaged geometrically over the two scalars, multiplies $\vartheta$, and
the calibration is repeated until the ratio converges to one. Convergence to $\sim 1\%$ is typically achieved in three to five iterations.

Because
the same Haar operator is applied to the profile and to the simulated
field, conventions defining the envelope amplitude, the Haar definition, and the interpolation compensation factors described below are absorbed
into $\vartheta$. For unknown reasons the converged values for $\vartheta$ depend on the value of the horizontal Hurst exponent. For the runs here $H_h = 0.45$ and $\vartheta = 0.25518$. 

The scale-dependent amplitudes
follow $C_{\Phi,\ell}(z) = n_c^{-1/\alpha}\, C_{\Phi,L}(z)(\ell/L)^{H_h}$
as described in Appendix A and are interpolated
to each size class's vertical grid.

The bound buffer widths are computed from the same amplitude profiles
as
\begin{align}
  b_{\Phi,\ell}(z) = n_b\, \frac{C_{\Phi,\ell}(z)}{1 - 2^{-H_h/n_c}},
\end{align}
the geometric sum, in closed form, of the mean absolute amplitudes of
the current class and of every class the cascade could ever add below
it. The sum conservatively extends past the classes a particular run
carries. Truncating it at a run's own finest class would make the
buffer, and with it the realized field, depend on where the run stops,
breaking the identity between a nested refinement and a deeper root
run (Section~\ref{sect:nested refinement}).

\paragraph[Working grids]{Working grids.}\label{ssect:working grids}

For computational efficiency, each size class works on its own grid
with $\Delta x_\ell = \ell/(2s_x)$ and $\Delta y_\ell = \ell/(2s_y)$ --- coarser
than the output grid for all but the finest classes. When $s = 1$, the
spacing is half the turbulon size and the turbulon is represented
coarsely, as shown in Fig.~\ref{fig:discretized turbulon}.
The horizontal dimensions $n_{x,\ell} = L_x/\Delta x_\ell$ and $n_{y,\ell} =
L_y/\Delta y_\ell$ are exact integers provided $L_x$ and $L_y$ are integer
multiples of $L$. Each horizontal domain dimension is enforced to be either an integer multiple of the outer scale or smaller than it.
Grid counts along a narrower dimension are rounded to at least one cell, and any turbulon
envelope wider than the extent is periodically folded onto it by summation.

The vertical grid requires more care because $\ell_z(z)$ varies with
height. To compute $\Delta z$, cells are accumulated upward from the surface with local
spacing $\ell_z(\ell, \ell_s(z))/(2s_z)$ until the accumulated grid first reaches
or overshoots the domain top, with the overshooting cell retained. The resulting spacings are then rescaled by a single common factor
$\le 1$ so that they tile $L_z$ exactly. For a constant spheroscale this reduces
to a uniform grid at least as fine as the Nyquist target.

As described below, at each cascade step the fields are trilinearly interpolated to the next working grid dimensions. The interpolation is cell-consistent: source and target cells tile the same extent, and values are sampled at cell centers. In $x$ and $y$ the grid is periodic. To enforce this the source is padded with wrapped cells and the margin cropped after the interpolation. In the vertical direction, the convolutions are not periodic and are instead zero-padded. 

Due to the coarseness of the grid, interpolation does not preserve the mean absolute value of the turbulon perturbations. A size class that is poorly resolved on the output grid introduces more variability than its $C_{\Phi,\ell}$ factor alone would imply. To account for this, each class carries a compensation factor $\digamma$ that depends on $\ell/\Delta x$, the turbulon size in units of the output grid spacing (Table \ref{tab:interpolation compensation}).
\begin{table}
\centering
\begin{tabular}{lccccccccc}
\hline
$\ell/\Delta x$ & 2 & 4 & 8 & 16 & 32 & 64 & 128 & 256 & 512 \\
\hline
$\digamma$ & 0.337 & 0.595 & 0.798 & 0.894 & 0.925 & 0.950 & 0.975 & 0.988 & 1 \\
\hline
\end{tabular}
\caption{Compensation factors for the resolution dependence of mean absolute turbulon amplitudes, where $\Delta x$ is the final output grid resolution in the horizontal direction. Factors are measured on the full algorithm (\texttt{tests/heavy/interpolation\_compensation.py}) and normalized to the $\ell/\Delta x = 512$ reference; intermediate values are interpolated linearly in $\log_2(\ell/\Delta x)$.}\label{tab:interpolation compensation}
\end{table}
The factor is applied when the output is composed, not inside the cascade itself, and therefore the gradient weighting and the bound factor are computed from the running, uncompensated state. 

To keep track of the compensation factors, the cascade accumulates a compensation deficit field for each scalar, $\sum_i (\digamma_i - 1)\,\Delta_i$, where $\Delta_i$ is the increment class $i$ actually added. The deficit is a three-dimensional field rather than a per-class constant because the correction must restore a fraction of each class's particular increment, and so inherits the increments' spatial structure. The deficit is interpolated between classes together with the state, and the output is the sum of the two as described in Section~\ref{ssect:final fields} below. The flux accumulates a deficit in the same way as well. The ``retention'' factors underlying $\digamma$, defined as the fraction of the mean absolute amplitude that survives one interpolation step, are measured by rerunning a single size class of the full algorithm at successively finer output resolutions and taking ratios of the resulting mean absolute perturbations.

Two effects contribute to the resolution dependence. First, at small $\ell/\Delta x$ the sampled points on the envelope curve land preferentially near the envelope peaks; at the extreme $s = 1$ the grid points coincide with the turbulon centers. A marginally resolved class therefore reads up to $1/0.337 \approx 3\times$ the amplitude of a well resolved one. Second, at large $\ell/\Delta x$, each regrid resamples the piecewise linear field at points that do not exactly coincide with the previous ones, especially in the vertical direction where upsampling occurs by a noninteger factor $2^{H_z}$. This removes 1--3\% of the amplitude per octave out to at least $\ell/\Delta x = 256$. There is consequently no fully converged reference resolution. The factors in Table \ref{tab:interpolation compensation} are instead normalized to $\ell/\Delta x = 512$, the largest class used by any simulation in this work. The choice of reference is a single overall constant and is absorbed into the calibration factor $\vartheta$ above.

Retentions differ between the isotropic and anisotropic regimes because of the different vertical upsampling factors ($2$ versus $2^{H_z}$ per step), and were therefore measured separately in the two regimes (Table \ref{tab:hop retention}).
\begin{table}
\centering
\begin{tabular}{lcccccccc}
\hline
$y$ & 2 & 4 & 8 & 16 & 32 & 64 & 128 & 256 \\
\hline
$r_{\mathrm{ani}}(y)$ & 0.5664 & 0.7456 & 0.8926 & 0.9667 & 0.9730 & 0.9746 & 0.9871 & 0.9880 \\
$r_{\mathrm{iso}}(y)$ & 0.5401 & 0.8362 & 0.9587 & 0.9839 & 0.9956 & --- & --- & --- \\
\hline
\end{tabular}
\caption{Per-regrid retention factors in the anisotropic and isotropic regimes, measured under the cell-consistent interpolation convention. Isotropic entries beyond $y = 32$ are unmeasured and treated as well resolved (the tabulated trend has converged to within measurement precision of 1).}\label{tab:hop retention}
\end{table}
A class's compensation factor is computed from the retention factors for all regrids that its perturbation field will eventually undergo, each taken from the regime of that regrid's destination grid at each height level. A cascade crossing the spheroscale therefore composes isotropic retentions below it and anisotropic retentions above it, with the crossing step approximated by the regime of its destination grid.

The specific rule is as follows. Write $r(y)$ for the retention of one regrid, with $y$ the class's value of $\ell/\Delta x$ on the grid the regrid starts from. A class's perturbation field is created at $y = 2$, on the class's own working grid, and $y$ doubles at each subsequent regrid. Intermediate $y$ (which arise for $n_c > 1$) are interpolated linearly in $\log_2 y$. A regrid at $y$ at or beyond twice the last tabulated value retains 1, the class being well resolved by then; between the last tabulated value and twice it, the last tabulated retention applies. At each level $z$, a class whose perturbation field has yet to traverse regrids $m = 1, 2, \dots$ carries
\begin{align}
  \digamma(z) = D_{512} \Big/ \prod_m r_{\rho_m(z)}(y_m), \qquad
  D_{512} = \prod_{y = 2, 4, \dots, 256} r_{\mathrm{ani}}(y) \approx 0.337,
\end{align}
where $\rho_m(z)$ selects the column of Table \ref{tab:hop retention} by the regime of the $m$-th regrid's destination grid at that level, and $D_{512}$ is the delivery of the reference chain. The regrids reduce the class's perturbation field by exactly the product being divided by, so once the deficit is added at the output, every class arrives with the same effective normalization $D_{512}$. For a run with no classes below the spheroscale this reduces to Table \ref{tab:interpolation compensation}.

\paragraph[Cascade loop]{Cascade loop.}

The perturbation fields $\sum_\ell h_\ell$ and $\sum_\ell q_{t,\ell}$ are
initialized to zero and the flux field $\mathcal{F}$ to one on the
coarsest grid. The algorithm then iterates over size classes from large
to small, performing the following sequence of operations at each class.

First, the perturbation and flux fields are trilinearly interpolated
from the previous class's grid to the current one, together with the
compensation deficit fields of Section~\ref{ssect:working grids} once
they exist. The mean profiles are linearly interpolated to the current
vertical grid. 

The flux is then advanced by one size class. A sparse field $\gamma$ of
independent extremal ($\beta = -1$) L\'evy $\alpha$-stable variables with
$\alpha = 1.8$ is drawn at the turbulon centers every $s_x, s_y, s_z$
cells. This corresponds exactly to every $\ell/2$ in physical space horizontally, but vertically the common rescaling of the
vertical grid (Section \ref{ssect:working grids}) makes them less than or equal to
$\ell_z/2$. No centers are placed within $2 s_z m$ vertical cells of the
surface or of the domain top, corresponding to approximately $m\, \ell_z(\ell)$ in physical space with $m$
defined in Table \ref{tab:steam inputs}. At the default $m = 0$ no centers are
excluded, and turbulons centered at the boundary cells are cut off by the
vertical zero padding. The L\'evy noise is scaled by
$\varpi\, n_c^{-1/\alpha}$ and shifted by the deterministic
constant $\varpi^\alpha/(n_c(\alpha-1))$, which enforces $\langle
e^\gamma \rangle = 1$ exactly. This can be seen by writing $\gamma_0$ for the unscaled
draw, $\ln\langle \exp(\varpi\, n_c^{-1/\alpha}\gamma_0) \rangle =
\varpi^\alpha/(n_c(\alpha-1))$ in closed form, the $\alpha$-stable
generalization of the lognormal $-\sigma^2/2$. The expectation is
finite because $\beta = -1$ places the heavy tail on the side the
exponential suppresses. 

Once L\'evy noise is generated and scaled, the flux is updated additively as
$\mathcal{F} \mathrel{+}= \mathfrak{T}_\ell * [(e^\gamma - 1)\mathcal{F}]$. After the addition, any negative values are set to zero. The clipped flux is then rescaled by a multiplicative factor that restores the volume
mean the field had before the perturbation was added.
The signed scalar noise $S_\ell$ is defined as the increment $(e^\gamma -
1)\mathcal{F}$, evaluated with the entering flux and normalized by the mean
absolute value of $e^\gamma - 1$ over the turbulon centers. The scalar noise is shared
between $h$ and $q_t$.

Next, the scalar amplitudes are computed from the noise, bound factor, and gradient weighing factor. The bound factor $g_\Phi$ is evaluated pointwise from the running field $\Phi_{>\ell}$ using the buffer width $b_{\Phi,\ell}(z)$ defined in Appendix A.
 
Gradients are taken with periodic
central differences in $x$ and $y$ and with one-sided differences at the
vertical boundaries (central in the interior). The gradient weight is computed as $W_{\Phi,\ell} =
|\nabla_h \Phi_{>\ell}| + (\ell_z/\ell)|\partial \Phi_{>\ell}/\partial z|$. The amplitude pattern $P_{\Phi,\ell} = g_\Phi W_{\Phi,\ell} S_\ell$ is divided
by its mean absolute value over the turbulon centers at each level.
Centers where the pattern is zero --- whether pinned to zero by the bound
taper or holding an exactly zero gradient weight --- are excluded from the
mean.
Finally, the normalized pattern
is multiplied by $C_{\Phi,\ell}(z)$, giving the amplitude field
$A_\Phi$.

The amplitude fields are then convolved with the turbulon envelope. The 3D
Mexican hat defined in Appendix A is constructed
with $\sigma = \ell/\pi$ isotropically in grid-index space and truncated at a distance $3\ell$ from its center along each
axis. If the kernel exceeds
the field extent horizontally it is periodically folded by summation onto the field
period. $A_\Phi$ is convolved with this kernel using a horizontally periodic and vertically zero-padded convolution to obtain a candidate perturbation field for
$\Phi_\ell$. 

Because the candidate perturbation field may cause the running field $\Phi_{\ge\ell}$ to exceed its bounds, it cannot be straightforwardly added to $\Phi_{>\ell}$. Instead, 
a bounded,
amplitude-preserving projection is used. At every level the added perturbation
must satisfy three conditions: (i) zero horizontal mean; (ii) mean
absolute amplitude as close as possible to that of the unclipped
candidate field, so that the amplitude at each scale is set by the
normalization factors alone; and (iii) $\Phi_{\min} \le \Phi \le
\Phi_{\max}$ pointwise. These criteria are met iteratively.
The added perturbation is taken from a two-parameter family: the
candidate field $\delta_0$ is scaled by $s$, shifted by $\mu$, and
clipped to the bounds,
\begin{align}
  \delta(s,\mu) = \min\!\big(\max(s\,\delta_0 - \mu,\; \Phi_{\min} -
  \Phi),\; \Phi_{\max} - \Phi\big).
\end{align}
The shift enforces condition (i). At fixed $s$, the level mean of
$\delta$ is a continuous, piecewise linear, and nonincreasing function
of $\mu$, so the value of $\mu$ that produces a zero mean is unique and
is found by bisection. The scale enforces condition (ii): the mean
absolute value of $\delta$ increases monotonically with $s$, and a few
multiplicative updates bring it to the target, each followed by
subtracting the level mean and reclipping.

The three conditions cannot always be met at once. A perturbation that
has zero mean and keeps $\Phi$ within its bounds can have mean absolute
amplitude no larger than $2\min(\langle\Phi\rangle - \Phi_{\min},\;
\Phi_{\max} - \langle\Phi\rangle)$, because positive and negative
excursions must balance and the smaller of the two headrooms limits
both. The target amplitude exceeds this limit wherever the level mean
sits too close to a bound. For $q_t$ this happens at dry upper levels,
where the summed amplitudes $\sum_\ell C_{q_t,\ell}$ can reach ten times the
level mean. On such levels the iteration stops at the largest amplitude
the bounds permit, and conditions (i) and (iii) still hold exactly.
Once accepted, the perturbation is added to the running field, so
subsequent size classes see the gradients of the bounded field.

\paragraph[Final fields and output files]{Final fields and output files.}\label{ssect:final fields}

After the finest class, the outputs are composed from the cascade
states and their compensation deficits. Each scalar output is 
$$\Phi =
\langle \Phi \rangle_t + \sum_\ell \Phi_\ell + \sum_i (\digamma_i - 1)
\Delta_{\Phi,i},$$
projected onto $[\Phi_{\min}, \Phi_{\max}]$ with each level's horizontal mean preserved. A clipped level is replaced by the closest in-bounds field with the same mean, using the shift-and-clip construction of the bounded projection above without its mean absolute amplitude condition. This final projection is needed because
adding the compensation $\sum_i (\digamma_i - 1)
\Delta_{\Phi,i}$ can push the sum past
bounds even after the per-class bounded projection. 

The flux output is the flux field plus its deficit, clipped
at zero and rescaled to the volume mean of the uncompensated field,
following the same correction rescaling method that is applied at each class. The composition is the last
operation and a descendant nested simulation does not read the compensated fields. A run whose classes are all well resolved ($\digamma \equiv 1$) skips
compensation entirely. Finally, temperature, water phases, and pressure are
recovered from the composed $h$ and $q_t$ by the saturation adjustment
of the main-text appendix.

Each simulation produces one NetCDF4 file, with the contents listed in
Table \ref{tab:output file}.
\begin{sidewaystable}
\centering
\footnotesize
\setlength{\tabcolsep}{4pt}
\begin{tabular}{llll}
\hline
Variable & Dimensions & Units & Description \\
\hline
\texttt{x}, \texttt{y}, \texttt{z} & $(x)$, $(y)$, $(z)$ & m & output-grid coordinates \\
\hline
\texttt{h} & $(x,y,z)$ & J\,kg$^{-1}$ & moist static energy \\
\texttt{qt} & $(x,y,z)$ & kg\,kg$^{-1}$ & total water mixing ratio \\
\texttt{flux} & $(x,y,z)$ & --- & dimensionless conserved flux $\mathcal{F}$ \\
\hline
\texttt{T} & $(x,y,z)$ & K & temperature$^{\ast}$ \\
\texttt{qv}, \texttt{qc}, \texttt{qi} & $(x,y,z)$ & kg\,kg$^{-1}$ & vapor, liquid, ice$^{\ast}$ \\
\texttt{p} & $(x,y,z)$ & Pa & pressure$^{\ast}$ \\
\hline
\texttt{z\_profile} & $(z_p)$ & m & input profile grid \\
\texttt{h\_profile}, \texttt{qt\_profile} & $(z_p)$ & J\,kg$^{-1}$, kg\,kg$^{-1}$ & input mean profiles \\
\texttt{spheroscale\_profile} & $(z_p)$ & m & input spheroscale profile \\
\hline
\texttt{k\_values}, \texttt{k\_z\_values} & $(\ell)$ & m & horizontal and vertical size classes \\
\texttt{C\_h\_k}, \texttt{C\_qt\_k} & $(\ell, z_\ell)$ & J\,kg$^{-1}$, kg\,kg$^{-1}$ & per-class amplitude profiles \\
\texttt{dz}, \texttt{spheroscale} & $(z)$ & m & cell heights and spheroscale, output grid \\
\hline
\texttt{p\_bottom} & $(x,y)$ & Pa & basal pressure (elevated nests only) \\
\hline
\texttt{h\_perturbation}, \texttt{qt\_perturbation} & $(x,y,z)$ & J\,kg$^{-1}$, kg\,kg$^{-1}$ & cascade states$^{\dagger}$ \\
\texttt{flux\_state} & $(x,y,z)$ & --- & flux cascade state$^{\dagger}$ \\
\texttt{class\_increments} & group; one subgroup per class & --- & added $h$, $q_t$, flux increments, each on its class's working grid$^{\dagger}$ \\
\hline
\end{tabular}
\caption{Contents of a STEAM output file. $^{\ast}$Diagnostic fields
recovered from $h$ and $q_t$ by the saturation adjustment after the
cascade completes, in a separate pass. $^{\dagger}$Stored only for
runs intended as refinement parents. These are the states as the
cascade left them, with no compensation deficit, no projection, and
the mean profile not added; a nest continues from them, because the
written fields, being composed, cannot be turned back into them.
Every class's row of \texttt{C\_h\_k} and \texttt{C\_qt\_k} is
interpolated onto the output $z$ grid before storage.}\label{tab:output file}
\end{sidewaystable}
The three-dimensional fields are single
precision. The input $z$ and spheroscale profiles are kept in double
precision given that the storage cost is negligible.
Compression is optional and off by default. When enabled,
the fields use the \verb|blosc-zstd filter| (byte-shuffled, compression level
1), and chunks smaller than 16\,KiB are stored uncompressed. The
inputs are stored alongside the outputs and scalar attributes record the full configuration:
grid dimensions and spacings, outer scale, domain height, bounds,
sparsity factors, cascade density, surface pressure, random seed,
$H_h$, $H_z$, $\vartheta$, the flux parameters $\varpi$ and $\alpha$, and
the envelope and anisotropy options. 

If nested simulations are to be performed, the flag \texttt{save\_for\_refinement} is required, which additionally stores the cascade states and per-class increments: \texttt{h\_perturbation}, \texttt{qt\_perturbation}, \texttt{flux\_state}, and \texttt{class\_increments}. This allows nested simulations to continue the cascade without reading the final compensated fields. Compensation is only done at final output resolution and not mid-cascade.

Nested refinements (next section) are stored as NetCDF groups inside
the parent's file (e.g. \texttt{refinements/r0}) with the same layout.
A nest's coordinates are world-absolute and attributes record its
parent group, its index range within the parent grid, and whether each
of its horizontal axes is periodic. A nest whose base is elevated
above the ground additionally stores the two-dimensional pressure
field at its base, taken from the parent's three-dimensional pressure,
from which its hydrostatic integration starts.

\paragraph[Nested refinement]{Nested refinement.}\label{sect:nested refinement}

A completed simulation can be ``refined'' to finer resolution over a portion of its domain by simply continuing the cascade within the desired region. The design requirement is that a nest be the same cascade carried further: a nest spanning the full parent domain is identical, cell for cell, to a root run taken directly to the nest's resolution. A nest covering part of the domain deviates from that identity only in that the global statistics (the weight normalization, the mean absolute multiplier noise, and the flux volume means) are evaluated over the nest's own region rather than the full domain. The nest therefore continues from the stored cascade states of Table \ref{tab:output file}, extracted over the nest region and interpolated trilinearly onto the nest's grids. A parent intended for refinement must therefore store intermediate cascade states, because the final projection is not always invertible. The
amplitude profiles $C_{\Phi,\ell}$ are rederived from the mean profile
by exactly the computation the root performed, with the root's outer
scale, $H_h$, and $\vartheta$. Because the Haar response is evaluated
exactly on the profile's own grid (Section~\ref{ssect:normalization}), this
reproduces the root's amplitudes class by class with no discontinuity
at the overlap scale. The
flux anomaly requires no special handling because the per-class
renormalization restores the volume mean the field entered with. For a
nest, flux fields are normalized to the regional flux mean, computed over the nested domain.

The interpolation compensation factors depend on the final output grid, so
the deficit belonging to the nest's output differs from the parent's.
The parent therefore stores each class's added increments ($h$, $q_t$,
and flux) on that class's own working grid, in the
\texttt{class\_increments} group of Table \ref{tab:output file}. Because the
working grids are at coarsened resolution, the storage cost is only approximately 
$1.14\times$ one output-grid field per stored field. 

From the stored
increments, the nest accumulates the deficit of its inherited classes
with $\digamma_j$ evaluated on its own output grid through $\sum_j
(\digamma_j - 1) \Delta_j$, through the same sequence of regrids the
cascade itself uses. Its cascade then continues the accumulation over
the nest's own classes, and the composition at the nest's output is
exactly a root's, because the composition is linear in the increments.

Parent fields are extracted including a buffer on each nonspanning horizontal side whose per-class physical width is equal to the turbulon support at that
class. Turbulons centered outside the nest but reaching into it are therefore still simulated, so that the component of such turbulons that extends into the interior of the domain is included.
The buffer is excluded from every statistic
evaluated over the simulated domain --- the advective-weight
normalization, the mean absolute multiplier noise, the flux volume
means, and the bounded projection --- and it is discarded on output.
Bounds are still enforced across the buffer because the next class
reads its gradients and its bound factor. Along a periodic axis that the
nest spans entirely, periodicity is preserved and no buffer is used.
A nest that spans, or whose buffer would cross, a nonperiodic parent
boundary (e.g. a parent that is itself a nested simulation) is rejected. Vertically, the buffer is
clipped at the ground and domain top.

By default a nest draws no new seed and its per-class seeds continue the
root's stream where the parent left off. A root run without a seed
draws a random one and records it, so any root can be continued.
Output is written as a NetCDF group inside the parent's file as
described in the previous section.

\subsubsection[Complete specification of the comparison analysis]{Complete specification\\* of the comparison analysis}
\label{sect:comparison analysis}

This section specifies the analysis behind the comparison of
STEAM to hydrodynamic simulations and to satellite observations, with
the intent that it could be reproduced exactly from this description
together with the two repositories. The hydrodynamic references are
eleven hosts: the elongated channel configurations (\texttt{RCE\_large300})
of nine models from the RCEMIP cloud-resolving ensemble
\citep{wing2018,wing2020}, and two SAM large-eddy simulations with boundary conditions derived from the TWP-ICE
and GATE field campaigns. These enter through matched profile statistics, single-level
distributions, and horizontal scaling functions. The cloud geometry
metrics are computed on a separate square-domain STEAM ensemble and
read against the two SAM cases and a MODIS retrieval.

\paragraph[Comparison data]{Comparison data.}
\label{ssect:comparison data}

Nine models provide usable three-dimensional snapshot output for the
300~K channel configuration: SAM, CM1, three Met Office Unified Model
variants (CASIM, RA1-T, RA1-T-nocloud), SCALE, UCLA-CRM, and two ICON
variants (LEM and NWP). Domains are channels of approximately
$6000\times 400$~km$^2$ at 3~km horizontal spacing ($1984$--$2048$
cells long, $128$--$144$ cells wide), with model-specific stretched
vertical grids ($74$--$98$ levels to $33$--$39.5$~km; through the free
troposphere the spacing is 500~m, or 250~m for the three UM variants).
We consider three statistically independent snapshots separated by ten days near the end
of the integration for each model. Mean profiles and comparison statistics are averaged or pooled across all three model snapshots. Two
archived models are excluded: ICON-AES, because its output contains
neither a height coordinate nor pressure, and MESONH, because its archived
three-dimensional specific humidity is a documented data error in the
RCEMIP archive (Known RCEMIP Bugs document, Section~17: the field is
``much too small in the lower troposphere'' by a height-dependent
factor).

All humidities are converted to mixing ratios before any further
computation. Most models archive specific quantities (mass per unit
moist air, CMOR names \texttt{hus}, \texttt{clw}, \texttt{cli}). SAM archives vapor as a
mixing ratio directly. Cloud liquid and ice are taken from
\texttt{clw} and \texttt{cli} (UCLA names ice \texttt{ice}).
Precipitating condensate is excluded everywhere. Total water is $q_t = r_v + r_c + r_i$ and
moist static energy is $h = c_p T + g z + L_v r_v$ with
$c_p = 1004$~J\,kg$^{-1}$\,K$^{-1}$, $g = 9.81$~m\,s$^{-2}$,
$L_v = 2.5\times 10^6$~J\,kg$^{-1}$.

Two archives require special handling. (i)~The ICON files carry only a
level \emph{index} (top-down) as their vertical coordinate; physical
height is recovered by inverting the frozen moist static energy field
distributed alongside them, $z = (f_{\mathrm{mse}} - 1004.64\,T -
2500800\,q_v + 333700\,q_i)/9.80665$ (the constants of that archive,
used only for this recovery), averaged horizontally at one snapshot;
the level-wise horizontal standard deviation of the recovered height
is required to be below 50~m. (ii)~UCLA-CRM includes a below-ground ghost level, which is
dropped. Where a model archives no pressure (SCALE, UCLA-CRM), the
RCEMIP analytic-sounding surface pressure of 1014.8~hPa is used to
initialize STEAM; otherwise the horizontal mean of the model's
lowest-level pressure is used. All reductions of single-precision
fields use double-precision accumulators, and the comparison
statistics cast kelvin-scale fields to double precision upon read.

The two large-eddy cases are SAM simulations on the same geometry,
$2048^2$ cells at 100~m horizontal spacing over 204.8~km, with
stretched vertical grids of ${\sim}255$ levels (50~m near the surface
through 100~m across the free troposphere). One snapshot is used from
each. For TWP-ICE, the last
archived snapshot (model day 20) is used, while for GATE, hour 23 is used,
the last hour with complete data. SAM archives its
(nonfrozen) moist static energy directly, in temperature units, so
$h$ is $c_p$ times the archived field.
Vapor and condensate are mixing ratios in g\,kg$^{-1}$, and $q_t$
excludes the archived precipitating species by the same convention as
above. GATE archives only the combined
condensate, so its liquid/ice split is reconstructed with a
linear partition that is all liquid at 0$^\circ$C and all ice at
$-38\,^\circ$C. This partitioning is applied at native resolution before
any coarsening. GATE's $h$ is built as
$c_p T + g z + L_v r_v$ from the archived temperature.

\paragraph[STEAM configuration]{STEAM configuration.}
\label{ssect:steam config}

With the exception of the mean profiles, no STEAM inputs are
calibrated to any individual host. The spheroscale is a constant 10~m,
with turbulons isotropic below it
(\texttt{piecewise\_isotropic\allowbreak\_below\allowbreak\_spheroscale}); the Hurst
exponents, $\vartheta$, the turbulon envelope, the cascade density, and
the flux noise parameters $\alpha$ and $\beta$ are those of
Section~\ref{ssect:steam algorithm spec}. The three values $\varpi \in \{0.02, 0.05, 0.17\}$ are used for the flux noise amplitude throughout the comparison. The simulated depth is 20~km and
the input profiles are sampled at 50~m. The run sets are listed in
Table~\ref{tab:steam runs}.

\begin{sidewaystable}
\centering
\small
\setlength{\tabcolsep}{4pt}
\begin{tabular}{llllll}
\hline
Set & Grid & $\Delta x$ & Extent & $L$ & Runs \\
\hline
Channel-matched & $1024\times 64$ & 6~km & $6144\times 384$~km$^2$ & 384 or 6144~km & 9 hosts $\times$ 3 amplitudes $\times$ 2 outer scales \\
LES-matched & $1024^2$ & 200~m & $204.8\times 204.8$~km$^2$ & 204.8~km & 2 hosts $\times$ 3 amplitudes $\times$ 5 members \\
Square ensemble & $2048^2$ & 1~km & $2048\times 2048$~km$^2$ & 2048~km & 10 members $\times$ 3 amplitudes \\
\quad nest A & $512^2$ & 62.5~m & $32\times 32$~km$^2$, full depth & --- & one per member \\
\quad nest B & $1024^2$ & 7.8125~m & $8\times 8$~km$^2$, $z = 1$--5~km & --- & one per member \\
Visualization & $2048\times 1024$ & 10~m & $20.48\times 10.24$~km$^2$ & 20.48~km & 3 amplitudes $\times$ 3 spheroscales \\
\quad nest & $512\times 1024$ & 2.5~m & $1.28\times 2.56$~km$^2$, full depth & --- & one per run \\
\hline
\end{tabular}
\caption{STEAM run sets of the comparison analysis. Nests continue
their parent's cascade and have no outer scale of their own. The
visualization runs reach only 4~km depth; their nests are the only
simulations whose finest size class, $2\Delta x = 5$~m, lies below the
spheroscale.}\label{tab:steam runs}
\end{sidewaystable}

The scalar bounds are anchored to a 300~K sea surface, prescribed by the RCEMIP channel configuration, and
applied to all hosts: $q_t \in [0,\
q_{\mathrm{sat}}(300\,\mathrm{K}, p_s)]$, since no parcel can acquire
more vapor than saturated contact with the sea surface provides, and
$h \in [\min_z \langle h \rangle(z) - c_p \Delta T_{\max},\
\max(c_p \cdot 300\,\mathrm{K} + L_v\,
q_{\mathrm{sat}}(300\,\mathrm{K}, p_s),\ \max_z \langle h
\rangle(z))]$, where the upper bound is the larger of the surface
saturation moist static energy and the mean-profile maximum. The
lower bound for $h$ allows a maximum temperature deficit of
$\Delta T_{\max} = 10$~K below the coldest point of the mean profile.

The square ensemble provides the cloud geometry statistics of Section~\ref{ssect:cloud geometry} using ten
members per value for $\varpi$, with mean profiles set by the UM RA1-T profile, which has the lowest cloud fraction of the hydrodynamic ensemble. The flux amplitude sets share seeds
member for member, so simulation sets are created that differ only in $\varpi$. For each square member, 
the two nested refinements of Table~\ref{tab:steam runs}
(Section~\ref{sect:nested refinement}) are also simulated for visualization. The archived product per member is
the condensate ($q_c$, $q_i$) of the parent and both nests, plus the
parent's two-dimensional vertically integrated optical depth
(Section~\ref{ssect:albedo masks}), stored unthresholded.

\paragraph[Profile statistics and distributions]{Profile statistics and distributions.}
\label{ssect:matched stats}

Profiles and distributions are compared at matched resolution. The
hydrodynamic host fields are coarsened in $2\times 2\times 2$ blocks
before any statistic is computed, and STEAM is configured so that its
horizontal and vertical resolutions nearly match the coarsened host
grids. Comparison levels are the coarsened host's own and are clipped
to STEAM's uppermost stored level, just below the 20~km domain top.

The per-level statistics are the horizontal standard deviations of
$h$, $q_t$, $q_c$, and $q_i$, and cloud fraction, defined as the areal
fraction of cells holding at least 0.01~g\,kg$^{-1}$ of
nonprecipitating condensate, with the threshold applied after
coarsening.

Distributions are drawn at single model levels nearest 5~km and 10~km. For $h$
and $q_t$, the samples at a level are histogrammed in 80 equal-width
bins spanning the pooled range of all compared cases and normalized to a density. For the
condensates, cells holding at least $0.01\,\mathrm{g/kg}$ of that
phase are binned using 50 logarithmically spaced bins upward from the $0.01\,\mathrm{g/kg}$
threshold.

\paragraph[Scaling functions]{Scaling functions.}
\label{ssect:scaling functions}

The horizontal scaling functions are computed at native resolution
for all models. The host and STEAM curves therefore begin at different
smallest lags --- 200~m against 400~m for TWP-ICE, 6~km against 12~km
for the channels --- and where they overlap they measure the same
physical scales. Scaling functions are computed for TWP-ICE and the
nine channels.

The fluctuation measure is the one dimensional first order Haar fluctuation, defined for lag
$\ell$ as the mean of the field over the upper half of an $\ell$-wide
window minus the mean over the lower half:
$$F_1(\ell) = \langle |\Delta\Phi_\ell| \rangle.$$
For $h$ and $q_t$ at
the level nearest 5~km and 10~km, the transform runs along the longer
horizontal axis with periodic boundary conditions. Ensemble means are computed over the orthogonal horizontal dimension and snapshot index. The level mean is subtracted first
and the computation is done in double precision. The Haar kernel is
zero-mean, so the subtraction changes nothing analytically but reduces numerical error. The local slope shown
alongside each curve is the least-squares slope of $\log_{10} F_1$
against $\log_{10}\ell$ over the half-decade window centered on each
lag. Haar computations use the \verb|scaleinvariance| function \verb|haar_fluctuation|.

\paragraph[Cloud masks and albedo thresholds]{Cloud masks and albedo thresholds.}
\label{ssect:albedo masks}

Cloud masks are thresholds of visual albedo. Column optical depth
$\tau$ is computed from the vertically integrated liquid and ice water
using a parameterization implemented in the
\texttt{cloudyview} package. Precipitating species are excluded. Optical depth is
converted to two-stream albedo for an
overhead sun using $A = \tau/(\tau + 2/(1-g))$ with asymmetry parameter
$g = 0.85$. A column is cloudy where $A$ exceeds a threshold $R$, at
each of $R = 0.1$, $0.2$, $0.3$. $A$ is monotonic in $\tau$,
so the three masks are equivalently $\tau > 1.48$, $3.33$, $5.71$.
Every metric is computed independently at each threshold on the model's native grid resolution.

\paragraph[Cloud geometry metrics]{Cloud geometry metrics.}
\label{ssect:cloud geometry}

Four metrics are computed from each mask set, as defined in the main
text: the individual fractal dimension $D_f$, from the scaling of
perimeter against size across objects; the ensemble fractal dimension
$D_e$, the correlation dimension of the cloud-edge point set; and the
area and nested-perimeter size-distribution exponents
$\tau_{\mathrm{area}}$ and $\tau_{\mathrm{per}}$. All four are
computed with \texttt{objscale} \citep{dewitt2026objscale} at its default
parameters throughout, following \citet{dewitt2024b} and \citet{dewitt2026}.

The ten members of an amplitude set are pooled into one ensemble and
passed to \texttt{objscale} in a single call per metric. 
The correlation integral sampling is thinned tenfold for the pooled
square set and a hundredfold for the MODIS archive, whereas the single-snapshot SAM cases are not
thinned. 

The observational reference is recomputed using the same dataset and methodology used in \citet{dewitt2026}, with the exception of a reflectivity adjustment for viewing angle, described below. 
Band~1 (0.645~$\mu$m) reflectance is read from the 250~m detectors
aggregated onto the 1~km grid. Geolocation, solar geometry, and
per-pixel footprints come from the matching \verb|MOD03| granules. A pixel at
the swath edge covers about $4.8\times 2.0$~km against $1\times 1$~km
at nadir, so footprints are passed to \texttt{objscale} per pixel,
measured as great-circle distances between neighboring pixel centers.
Along-track distances are taken only between rows of the same
ten-detector scan. The swath is trimmed at $60^\circ$
sensor zenith. The ensemble estimators take a single footprint grid
for every array, so the per-granule footprints are averaged over the
72 granules. The scan geometry repeats granule to granule, and the
spread about that mean is measured and found to be negligible.

Saturated pixel detections, where pixels are brighter than the device is capable of recording, are set to the maximum allowable reflectivity value. With the exception of the correlation dimension, for which missing data are set to zero, missing data are passed to the \verb|objscale| function as \verb|nan|.

The stored L1B reflectance is $R\cos\theta_0$ where $R$ is reflectivity, and so we obtain $R$ by dividing the data product by the pixel-by-pixel cosine of the solar zenith angle. This is a departure from \citet{dewitt2026} who did not divide by this factor. The exponents move by at most 0.04 between the two methodologies,
while cloud cover moves by up to a factor of 2.1. All granules have a solar zenith angle of at most $69^\circ$.

We do not report exponents for cases failing the statistical validity criteria proposed by \citet{dewitt2024b,dewitt2026}. These cases are simply identified when \verb|objscale| emits warnings.

\subsubsection{Computational cost}
\label{sect:computational cost}

This section compares the computational cost of STEAM to that of CM1, 
measured on a single desktop machine
(16-core AMD Ryzen 9950X with 64~GB memory). We consider the cost per independent snapshot.
Because STEAM is not time-dependent, it generates independent snapshots directly, in contrast to
a hydrodynamic model which must simulate a large number of intermediate states. For statistical analysis of cloud fields like what is shown in the main text, this is an advantage of STEAM rather than a limitation, because independent snapshots are most useful for computing statistics. For other cases such as weather forecasting, the fact that STEAM is not time-dependent becomes a limitation.

CM1 serves as the reference
host because its exact RCEMIP release is public. The measurement is
archived in the analysis repository (\texttt{cm1-cost-comparison/})
and reproduces from two scripts.
The RCEMIP CM1 submission used CM1 release 19.6
\citep{wing2020}. We build that release unmodified except
for one patch, which uncomments the values for 
insolation, the ozone profile, and trace gas concentrations that were used for RCEMIP
\citep{wing2018}. The model is then configured as the
\texttt{RCE\_large300} channel of Section~\ref{ssect:comparison data},
following the archived CM1 model documentation: $2016 \times 134
\times 74$ points at 3~km spacing, the \citet{wing2018} vertical
grid, RRTMG radiation, Morrison microphysics, and the Bryan--Rotunno
boundary-layer scheme. We integrate one model hour at a fixed 18~s
time step, the measured average step of the production run (CM1 model documentation form, RCEMIP archive), on all 16 cores.

One model hour of the channel takes 1058~s of wall time. This rate
corresponds to 7.1~hours per model day, or $2.5\times 10^{6}$~s
(29~days) for the 100-day production integration. The measurement
starts from the quiescent analytic sounding, which favors CM1
slightly given that clouds have yet to form. 
Active microphysics in a convecting state would add a few
percent to the per-timestep cost. Radiation is computed on a fixed model-time interval and contributes
158~s per model hour independent of the time step.

It is not clear whether the most fair comparison between CM1 and STEAM would
be for matched horizontal resolutions. In the main text, we argued that hydrodynamic
simulations should be first coarsened by at least a factor of 2 to account for 
numerical artifacts, while STEAM does not. The matched STEAM configuration
used in the main text ($1024 \times 64 \times 78$; Section~\ref{ssect:comparison data})
takes 1.7~s on the same 16 cores. A realization instead simulated at the host's native
3~km spacing ($2048 \times 128 \times 115$, more grid points than the
CM1 grid itself) takes 3.4~s. Both figures are CPU-only. 

We estimate the order-of-magnitude decorrelation timescale for RCEMIP by dividing the channel length by the mean wind speed at 10~km.
For the three CM1 snapshots considered here, the 10~km mean for $u$ is $3.8$~m/s, indicating a decorrelation time of about 18 days. This timescale is generous given that \citet{wing2018} found that the largest convective clusters took approximately 40 days to develop.
Assuming independent snapshots are separated by this timescale,
the host cost comes to $4.6\times
10^{5}$~s per independent field. Each STEAM realization is an independent field
by construction. The resulting ratio is $2.6\times 10^{5}$ at the
paper's matched resolution, or $1.4\times 10^{5}$ at the host's own
grid spacing. These ratios double if the 40-day spinup period is used instead.
Overall, we estimate the computational advantage per simulated independent snapshot to be between approximately
$10^5$ and $10^6$.

\setlength{\bibhang}{0in}
\bibliographystyle{ametsocV6}
\bibliography{sources}

\end {document}